\documentclass[zpreprint,zbstnp]{zeus_paper}
\usepackage[english]{babel}

\newcommand{\ZcoosysB}{%
The ZEUS coordinate system is a right-handed Cartesian system, with the $Z$
axis pointing in the proton beam direction, referred to as the ``forward
direction'', and the $X$ axis pointing left towards the centre of HERA.
The coordinate origin is at the nominal interaction point.\xspace}

\newcommand{\ZcoosysfnB}{\footnote{\ZcoosysB}}

\newcommand{\Zctddesc}[1]{%
Charged particles are tracked in the central tracking detector (CTD)~\citeCTD,
which operates in a magnetic field of $1.43\Tesla$ provided by a thin 
superconducting coil. The CTD consists of 72~cylindrical drift chamber 
layers, organised in nine~superlayers covering the polar-angle#1 region 
\mbox{$15^\circ<\theta<164^\circ$}. The transverse-momentum resolution for
full-length tracks is $\sigma(p_T)/p_T=0.0058p_T\oplus0.0065\oplus0.0014/p_T$,
with $p_T$ in $\Gev$.}

\chardef\usc=95
\chardef\til=126
\catcode`\@=11 
\DeclareRobustCommand\xdotspace{\futurelet\@let@token\@xdotspace}
\def\@xdotspace{%
  \ifx\@let@token.\else
  \ifx\@let@token\bgroup.\else
  \ifx\@let@token\egroup.\else
  \ifx\@let@token\/.\else
  \ifx\@let@token\ .\else
  \ifx\@let@token~.\else
  \ifx\@let@token!.\else
  \ifx\@let@token,.\else
  \ifx\@let@token:.\else
  \ifx\@let@token;.\else
  \ifx\@let@token?.\else
  \ifx\@let@token/.\else
  \ifx\@let@token'.\else
  \ifx\@let@token).\else
  \ifx\@let@token-.\else
  \ifx\@let@token\@xobeysp.\else
  \ifx\@let@token\space.\else
  \ifx\@let@token\@sptoken.\else
   .\space
   \fi\fi\fi\fi\fi\fi\fi\fi\fi\fi\fi\fi\fi\fi\fi\fi\fi\fi}
\catcode`\@=12 

\newcommand{\stru}[2]{%
   \relax\ifmmode\hbox{\vrule height#1 depth#2 width0pt}%
   \else\vrule height#1 depth#2 width0pt\fi}

\newcommand{\Ronum}[1]{\uppercase\expandafter{\romannumeral#1}}
\newcommand{\ronum}[1]{\expandafter{\romannumeral#1}}
\DeclareRobustCommand{\LaTeXZ}{%
  \LaTeX\kern-.05em4\kern-.1em
  {\raisebox{-0.2ex}{$\scriptstyle\text{ZEUS}$}}\xspace}

\DeclareMathAlphabet{\mathbf}{OT1}{cmr}{bx}{sl}
\newcommand{\eVdist}{\kern-0.06667em}

\newcommand{\Gev}{{\text{Ge}\eVdist\text{V\/}}}

\newcommand{\Tesla}{\,\text{T}}

\newcommand{\slashfrac}[2]{%
  \raisebox{0.5ex}{\ensuremath #1}\kern-0.12em/\kern-0.08em
  \raisebox{-.8ex}{\ensuremath #2}}

\newcommand{\sqr}[3]{%
    {\vcenter{\hrule height.#3ex\hbox{\vrule width.#2ex height#1ex
     \kern#1ex\vrule width.#3ex}\hrule height.#2ex}}}

\catcode`\@=11 
\newcommand{\parenbar}{\mathpalette\p@renb@r}
\def\p@renb@r#1#2{\vbox{%
  \ifx#1\scriptscriptstyle \dimen@.7em\dimen@ii.2em\else
  \ifx#1\scriptstyle \dimen@.8em\dimen@ii.25em\else
  \dimen@1em\dimen@ii.4em\fi\fi \offinterlineskip
  \ialign{\hfill##\hfill\cr
    \vbox{\hrule width\dimen@ii}\cr
    \noalign{\vskip-.3ex}%
    \hbox to\dimen@{$\mathchar300\hfil\mathchar301$}\cr
    \noalign{\vskip-.3ex}%
    $#1#2$\cr}}}
\catcode`\@=12 

\newcommand{\IP}{{\rm I$\kern-0.01667em$P}\xspace}

\newcommand{\tot}{{\rm tot}}

\mathchardef\qsm=63
\mathchardef\pls=43
\mathchardef\mns=512
\mathchardef\plm=518
\mathchardef\eql=61
\mathchardef\smallleft=300
\mathchardef\smallright=301
\mathchardef\les=316
\mathchardef\gre=318
\mathchardef\leq=532
\mathchardef\grq=533

\catcode`\@=11 
\newcounter{pict@width}
\newcounter{pict@height}
\newlength{\pict@scale}
\newcommand{\psfigadd}[4]{%
\setcounter{pict@width}{1*\ratio{#2+\pict@scale/2}{\pict@scale}}
\setcounter{pict@height}{1*\ratio{#3+\pict@scale/2}{\pict@scale}}
\setlength{\unitlength}{\pict@scale}
\hbox to #2{\hspace{-\fill}\begin{picture}(\thepict@width,\thepict@height)
\put(0,0){\psfig{figure=#1,width=#2,height=#3,clip=}}
\SetScale{0.283466457}
\SetWidth{1.763889}
{#4}
\end{picture}}
}
\newcounter{pict@widthfst}
\newcounter{pict@widthscd}
\newcounter{pict@widthtot}
\newcommand{\psfigaddtwo}[7]{%
\setcounter{pict@widthfst}{1*\ratio{#2+\pict@scale/2}{\pict@scale}}
\setcounter{pict@widthscd}{1*\ratio{#2+#4+\pict@scale/2}{\pict@scale}}
\setcounter{pict@widthtot}{1*\ratio{#2+#4+#6+\pict@scale/2}{\pict@scale}}
\setcounter{pict@height}{1*\ratio{#3+\pict@scale/2}{\pict@scale}}
\setlength{\unitlength}{\pict@scale}
\hbox{\hspace{-\fill}\begin{picture}(\thepict@widthtot,\thepict@height)
\put(0,0){\psfig{figure=#1,width=#2,height=#3,clip=}}
\put(\thepict@widthscd,0){\psfig{figure=#5,width=#6,height=#3,clip=}}
\SetScale{0.283466457}
\SetWidth{1.763889}
{#7}
\end{picture}}
}
\newcommand{\psfigror}[4]{%
\setcounter{pict@width}{1*\ratio{#2+\pict@scale/2}{\pict@scale}}
\setcounter{pict@height}{1*\ratio{#3+\pict@scale/2}{\pict@scale}}
\setlength{\unitlength}{\pict@scale}
\hbox{\begin{picture}(\thepict@width,\thepict@height)
\put(0,\thepict@height){\psfig{figure=#1,width=#3,height=#2,clip=,angle=270}}
\SetScale{0.283466457}
\SetWidth{1.763889}
{#4}
\end{picture}}
}
\newcommand{\psfigrol}[4]{%
\setcounter{pict@width}{1*\ratio{#2+\pict@scale/2}{\pict@scale}}
\setcounter{pict@height}{1*\ratio{#3+\pict@scale/2}{\pict@scale}}
\setlength{\unitlength}{\pict@scale}
\hbox{\begin{picture}(\thepict@width,\thepict@height)
\put(0,0){\psfig{figure=#1,width=#3,height=#2,clip=,angle=90}}
\SetScale{0.283466457}
\SetWidth{1.763889}
{#4}
\end{picture}}
}
\catcode`\@=12 
\newlength\listtextwidth

\catcode`\@=11 
\newlength{\@tabfninsert}
\newlength{\@tabfnwidth}
\newcommand{\tabfootnote}[2]{%
  \setlength{\@tabfninsert}{0.8em}
  \setlength{\@tabfnwidth}{\textwidth}
  \addtolength{\@tabfnwidth}{-\@tabfninsert}
  \addtolength{\@tabfnwidth}{-0.4em}
  \noindent\makebox[\@tabfninsert][r]{\footnotesize$^{#1}$\hfil}\hfill%
  \parbox[t]{\@tabfnwidth}{\footnotesize #2\hfill}}
\catcode`\@=12 

\newcommand {\pom} {I\!\!P}
\newcommand {\reg} {I\!\!R}
\newcommand {\xpom} {x_{\pom}}

\def\citeCTD{{\cite{%
nim:a279:290,*npps:b32:181,*nim:a338:254%
}}\xspace}
\def\citeCAL{{\cite{%
nim:a309:77,*nim:a309:101,*nim:a321:356,*nim:a336:23%
}}\xspace}

\def\citediff0{{\cite{%
zeusdiff0,*h1diff0%
}}\xspace}

\def\bpc1{{\cite{%
bpc01,*bpc02,*bpc03%
}}\xspace}

\begin{document}
\prepnum {DESY-04-131}
\def\lsim{\raisebox{-.65ex}{\rlap{$\sim$}} \raisebox{.4ex}{$<$}}
\def\gsim{\raisebox{-.65ex}{\rlap{$\sim$}} \raisebox{.4ex}{$>$}}

\title{
Dissociation of virtual photons in events with a leading proton at HERA
}                                                       
                    
\author{ZEUS Collaboration}

\abstract{
The ZEUS detector has been used to study
dissociation of virtual photons in events with a leading proton, 
$\gamma^{\star} p \to X p$, in $e^+p$ collisions at HERA. 
The data cover photon virtualities in two ranges, 
$0.03<Q^2<0.60$~GeV$^2$ and $2<Q^2<100$~GeV$^2$, with
$M_X>1.5$ GeV, where $M_X$ is the mass of the hadronic final state, 
$X$. Events were required to have a leading proton, detected in the ZEUS 
leading proton spectrometer, carrying at least 90\% 
of the incoming proton energy. The cross section is presented as a function of 
$t$, the squared four-momentum transfer at the proton vertex, $\Phi$, the 
azimuthal angle between the positron scattering plane and the proton 
scattering plane, and $Q^2$. The data are presented in terms of the 
diffractive structure function, $F_2^{D(3)}$. A next-to-leading-order QCD 
fit to the higher-$Q^2$ data set and to previously published diffractive
charm production data is presented.
}

\makezeustitle

\def\3{\ss}                                                                                        
\pagenumbering{Roman}                                                                              
                                                   %
\begin{center}                                                                                     
{                      \Large  The ZEUS Collaboration              }                               
\end{center}                                                                                       
  S.~Chekanov,                                                                                     
  M.~Derrick,                                                                                      
  J.H.~Loizides$^{   1}$,                                                                          
  S.~Magill,                                                                                       
  S.~Miglioranzi$^{   1}$,                                                                         
  B.~Musgrave,                                                                                     
  \mbox{J.~Repond},                                                                                
  R.~Yoshida\\                                                                                     
 {\it Argonne National Laboratory, Argonne, Illinois 60439-4815}, USA~$^{n}$                       
\par \filbreak                                                                                     
  M.C.K.~Mattingly \\                                                                              
 {\it Andrews University, Berrien Springs, Michigan 49104-0380}, USA                               
\par \filbreak                                                                                     
  N.~Pavel \\                                                                                      
  {\it Institut f\"ur Physik der Humboldt-Universit\"at zu Berlin,                                 
           Berlin, Germany}
% ...
\newpage                                                                                           
$^{\    1}$ also affiliated with University College London, UK \\                                  
$^{\    2}$ retired \\                                                                             
% ...
 \par         
                                                           %
                                                           %
\begin{tabular}[h]{rp{14cm}}                                                                       
$^{a}$ &  supported by the Natural Sciences and Engineering Research Council of Canada (NSERC) \\  
$^{b}$ &  supported by the German Federal Ministry for Education and Research (BMBF), under        
          contract numbers HZ1GUA 2, HZ1GUB 0, HZ1PDA 5, HZ1VFA 5\\                                
$^{c}$ &  supported in part by the MINERVA Gesellschaft f\"ur Forschung GmbH, the Israel Science   
          Foundation (grant no. 293/02-11.2), the U.S.-Israel Binational Science Foundation and    
          the Benozyio Center for High Energy Physics\\                                            
$^{d}$ &  supported by the German-Israeli Foundation and the Israel Science Foundation\\           
$^{e}$ &  supported by the Italian National Institute for Nuclear Physics (INFN) \\                
$^{f}$ &  supported by the Japanese Ministry of Education, Culture, Sports, Science and Technology 
          (MEXT) and its grants for Scientific Research\\                                          
$^{g}$ &  supported by the Korean Ministry of Education and Korea Science and Engineering          
          Foundation\\                                                                             
$^{h}$ &  supported by the Netherlands Foundation for Research on Matter (FOM)\\                   
$^{i}$ &  supported by the Polish State Committee for Scientific Research, grant no.               
          620/E-77/SPB/DESY/P-03/DZ 117/2003-2005\\                                                
$^{j}$ &  partially supported by the German Federal Ministry for Education and Research (BMBF)\\   
$^{k}$ &  supported by RF President grant N 1685.2003.2 for the leading scientific schools and by  
          the Russian Ministry of Industry, Science and Technology through its grant for           
          Scientific Research on High Energy Physics\\                                             
$^{l}$ &  supported by the Spanish Ministry of Education and Science through funds provided by     
          CICYT\\                                                                                  
$^{m}$ &  supported by the Particle Physics and Astronomy Research Council, UK\\                   
$^{n}$ &  supported by the US Department of Energy\\                                               
$^{o}$ &  supported by the US National Science Foundation\\                                        
$^{p}$ &  supported by the Polish Ministry of Scientific Research and Information Technology,      
          grant no. 112/E-356/SPUB/DESY/P-03/DZ 116/2003-2005\\                                    
\end{tabular}                                                                                      
                                                           %
                                                           %

\pagenumbering{arabic} 
\pagestyle{plain}

\section{Introduction}
\label{sec-int}

In diffractive processes in hadron-hadron or photon-hadron collisions, the
initial state particles undergo a ``peripheral"
collision, in which either the particles stay intact (elastic
scattering), or they dissociate into low-mass states (diffractive 
dissociation).
The scattered hadrons (or the low-mass states in the dissociative case) 
have energy equal, to within a few per cent, to that of the incoming 
hadron, and very small transverse momentum. Diffractive 
interactions can be parameterised in the framework of Regge 
phenomenology, where they are ascribed to the exchange of a trajectory 
with the vacuum quantum numbers, the Pomeron 
trajectory~\cite{regge}. In the same framework, events in which the proton 
loses a more substantial fraction of its energy are ascribed to the 
exchange of subleading trajectories. 

Significant progress has recently been made in understanding diffraction 
in terms of QCD, notably by studying the diffractive dissociation
of virtual photons in electron-proton or positron-proton collisions at 
HERA. In fact, 
diffraction has proven to be a tool to study QCD and the low-$x$ 
structure of the proton~\cite{recent_review, rev}.
In the proton's rest frame, diffractive $ep$ scattering, $ep
\rightarrow eXp$, proceeds 
from the fluctuation of the virtual photon emitted by the electron (or 
by the positron) into a 
colour dipole, such as a quark-antiquark pair or a quark-antiquark-gluon 
system. The dipole interacts 
hadronically with the proton via the exchange of an object with 
vacuum quantum numbers -- a gluon pair, in leading-order QCD -- and 
then 
dissociates into the hadronic state, $X$. The dipole has transverse 
dimensions which decrease as the photon virtuality, $Q^2$, increases. It is 
thus possible to study diffractive interactions in a regime 
where one of the two interacting hadrons is so small that the strong interaction 
can be treated perturbatively. Alternatively, in a frame in which the proton
is fast, the reaction can be seen as the deep inelastic scattering (DIS) of 
a pointlike virtual photon off the exchanged object. This gives access to 
the diffractive parton distribution functions (PDF),
for which a QCD factorisation theorem has been 
proven~\cite{trentadue,qcdf,*qcdf1,berera}. Diffractive 
PDFs are defined as the proton PDFs probed when the vacuum quantum numbers 
are exchanged and the proton emerges intact from 
the interaction, suffering only a small energy loss. 
In the context of QCD, Pomeron exchange should then be understood  as a 
synonym for exchange  of partons from the proton with the vacuum quantum numbers.

This paper presents new measurements of the reaction $e^+p 
\rightarrow e^+Xp$ in the regions $0.03 < Q^2 <0.60$~GeV$^2$ (low-$Q^2$ 
sample) and $2<Q^2 <100$ GeV$^2$ (high-$Q^2$ sample)~\footnote{In the following, 
for simplicity, the symbol $e$ will be used to 
denote both electrons and positrons.}. 
The measurements were made using the ZEUS detector at the HERA $ep$ collider. 
The events were selected by requiring the detection 
in the ZEUS leading proton spectrometer (LPS) of a scattered proton,
carrying a fraction $x_L$  of the incoming proton momentum of at least 
$0.9$; such fast protons are referred to as leading. This $x_L$ 
range includes the so-called diffractive peak, the 
narrow peak in the cross section at $x_L \approx 1$ ascribed to 
Pomeron exchange, as well as the transition to the lower $x_L$ region in 
which subleading, mesonic 
exchanges, notably Reggeons, are thought to dominate~\cite{regge,rev}. 
The measurement covers the region $0.075 <|t| <0.35$~GeV$^2$,
where $t$ is the square of the four-momentum transferred at the proton vertex.

Sections~\ref{sec-kin}-\ref{sec-sys} present the experimental set-up and 
the 
details of the analysis. Section~\ref{sec-res} gives the results. 
The $t$ and $\Phi$ dependences of the cross section are discussed first, 
where $\Phi$ is the azimuthal angle between the positron and the proton 
scattering planes in the $\gamma^*p$ rest frame. The distribution of 
$\Phi$ is sensitive to the interference between the amplitudes for scattering 
of longitudinally and transversely polarised photons, and thus to the helicity 
structure of the interaction. The $Q^2$ dependence of the photon-proton 
differential cross-section $d\sigma^{\gamma^*p \rightarrow Xp}/dM_X$ is then 
studied for different values of the photon-proton centre-of-mass 
energy, $W$. The data are also discussed in terms of the diffractive structure 
function, $F_2^{D(3)}$. The dependence of $F_2^{D(3)}$ on $\xpom$ is 
studied, where $\xpom$ is the fraction of the proton momentum carried by
the object (the Pomeron or the Reggeon, in the Regge framework)
exchanged between the virtual photon and the proton, $\xpom \simeq 1-x_L$. 
The $Q^2$ and $\beta$ dependences of $F_2^{D(3)}$ for different values of 
$\xpom$ are investigated, and the behaviour of $F_2^{D(3)}$ is compared to 
that of the inclusive proton structure function, $F_2$. The variable $\beta$  
is the Bjorken variable defined with respect to the four-momentum of the 
exchanged object. The results are compared to theoretical predictions 
based on the colour-dipole approach outlined above. Finally, a 
next-to-leading-order (NLO) QCD fit to the higher-$Q^2$ data is presented.

The present data correspond to an integrated luminosity about a factor 
four larger than that of the previous ZEUS-LPS analysis~\cite{lps95}. The 
low-$Q^2$ results have a wider $Q^2$ and $W$ coverage than that studied so 
far with the LPS or other methods~\cite{lps95,low-xl}. The $\Phi$ distribution 
is investigated for the first time. The cross section is measured up 
to $M_X$ values of 40~GeV, so far unexplored, and $F_2^{D(3)}$ is 
presented up to $\xpom$ values of 0.07, thus covering the diffractive-peak 
region ($\xpom~\lsim~0.02$) 
and the transition to the non-diffractive 
region that is dominated by subleading exchanges.

It is also possible to select diffractive events  without detecting
the scattered proton. In a previous paper~\cite{zeusdiff}, a method  
based on features of the shape of the mass spectrum of 
the hadronic final-state $X$ ($M_X$ method) was applied. A discussion of 
the two approaches is presented, along with a comparison of the 
corresponding results.

\section{Kinematics and cross sections}
\label{sec-kin}
\begin{figure}[p]
\vfill
\begin{center}
\includegraphics[width=9cm,height=9cm]{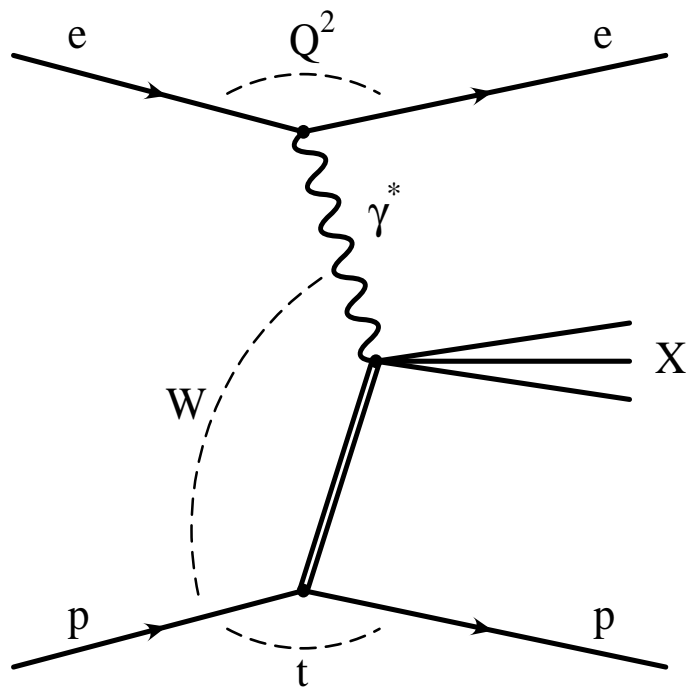}
\end{center}
\vspace{3.0cm}
\caption{
Schematic diagram of the reaction $ep \rightarrow eXp$.
}
\label{fig-contfey}
\vfill
\end{figure}

Figure~\ref{fig-contfey} shows a schematic diagram of the  
process
$ep \to e X p$. The kinematics of this reaction is described by
the variables:
\begin{itemize}
\item
$Q^2=-q^2=-(k-k')^2$, the negative four-momentum squared of the virtual 
photon, where $k$ $(k')$ is the four-momentum of the incident (scattered) 
positron;
\item
$W^2=(q+P)^2$, the squared centre-of-mass energy of the photon-proton system, 
where $P$ is the four-momentum of the incident proton;
\item 
$x=Q^2/(2P\cdot q)$, the fraction of the proton 
momentum carried by the quark struck by the virtual photon in the 
infinite momentum frame (the Bjorken variable);
\item
$M^2_X=(q+P-P')^2$, the squared mass of the system $X$,
where $P'$ is the four-momentum of the scattered proton;
\item
$t=(P-P')^2$, the squared four-momentum transfer at the proton vertex;
\item
$\Phi$, the angle between the positron scattering plane and the proton
scattering plane in the $\gamma^{\star}p$ centre-of-mass frame.
\end{itemize}

\noindent The variables $Q^2$, $W$ and $x$ are related by 
$x=Q^2/(Q^2+W^2-M_p^2)$, where $M_p$ is the proton mass.

The differential cross sections for the reactions
$ep \to eXp$ and  $\gamma^{\star}p \to Xp$ are related by
\begin{eqnarray}
\frac{d\sigma^{ep \rightarrow eXp}}{d\ln{W^2} dQ^2dM_{X}dt} 
&=& 
\frac{\alpha}{2\pi} \frac{1+(1-y)^2}{Q^2} 
\frac{d(\sigma_T^{\gamma^{\star}p\rightarrow Xp}+\sigma_L^{\gamma^{\star}p\rightarrow Xp})}{dM_Xdt}
\left[ 1-\frac{y^2}{1+(1-y)^2}\frac{R^D}{1+R^D}\right]
\label{sigma-0} \nonumber\\
&\simeq& 
\frac{\alpha}{2\pi} \frac{1+(1-y)^2}{Q^2} 
\frac{d\sigma^{\gamma^{\star}p\rightarrow Xp}}{dM_{X}dt},
\label{sigma-1}
\end{eqnarray}
where $\alpha$ is the fine structure constant and $y=(P\cdot q)/(P\cdot 
k)$ is the fraction of the positron energy transferred to the proton in its 
rest frame. The quantity
$R^D= \sigma_L^{\gamma^* p \rightarrow Xp}/\sigma_T^{\gamma^* p 
\rightarrow Xp}$ 
is the ratio of the cross sections for longitudinally and transversely 
polarised virtual photons. In the region covered by the present data, the 
term within the square brackets is taken to be unity
since $R^D$ is expected to be small~\cite{rsmall}.

The kinematics of the reaction $ep \to eXp$ can also be described by 
$Q^2$, $t$ and $\Phi$, in conjunction with the two
dimensionless variables $x_{\pom}$ and $\beta$ introduced in 
Section~\ref{sec-int} given by
\begin{equation}
x_{\pom}=\frac{(P-P')\cdot q}{P\cdot q} = \frac{Q^2+M_X^2-t}{Q^2+W^2-M_p^2},
\label{eq-xpom}
\end{equation}
\begin{equation}
\beta=\frac{Q^2}{2(P-P')\cdot q} = \frac{Q^2}{Q^2+M_X^2-t}.
\label{eq-beta}
\end{equation}
The quantities $x_{\pom}$ and $\beta$ are related to $x$ by 
$x_{\pom} \beta = x$.

The cross section for the reaction $ep \to eXp$ can be
expressed in terms of the structure function $F_2^{D(4)}$,
which is defined by the equation
\begin{eqnarray}
\frac{d\sigma^{ep \rightarrow eXp}}{
d\beta dQ^2dx_{\pom}dt} & = &
\frac{4\pi\alpha^2}{\beta Q^4}\biggl[1-y+\frac{y^2}{2(1+R^{D})}\biggr]
F_2^{D(4)}(\beta,Q^2,x_{\pom},t)  \nonumber \\
& \simeq &
\frac{4\pi\alpha^2}{\beta Q^4}\biggl[1-y+\frac{y^2}{2}\biggr]
F_2^{D(4)}(\beta,Q^2,x_{\pom},t).
\label{sigma-2}
\end{eqnarray}

\noindent
The structure function $F_2^{D(3)}(\beta, Q^2,\xpom)$ is obtained by 
integrating $F_2^{D(4)}$ over $t$:

\begin{equation}
F_2^{D(3)}(\beta, Q^2,\xpom)=\int{F_2^{D(4)}(\beta,Q^2,x_{\pom},t) dt}.\nonumber
\label{f2d3}
\end{equation}
\noindent
For the results presented in this paper, the integration was performed in 
the range $0<|t|<1$~GeV$^2$.

For unpolarised positrons and protons, the cross section can
also be decomposed as
\begin{equation}
\frac{d\sigma^{ep\rightarrow eXp}}{d\Phi} \propto 
\sigma_T^{\gamma^*p \rightarrow Xp}+
\epsilon\sigma_L^{\gamma^*p \rightarrow Xp}-
2\sqrt{\epsilon(1+\epsilon)}\sigma^{\gamma^*p \rightarrow Xp}_{LT}\cos{\Phi}-
\epsilon\sigma^{\gamma^*p \rightarrow Xp}_{TT}\cos{2\Phi},
\label{fullsigma}
\end{equation}
where $\sigma^{\gamma^*p \rightarrow Xp}_{LT}$ is the interference term 
between the amplitudes for longitudinal and transverse
photons and  $\sigma^{\gamma^*p \rightarrow Xp}_{TT}$ is the interference 
term between the amplitudes for the two transverse polarisations. The 
polarisation parameter 
$\epsilon$ is defined as $\epsilon = 2(1-y)/[1+(1-y)^2]$.

\section{Selection of diffraction at HERA}

The kinematics of diffractive scattering, $\gamma^*p \rightarrow Xp$, 
implies that three features should be present in the final state:

\begin{enumerate}

\item the proton suffers only a small perturbation and emerges from the 
interaction carrying a large fraction, $x_L$, of the incoming proton momentum. 
Diffractive events appear as a peak at $x_L\approx 1$, the
diffractive peak, which at HERA approximately covers the region 
$0.98<x_L<1$. 
The absolute value of the four-momentum transfer squared $t$ is typically much 
smaller than 1 GeV$^2$, with $\langle |t| \rangle \approx 0.15$~GeV$^{2}$;

\item conservation of momentum implies that any other produced system 
($X$) must have a small mass ($M_X$) with respect to the photon-proton
centre-of-mass energy (since $1-x_L~\gsim~M_X^2/W^2$);

\item the difference in rapidity between the outgoing proton and the 
system $X$ is $\Delta \eta \approx \ln{(1/\xpom)}$~\cite{rev}. This, 
combined with the peaking of the cross section at small values of $\xpom$, 
leads to a large separation in rapidity between  
the outgoing proton  and any other hadronic activity in the event.

\end{enumerate}

There are two basic ways to select inclusive diffractive events. The
first is the proton-tagging method (exploiting the first signature
above), used in the present study and in earlier ones~\cite{lps94a, 
lps94b, lps95, low-xl, h1fps}.
The second exploits the different characteristics of the system $X$ in 
diffractive and non-diffractive events:
\begin{itemize}
\item in non-diffractive DIS, both the hadronic system associated
with the struck quark, which is largely measured in the
detector, and that of the proton remnant, which largely escapes
down the beam-pipe, are coloured states.  In this case,
the distribution of the final-state particles is governed by conventional
QCD fragmentation and particles are expected to be emitted uniformly
in rapidity along the $\gamma^*$-$p$ axis. This leads to
a suppression of rapidity gaps as well as a suppression of small masses of 
the hadronic system observed in the detector;

\item in contrast,  small masses of the system $X$ and large rapidity gaps
are signatures of diffractive processes (the second and third signatures 
above). At HERA, 
diffractive analyses based on the hadronic methods have been made with 
event selections based both
on the presence of large rapidity gaps (rapidity-gap method, see 
e.g.~\cite{h1diff,recent_review} and references therein) and 
small masses of the system $X$ ($M_X$ method)~\cite{zeusdiff, lps95}.

\end{itemize}
The two basic approaches for the selection of diffractive
events, the proton-tagging method and the hadronic
methods, are complementary:

\begin{itemize}
\item in the hadronic methods, high $M_X$ values are not accessible since 
the non-diffractive background grows with $M_X$. Also, the estimation of 
the non-diffractive background relies on models  
of fragmentation.  Furthermore, the measured cross section 
includes a contribution from proton-dissociative events, 
$ep \rightarrow eXN$. In these events, the mass of the forward system 
($N$) enters as another variable, and the observed particles must be 
assigned either to the system $N$ or $X$. The number of events
in which no particle from $N$ is observed must be estimated from
a Monte Carlo simulation.  While these limitations add to
the systematic uncertainty of the hadronic methods, the
statistical precision of the results tends to be good due
to the high acceptance of the central detector.  Also, the
acceptance is not limited in $t$, although no measurement of $t$ is 
possible;

\item conversely, samples selected in the proton-tagging method
have little or no background from proton-dissociative events or
from non-diffractive DIS. They also allow a direct measurement of the
variables $t$, $\Phi$ and $\xpom$ (at large values of $\xpom$), and give
access to higher values of $M_X$.  The statistical precision,
however, is poorer than for the results obtained using hadronic methods
due to the small acceptance of the LPS -- approximately 2\%
in the diffractive peak region.

\end{itemize}

Section~\ref{sec-comparison} presents a comparison between the results 
obtained with the LPS and $M_X$ methods. The results are also compared to 
measurements made by the H1 collaboration~\cite{h1diff} in which 
diffraction was selected with the rapidity-gap method.

\section{Experimental set-up}
\label{sec-exp}
The measurements were carried out at the HERA collider in 1997 using
the ZEUS detector. At that time, HERA operated with 820 GeV protons  
and 27.5 GeV positrons. The data used in this analysis
correspond to integrated luminosities of $3.60\pm0.06$~pb$^{-1}$ (low-$Q^2$
sample) and $12.8\pm0.2$ pb$^{-1}$ (high-$Q^2$ sample).

A detailed description of the ZEUS detector can be found 
elsewhere~\cite{bluebook}. A brief outline of the components that are most 
relevant for this analysis is given below.


\Zctddesc\ZcoosysfnB


The high-resolution uranium-scintillator calorimeter (CAL)~\citeCAL 
consists of three parts: the forward (FCAL), the barrel (BCAL) and the 
rear (RCAL) calorimeters. The CAL energy resolutions, as measured under 
test beam conditions, are $\sigma(E)/E=0.18/\sqrt{E}$ for electrons and 
$\sigma(E)/E=0.35/\sqrt{E}$ for hadrons ($E$ in GeV).

Low-$Q^2$ events were selected by identifying and measuring the scattered 
positron in the beam-pipe calorimeter (BPC)~\cite{f2bpc}  and beam-pipe 
tracker 
(BPT)~\cite{f2bpt}. The BPC was a
tungsten-scintillator sampling calorimeter, located 3 m from the 
interaction point and covered positron scattering angles relative to the 
incident direction of 15 to 34 mrad. The 
BPT was a silicon-microstrip tracking device situated immediately in front 
the BPC. In 1997, it was equipped with two detector planes to measure the 
$X$ coordinate.

For the high-$Q^2$ sample, the impact point of the scattered positron was 
determined with the small-angle rear tracking detector (SRTD)~\cite{srtd} 
or the CAL. The SRTD is attached to the front face of the RCAL and consists 
of two planes of scintillator strips, 1 cm wide and 0.5 cm thick, arranged 
in orthogonal orientations and read out via optical fibres and photomultiplier 
tubes. It covers a region $68 \times 68$~cm$^2$ in $X$ and $Y$, excluding a 
$10 \times 20$~cm$^2$ hole at the centre for the beam-pipe. The 
corresponding angular coverage is between $4^{\circ}$ and $18^{\circ}$
around the beam-pipe.
 Ambiguities in 
SRTD hits were resolved with the help of the hadron-electron separator 
(HES)~\cite{hes}, which consists of a layer of 10\,000, $2.89 \times 
3.05$~cm$^2$ silicon-pad detectors inserted in the CAL at a depth of 3.3 
radiation lengths. 

The LPS~\cite{lps} detected positively charged particles scattered
at small angles and carrying a substantial fraction, $x_L$, of the
incoming proton momentum; these particles remain in the beam-pipe
and their trajectory was measured by a system of silicon microstrip
detectors that could be inserted very close (typically a few mm) to
the proton beam. The detectors were grouped in six stations, S1 to S6,
placed along the beam line in the direction of the proton beam,
between 23.8 m and 90.0 m from the interaction point. The particle
deflections induced by the magnets of the proton beam-line allowed
a momentum analysis of the scattered proton. For the present
measurements, only stations S4, S5 and S6 were used. The resolutions
were about $0.5\%$ on the longitudinal momentum and about 5 MeV on
the transverse momentum. The effective transverse-momentum
resolution is dominated by the intrinsic transverse-momentum
spread of the proton beam at the interaction point,
which is about 40~MeV in the horizontal plane and about 90 MeV
in the vertical plane. The LPS acceptance is approximately 2\% and
$x_L$-independent in the diffractive-peak region, $0.98<x_L<1$; it increases 
smoothly to about 10\% as $x_L$ decreases to 0.9.

The luminosity was measured from the rate of the bremsstrahlung process 
$ep \rightarrow e\gamma p$. The photon was measured in a 
lead-scintillator calorimeter~\cite{lumi1,*lumi2,*lumi3} placed in the 
HERA tunnel at $Z=-107$~m.

\section{Reconstruction of the kinematic variables}
\label{sec-rec}
In the low-$Q^2$ analysis, $0.03 <Q^2 <0.60$~GeV$^2$, the 
scattered positron was measured in the BPC/BPT.
The energy and angle of the scattered positron were used 
(``electron method'') to determine $Q^2$ and $W$.
For the high-$Q^2$ data ($2<Q^2<100$~GeV$^2$) the identification of the 
scattered positron
was based on a neural network~\cite{sira} which uses
information from the CAL.
The variables $W$ and $Q^2$ were reconstructed using a combination of the 
electron method and the double angle
method~\cite{da}.

The longitudinal ($p_Z$) and transverse ($p_X, p_Y$) momenta of the 
scattered proton were measured with the LPS. 
The fractional energy of the outgoing proton, $x_L$, was
defined as $x_L=p_Z/E_p$, where $E_p$ is the incoming proton energy. The
variable $t$ is given by
\begin{equation}
t=-\frac{p_T^2}{x_L} -\frac{(1-x_L)^2}{x_L}M_p^2,
\label{tpt2}
\end{equation}
where $p_T$ is the transverse momentum of the proton with respect to the 
incoming beam direction. 
The $t$ resolution is approximately 
$\sigma(|t|)=0.14~{\rm GeV} \sqrt{|t|}$ ($t$ in GeV$^2$) and is 
dominated by the angular spread of the beam. The proton and the positron 
momenta were used to determine $\Phi$, the azimuthal angle between the 
positron and proton scattering planes in the $\gamma^*p$ frame. The 
resolution on $\Phi$ is approximately 0.2 rad.

The four-momentum of the system $X$ was determined from
calorimeter and tracking information. The energy deposits in the CAL
and the track momenta measured in the CTD were combined in energy
flow objects (EFOs)~\cite{zeusdiff,gennady} to obtain the best momentum
resolution. The mass $M_X$ was evaluated as
\begin{equation}
M_{X, EFO}^2=\biggl(\sum{E_i}\biggr)^2-\biggl(\sum{p_{Xi}}\biggr)^2-\biggl(\sum{p_{Yi}}\biggr)^2-\biggl(\sum{p_{Zi}}\biggr)^2,\nonumber
\end{equation}
where the sums run over all EFOs not assigned to the scattered positron.
The mass $M_X$ can also be determined from the outgoing proton momentum 
as reconstructed in the LPS:
\begin{equation}
M_{X, LPS}^2 \approx [1-x_L(1+x)] W^2. \nonumber
\end{equation}
The best resolution on $M_X$ is obtained with $M_{X, EFO}$ when $M_X$ is 
small and with $M_{X, LPS}$ for large values of $M_X$, so $M_X$ was 
reconstructed as
\begin{equation}
M^2_{X} = w_{EFO} M^2_{X,EFO} + w_{LPS} M^2_{X,LPS},  
\label{mx}
\end{equation}
where the weights $w_{EFO}$ and $w_{LPS}$ are inversely proportional to 
the appropriate resolutions, and 
$w_{EFO}+w_{LPS}=1$. The resulting 
resolution is $\sigma(M_X)/M_X=0.35/\sqrt{M_X} 
+0.08$, with $M_X$ in GeV. 

The variables $\xpom$ and $\beta$ were obtained from 
Eqs.~(\ref{eq-xpom}) and~(\ref{eq-beta}), using  the value of $M_X$ from 
Eq.~(\ref{mx}). For the determination of $\xpom$, this procedure
 is equivalent, at large $M_X$, to evaluating $\xpom$ 
as $\xpom=1-x_L$.

The variable $y$ was reconstructed 
as $y_{JB}= \sum{\frac{(E_i-p_{Zi})_{\rm had}}{2E_e} }$, where the 
sum is over all EFOs not associated to the scattered positron and $E_e$ is 
the energy of the incident positron (``Jacquet-Blondel method''~\cite{jb}).

\section{Event selection}
\label{sec-sel}
The data used for the analysis were selected at the trigger level
by requiring the presence of a scattered positron in the BPC or CAL
and a scattered proton in the LPS. 
In the offline selection the following cuts were imposed, closely 
following  those used in the $F_2$ analyses at low $Q^2$~\cite{f2bpt} and 
high $Q^2$~\cite{f2dis}:
\begin{itemize}
\item
the energy of the  scattered positron, if measured in the BPC, was 
required to be between  3 and 7 GeV for the events with $W>260$ GeV and 
greater than 7 GeV for $W<260$ GeV, reflecting the trigger selection. If 
measured in the CAL, the scattered positron energy was required to be larger 
than 10 GeV.

The position of the scattered positron was required to be within the 
fiducial regions of the BPC or the CAL.
In addition, in case the positron was found in the BPC, 
the impact position  at 
the BPC front face, as extrapolated from the BPT measurement, was required 
to match with that of the BPC shower. Furthermore, in order to identify 
electromagnetic showers and to reject hadrons, the transverse size (energy 
weighted root mean square) of the shower in the BPC was required to be 
less than 0.8~cm;

\item
the requirements $30<(E-p_Z)<65$ GeV and $40<(E-p_Z)<65$ 
GeV  were imposed for 
the low- and high-$Q^2$ samples, respectively, 
where $E-p_Z = \sum{(E_i-p_{Zi})}$ 
and the summation runs over the energies and longitudinal momenta of 
the final-state
positron and all EFOs.
This cut reduces the size of the QED radiative corrections and the
photoproduction background, i.e. the $Q^2 \approx 0$ events where the 
scattered positron escapes undetected in the rear beam-hole;
\item
in order to limit event migrations from low $y$, the variable $y_{JB}$ was
required to be greater than 0.06;
\item
the $Z$ coordinate of the interaction vertex was required to be in the 
range
$-90<Z<90$~cm for the low-$Q^2$ sample and  
$-50<Z<50$ cm for the high-$Q^2$ sample. 
Events without a vertex reconstructed using BPT tracks were discarded 
in the low-$Q^2$ analysis. In the high-$Q^2$ sample, events without a 
measured vertex were assigned to the nominal interaction point.

\end{itemize}

The following requirements were used to select 
the scattered proton measured in the LPS:
\begin{itemize}
\item
the candidate proton was tracked along the beam 
line and was rejected if, at any point, the distance of approach to the beam 
pipe was less than 0.2 cm for $p_X<0$, or less than 0.3 cm for $p_X>0$.
This cut reduces the sensitivity of the acceptance to the uncertainty in 
the position of the beam-pipe apertures;
\item
the variable $t$ was required to be in the range $0.075<|t|<0.35$~GeV$^2$. 
This cut eliminated regions where the LPS acceptance was small or rapidly 
changing;

\item
beam-halo background is caused by  scattered protons, with energy close to
that of the beam, originating from the interaction of a beam proton with 
the residual gas in the beam-pipe or with the beam collimators. 
A beam-halo proton may overlap with a standard non-diffractive DIS event.
In this case,
the proton measured in the LPS is uncorrelated with the activity in the
central ZEUS detector. This background was suppressed by the requirement 
that the sum of the energy and the longitudinal component of the total
momentum measured in the CAL, the BPC and the LPS be less than the 
kinematic limit of twice the incoming proton energy:
$E+p_Z=(E+p_Z)_{\rm CAL}+ (E+p_Z)_{\rm BPC} +2 p_Z^{\rm LPS}
<1655$~GeV. This cut takes into
account the resolution of the measurement of $p_Z^{\rm LPS}$.
The residual beam-halo background and its subtraction are discussed in 
Section~\ref{sec-bkg}.
\end{itemize}

The low-$Q^2$ analysis was further limited to the kinematic region $0.03<Q^2<0.6$ 
GeV$^2$, $63<W<280$ GeV, $M_X>1.5$ GeV and $x_L>0.9$. The average $Q^2$ 
value for this sample is 0.23 GeV$^2$. The high-$Q^2$ analysis was restricted 
to $2<Q^2<100$ GeV$^2$, $25<W<240$~GeV, $M_X>1.5$ GeV and $x_{\pom}<0.07$; 
the average $Q^2$ value is 10.5 GeV$^2$.  These selections 
yielded 334 events in the low-$Q^2$ sample and 5945 events in the 
high-$Q^2$ sample.

\section{Monte Carlo simulation and acceptance corrections}
\label{sec-mc}
Monte Carlo simulations were used to correct the data for 
acceptance and detector effects. In the low-$Q^2$ analysis, events of the 
type  $ep \to eXp$ 
were simulated with the generator EPSOFT 2.0~\cite{epsoft2,epsoft3}, 
based on the triple-Regge formalism~\cite{regge,rev}, in which the 
cross section can be expressed in terms of three trajectories.
If all the trajectories are Pomerons ($\pom \pom \pom$),
the cross-section $d\sigma^{\gamma^*p \rightarrow Xp}/dM_{X}^{2}$
is approximately proportional to $1/M_{X}^{2}$.
If one of the trajectories is a Reggeon ($\pom \pom \reg$),
the cross-section $d\sigma^{\gamma^*p \rightarrow Xp}/dM_{X}^{2}$ falls as 
$\sim 1/M_{X}^{3}$.

In the high-$Q^2$ analysis, the reaction $ep \to eXp$ was
modelled with RAPGAP 2.08/06~\cite{rapgap}, which is based on the
model of Ingelman and Schlein~\cite{ingelman}. In RAPGAP,
the structure function $F_2^{D(4)}$ is expressed
as the sum of Pomeron and Reggeon contributions:
\begin{equation}
F_2^{D(4)}(x_{\pom},t,\beta,Q^2)=f_{\pom}(x_{\pom},t)F_2^{\pom}(\beta,Q^2)+f_{\reg}(x_{\pom},t)F_2^{\reg}(\beta,Q^2).
\label{i-s}
\end{equation}
The Pomeron and Reggeon fluxes, $f_{\pom, \reg}(x_{\pom},t)$, 
were parameterised~\cite{regge} as
\begin{equation}
f_{\pom, \reg}(x_{\pom},t)=\frac{e^{b_0^{\pom, 
\reg}t}}{x_{\pom}^{2\alpha_{\pom,\reg}(t)-1}},
\label{flux}
\end{equation}
with linear trajectories
$\alpha_{\pom,R}(t)=\alpha_{\pom, \reg}(0)+\alpha_{\pom, \reg}'t$, and
the values of the parameters were taken from hadron-hadron
data. The Pomeron structure function $F_2^{\pom}(\beta,Q^2)$ was taken 
from the H1 measurements~\cite{h1diff} (fit 2).
The Reggeon trajectory 
includes the $\rho$, the $\omega$, the $f_2$ and $a_2$ mesons. 
Their structure functions are unknown and were approximated with that of the 
pion~\cite{pion}.
The assumption that $F_2^{D(3)}$ can be expressed as the product of a 
flux, depending only on $\xpom$ and $t$, and the structure function of a 
particle-like object (see Eq.~(\ref{i-s})) is known as the ``Regge 
factorisation" hypothesis. 
It gives a fair description of the data, although it has 
no justification in QCD, where only the concept of 
diffractive PDFs, which are functions of ($\xpom, t, \beta, Q^2$), has a 
firm basis. 

The Monte Carlo generator DIFFVM 1.0~\cite{diffvm} was used to simulate
the double-dissociative reaction, $ep \to eXN$, where the proton
diffractively dissociates into the state $N$.

Initial- and final-state QED radiation were simulated by using EPSOFT and 
RAPGAP in conjunction with HERACLES 4.6~\cite{heracles}. 

\begin{figure}[p]
\vfill
\begin{center}
\includegraphics[width=15cm,height=15cm]{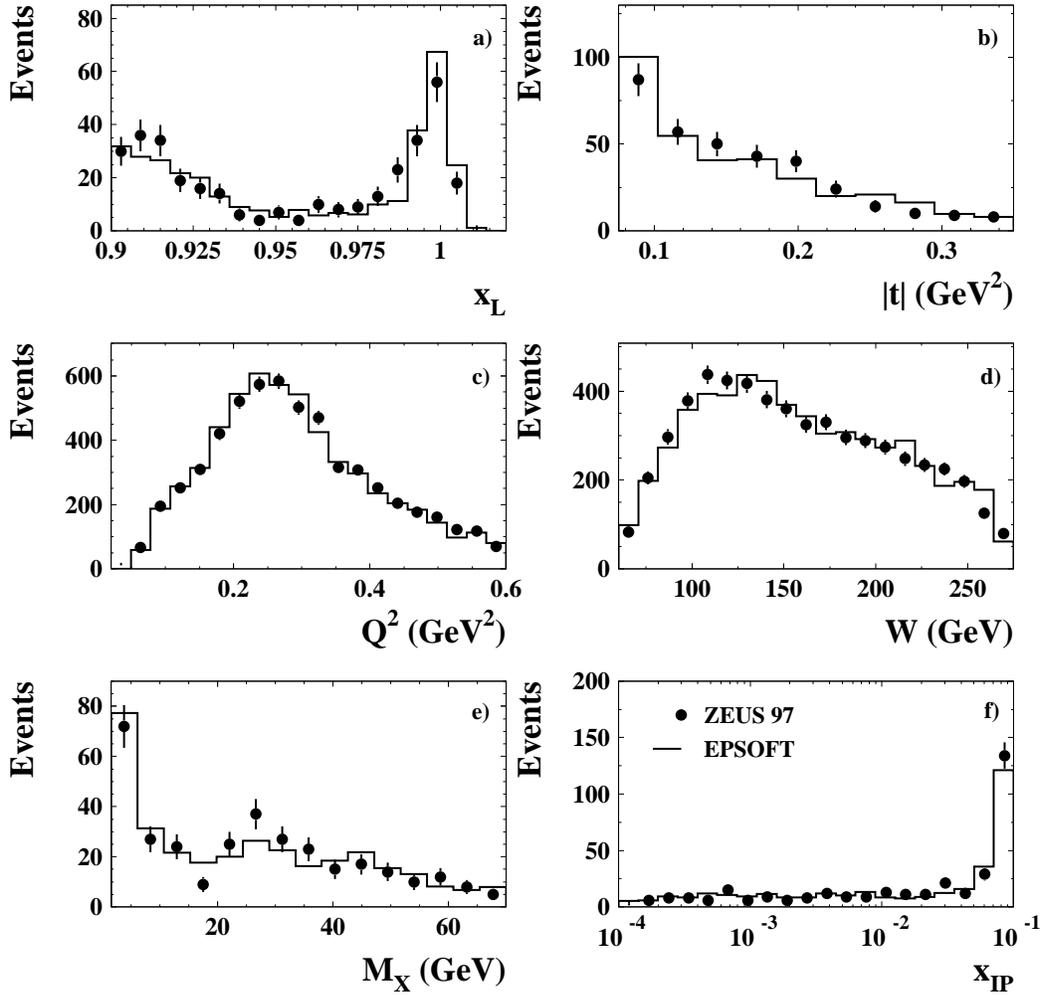}
\end{center}
\vspace{3.0cm}
\caption{
Comparison of the measured (points) and Monte-Carlo simulated (histograms)
distributions for $x_L$, $|t|$, $Q^2$, $W$, $M_X$ and $x_{\pom}$
in the low-$Q^2$ analysis. 
The $Q^2$ and $W$ distributions were obtained 
without the LPS requirement (see text). 
}
\label{fig-data-mc-bpc}
\vfill
\end{figure}

\begin{figure}[p]
\vfill
\begin{center}
\includegraphics[width=15cm,height=15cm]{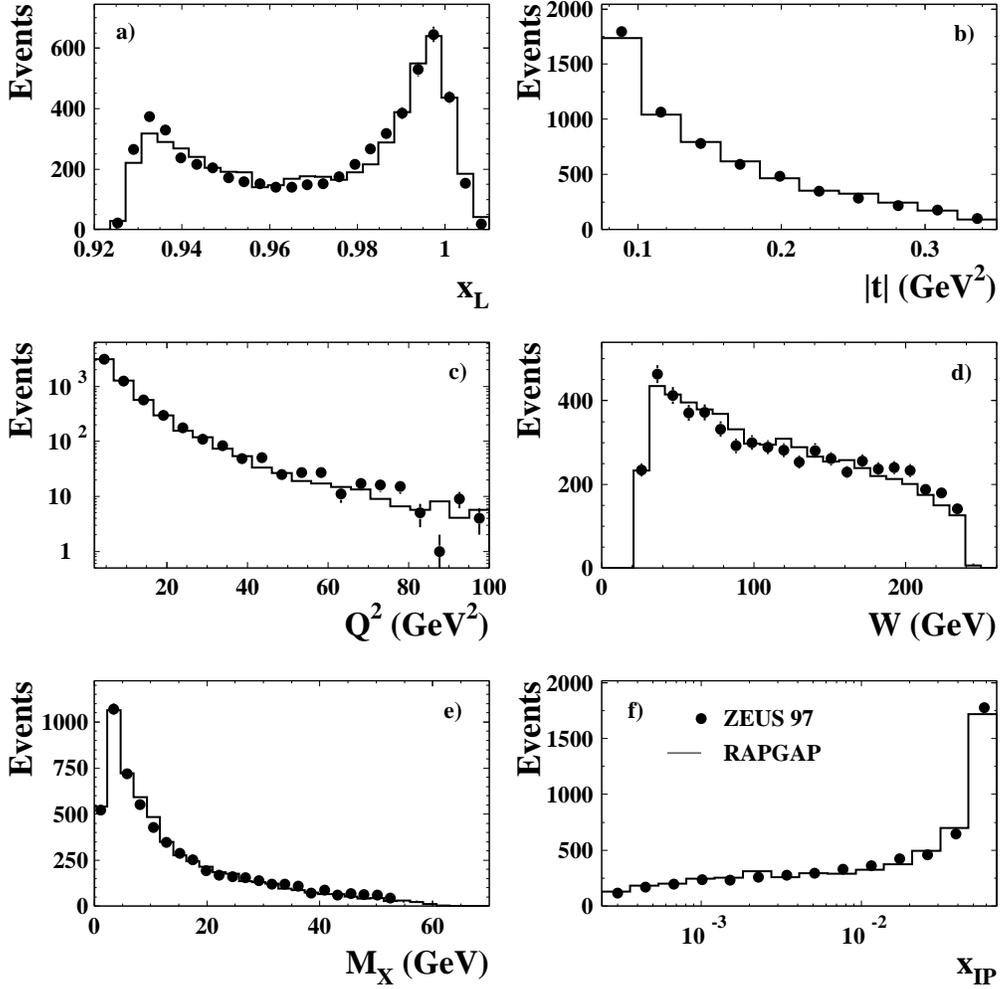}
\end{center}
\vspace{3.0cm}
\caption{
Comparison of the measured (points) and Monte-Carlo simulated (histograms)
distributions for $x_L$, $|t|$, $Q^2$, $W$, $M_X$ and $x_{\pom}$
in the high-$Q^2$ analysis. 
}
\label{fig-data-mc-dis}
\vfill
\end{figure}

The generated EPSOFT and RAPGAP events were reweighted in  $x_L$ and $t$ 
so that the measured distributions were well described.
All generated events were passed through the standard ZEUS detector
simulation, based on the GEANT program~\cite{geant}, and the trigger
simulation package. A comparison 
of data and MC simulations is presented in Figs.~\ref{fig-data-mc-bpc} 
and~\ref{fig-data-mc-dis} for the variables $x_L$, $t$, $Q^2$, $W$, $M_X$, 
and $x_{\pom}$. In Figs.~\ref{fig-data-mc-bpc}c 
and~\ref{fig-data-mc-bpc}d, no LPS 
requirement was imposed, so as to reduce the statistical fluctuations.
The simulations reproduce the data satisfactorily. The 
diffractive peak is evident in  
Figs.~\ref{fig-data-mc-bpc}a and~\ref{fig-data-mc-dis}a.

In the low-$Q^2$ analysis, the measured number of events was
corrected for acceptance bin-by-bin. Bin-centring corrections were applied 
in $W$, $\xpom$ and $M_X$ assuming 
$\sigma^{\gamma^*p \rightarrow Xp} \propto 
W^{2(2\bar{\alpha}_{\pom}-1)}$,
$F_2^{D(3)} \propto 1/\xpom^{2\bar{\alpha}_{\pom}-1}$, and
$d\sigma^{\gamma^*p \rightarrow Xp}/dM_{X}^{2}\propto 
1/M_X^{2(2\bar{\alpha}_{\pom}-1)}$, where $\bar{\alpha}_{\pom}$ is the 
$t$-averaged value of the Pomeron trajectory as obtained from the 
present measurement (Section~\ref{intercept}).
In the high-$Q^2$ analysis, the cross section for the dissociation of 
virtual photons at a given point within a bin was obtained from the ratio
of the measured number of events to the number of events in 
that bin predicted from the MC simulation, multiplied by the $\gamma^*p 
\rightarrow X p$ cross section calculated by the Monte Carlo 
generator. Both the acceptance and the bin-centring corrections were thus
taken from the MC simulation.

The cross section was 
directly measured only in a limited $t$ region and extrapolated to 
$0<|t|<1$~GeV$^2$ using the $t$ dependence assumed in the 
Monte Carlo generator, which was reweighted to the measured value of $b$ 
(see Section~\ref{sec-t}). The effect of the extrapolation is to increase 
the 
cross section by a factor of about two (to within 3\%); this factor is 
largely  
independent of the measured kinematic variables. 
In the region covered by the present measurements, the extrapolation 
is performed assuming an exponential dependence on $t$, 
$d\sigma^{ep \rightarrow eXp}/dt \propto \exp{(bt)}$, with $b 
\simeq 7.9$~GeV$^{-2}$ at low $\xpom$. 
Data from elastic and $p$-dissociative $pp$ and 
$\bar{p}p$ scattering indicate that the $t$ distribution is better described 
by the function $\exp{(bt+ct^2)}$. For example, fits to the $\bar{p}p$ 
data at $\sqrt{s}=546$~GeV~\cite{sps} yield $c=2.3 \pm 0.1$~GeV$^{-4}$. The 
present data are only weakly sensitive to the value of $c$. In the 
extrapolation to the range $0<|t|<1$~GeV$^2$, $c$ was taken to be zero, but 
was varied by up to 4~GeV$^{-4}$, yielding changes in the extrapolated 
cross section of up to 7\%. This was included in the 
normalisation uncertainty discussed in Section~\ref{sec-sys}.

The results presented in this paper were corrected to the Born level and 
to the 
following kinematic region:

\begin{itemize}
\item low-$Q^2$ data: 
$0.03<Q^2<0.6$ GeV$^2$, $63<W<280$ GeV, $M_X>1.5$ GeV, $x_L>0.9$ and
$0<|t|<1$~GeV$^2$;

\item high-$Q^2$ data: $2<Q^2<100$ GeV$^2$, $25<W<240$ GeV,
$M_X>1.5$ GeV, $x_L>0.9$ and $0<|t|<1$~GeV$^2$. 

\end{itemize}

\label{sec-bkg}
\section{Backgrounds}
\label{sec-bg}
%
%
The main background contribution 
is given by proton beam-halo events. In such events, the proton
detected in the LPS is not correlated with the measurements
in the central detector. To estimate the size of this background, the 
variable $E+p_Z$ (see Section~\ref{sec-sel}) was used. 
For a signal event, this quantity should be equal to twice the initial
proton energy, 1640 GeV, whereas for a beam-halo event it
can exceed this value.

\begin{figure}[p]
\vfill
\begin{center}
\includegraphics[width=15cm,height=15cm]{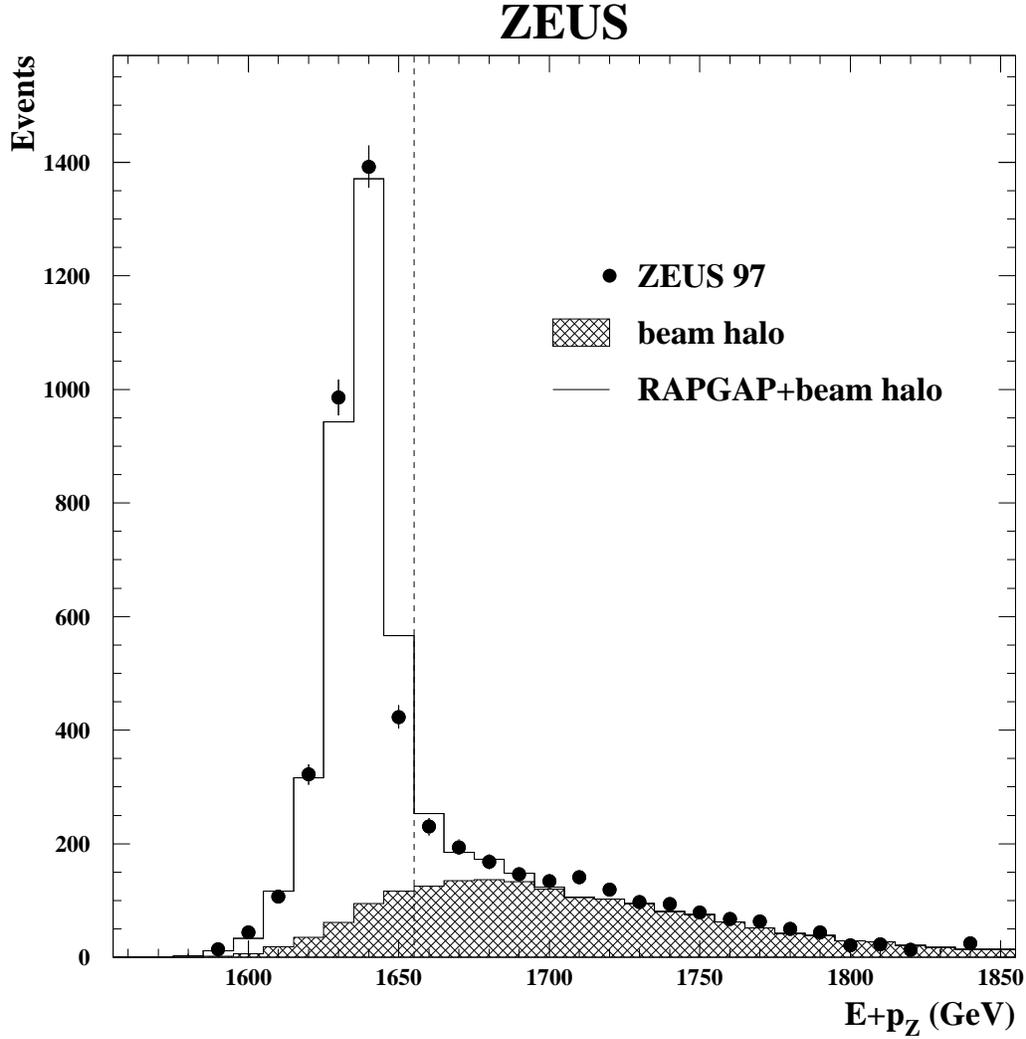}
\end{center}
\vspace{3.0cm}
\caption{
Distribution of $E+p_Z$ for the high-$Q^2$ events.
The hatched histogram represents the beam-halo sample obtained as 
discussed in the text. The empty histogram is the sum of the RAPGAP Monte 
Carlo and the beam-halo contribution. The vertical dashed line 
is at $E+p_Z=1655$~{\rm GeV}, the value of the cut used to suppress 
beam-halo 
events.} 
\label{fig-halo}
\vfill
\end{figure}

The $E+p_Z$ spectrum for the beam-halo events was constructed as
a random combination of generic DIS events (without the requirement
of a track in the LPS) and a beam-halo track measured in the LPS.
The resulting distribution, shown in Fig.~\ref{fig-halo} as the hatched 
histogram, was normalised to 
the data for $E+p_Z>1685$ GeV, which contain beam-halo events only (see 
Fig.~\ref{fig-halo}).
The background remaining after the $E+p_Z<1655$ GeV cut averaged 
$(10.2\pm0.7~(\rm stat.)) \%$ and was a function of  
$\xpom$ and $t$. All results presented in this paper were 
corrected for this background.

The contribution from proton-dissociative events, $ep \to eXN$, 
was studied with the Monte Carlo generator DIFFVM for the DIS sample and 
was found to be less than $4\%$ in all bins. This contribution was 
neglected.

\section{Systematic uncertainties}
\label{sec-sys}
The systematic uncertainties were calculated by varying the cuts 
and by modifying the analysis procedure.  The variations of the cuts were 
typically commensurate with the resolutions of the relevant variables, and 
in general the changes were similar to those made in earlier 
analyses~\cite{f2bpt,lps, lps94a, lps94b}.

The following systematic checks were
performed~\footnote{The corresponding average effect on the cross section in the 
measured bins is indicated using the notation $^{+a}_{-b}$: given a 
systematic check which produces an increase of the cross section in some
bins and a decrease in some other bins, $a$ is the average increase and $b$ is
the average decrease.}:

\begin{itemize}

\item
to evaluate the uncertainties due to the measurement of the scattered 
positron:

\begin{itemize}
\item low-$Q^2$ analysis: the checks performed in a previous 
publication~\cite{f2bpt} 
were  repeated, and consistent results obtained. Since the present data are 
a subsample of those used earlier, the systematic uncertainties 
found previously~\cite{f2bpt} were used (typically smaller than $\pm 
1.5\%$);

\item high-$Q^2$ analysis: the fiducial region for the impact position of 
the positron was modified ($^{+3.3}_{-2.9}\%$); the minimum energy of the 
positron was increased to 12 GeV ($^{+1.6}_{-1.4}\%$);
\end{itemize}

\item
to evaluate the uncertainties due to the reconstruction of the final-state 
proton, the checks described below were performed. Consistent results were 
obtained for the low- and the high-$Q^2$ samples; because of the larger 
statistical fluctuations of the low-$Q^2$ sample, the uncertainties quoted 
are those determined from the high-$Q^2$ events;

\begin{itemize}
\item 
the cut on the minimum distance of approach to the beam-pipe was
increased by 0.03 cm ($^{+4.1}_{-2.4}\%$);
the $t$ range was enlarged to $0.07<|t|<0.4$~GeV$^2$
($^{+2.8}_{-1.1}\%$);
the amount of the subtracted beam-halo background was varied by $\pm20\%$
($^{+2.0}_{-1.4}\%$); 
\end{itemize}

\item
sensitivity to the other selection cuts:

\begin{itemize}

\item low-$Q^2$ analysis:
the checks performed in a previous publication~\cite{f2bpt} were repeated 
and consistent results found, notably for the sensitivity to the 
selections on $y_{JB}$, $E-p_Z$ and the $Z$ coordinate of the vertex.
The systematic uncertainties determined in~\cite{f2bpt} were used (typically 
smaller than $\pm 1.5$\%);

\item high-$Q^2$ analysis: the minimum value of $E-p_Z$ was raised to 45 
GeV 
($^{+1.3}_{-2.1}\%$);
the cut on the $Z$ coordinate of the vertex was restricted to 
$-40<Z<40$~cm ($^{+2.2}_{-1.3}\%$);
the systematic error due to the uncertainty in the absolute calorimeter
energy calibration was estimated by changing the energy scale by $\pm2\%$
($^{+2.6}_{-1.5}\%$);
the minimum value of $M_X$ was decreased to 1.3 GeV ($^{+1.7}_{-1.4}\%$);

\end{itemize}

\item
sensitivity to the Monte Carlo simulations:

\begin{itemize}

\item low-$Q^2$ analysis: 
no reweighting of the $x_L$ distribution was applied in the Monte Carlo 
simulation ($^{+3.2}_{-6.3}\%$); the value of the $t$ slope was changed by 
$\pm1.5$~GeV$^{-2}$ ($^{+4.4}_{-0.7}\%$);

\item high-$Q^2$ analysis: the $x_{\pom}$ distribution was reweighted by a 
factor $(1/x_{\pom})^k$, with $k$
varying between $-0.05$ and +0.05 ($^{+1.7}_{-1.1}\%$);
the value of the $t$ slope was changed by $\pm1.5$~GeV$^{-2}$ 
($^{+6.2}_{-2.1}\%$);
the $\Phi$ distribution was reweighted by a factor $(1+k\cos{\Phi})$, with $k$
varying between $-0.15$ and +0.15 ($^{+1.8}_{-1.2}\%$);
the intrinsic transverse-momentum spread of the proton beam at the interaction
point was increased by 10 MeV in the horizontal plane and 20 MeV in the
vertical plane ($^{+2.7}_{-2.1}\%$);

\end{itemize}

\item the $t$ slope was determined with an alternative method~\cite{lps} 
based on expressing the measured $t$ distribution as a convolution of 
an exponential distribution, $d\sigma/dt \propto e^{-b|t|}$, and a 
two-dimensional Gaussian distribution representing the transverse momentum 
distribution of the beam. This led to changes in the value of the $t$ 
slope by up to $+4\%$.

\end{itemize}

The total systematic uncertainty for each bin was determined by 
quadratically adding the individual contributions. 
The quoted uncertainties do not include an overall normalisation
uncertainty of $\pm 10\%$ which originates mostly from the uncertainty 
of the simulation of the proton-beam optics; this uncertainty is largely 
independent of the kinematic variables and was therefore taken as a 
normalisation uncertainty. The $\pm 10\%$ normalisation uncertainty also 
includes the uncertainty on the integrated luminosity ($\pm 1.6\%$). In 
addition, a $+7\%$ uncertainty is present in the cross section and 
structure function 
results, except those for $d\sigma^{ep \rightarrow eXp}/dt$, due 
to the extrapolation from the measured $t$ region to $0<|t|<1$~GeV$^2$ 
(see Section~\ref{sec-mc}). Thus the overall normalisation 
uncertainty is $^{+12}_{-10}\%$.

\section{Results}
\label{sec-res}

The results in this section are presented as follows.
The differential cross-section $d\sigma^{ep\rightarrow eXp}/dt$ in the 
region
$0.075<|t|<0.35$~GeV$^2$ is presented first. The data are then 
integrated over $t$ and extrapolated to the range $0<|t|<1$~GeV$^2$, as 
discussed in Section~\ref{sec-mc}. The resulting cross sections are 
presented as a function of $\Phi$ in Section~\ref{sec-phi}, where the 
sensitivity of the present data to the helicity structure of the reaction 
$ep\rightarrow eXp$ is discussed. The dependence of the cross section 
on $Q^2$ is presented in Section~\ref{sec-q2}, and is compared with that 
of the total photon-proton cross section.

In Section~\ref{sec-f2} the data are presented in terms of the 
diffractive structure function, $F_2^{D(3)}$. This allows an
interpretation based on the diffractive PDFs of the proton. The 
present 
data in the diffractive-peak region are compared to those obtained 
with the $M_X$ method by ZEUS and with the rapidity-gap method by H1 in 
Section~\ref{sec-comparison}. 

In Section~\ref{intercept}, the $\xpom$ dependence of $F_2^{D(3)}$ is used 
to extract the intercept of the Pomeron trajectory, $\alpha_{\pom}(0)$,  
the quantity that, in Regge phenomenology, determines the energy 
dependence of the total hadron-hadron cross section~\cite{regge}. 
It is interesting to see if the value of $\alpha_{\pom}(0)$ in $ep$ 
diffractive scattering at high $Q^2$ is larger than that 
measured in 
hadron-hadron collisions, as expected in the framework introduced earlier, 
in which diffraction is due to the exchange of partons from the proton.
In Section~\ref{comp-models} the results are compared to some perturbative QCD
(pQCD) models based on the dipole picture outlined in 
Section~\ref{sec-int}. Finally,  an NLO QCD fit was performed.

The results of this paper extend up to $\xpom \simeq 0.07$. In the following, the data 
for the diffractive-peak region are often contrasted with those at high 
$\xpom$. For this purpose, the value $\xpom=0.01$ is chosen as the 
transition between the high- and low-$\xpom$ bins, such that the 
low-$\xpom$ bins are dominated by diffractive-peak events. This choice 
is somewhat restrictive, since the diffractive peak extends to $\xpom 
\simeq 0.02$, see Figs.~\ref{fig-data-mc-bpc}a and~\ref{fig-data-mc-dis}a.
In the region $\xpom <0.01$, 
the contribution from non-Pomeron exchanges is less than 10\%. The average 
value of $\xpom$ is 0.003 for $\xpom<0.01$ and 0.043 for  $\xpom>0.01$.

Preliminary versions of the present results, along with details of the 
analysis, have been presented 
earlier~\cite{ruspa,smalska,mastroberardino}.

\subsection{\boldmath{$t$} dependence}
\label{sec-t}

\begin{figure}[p]
\vfill
\begin{center}
\includegraphics[width=16cm,height=16cm]{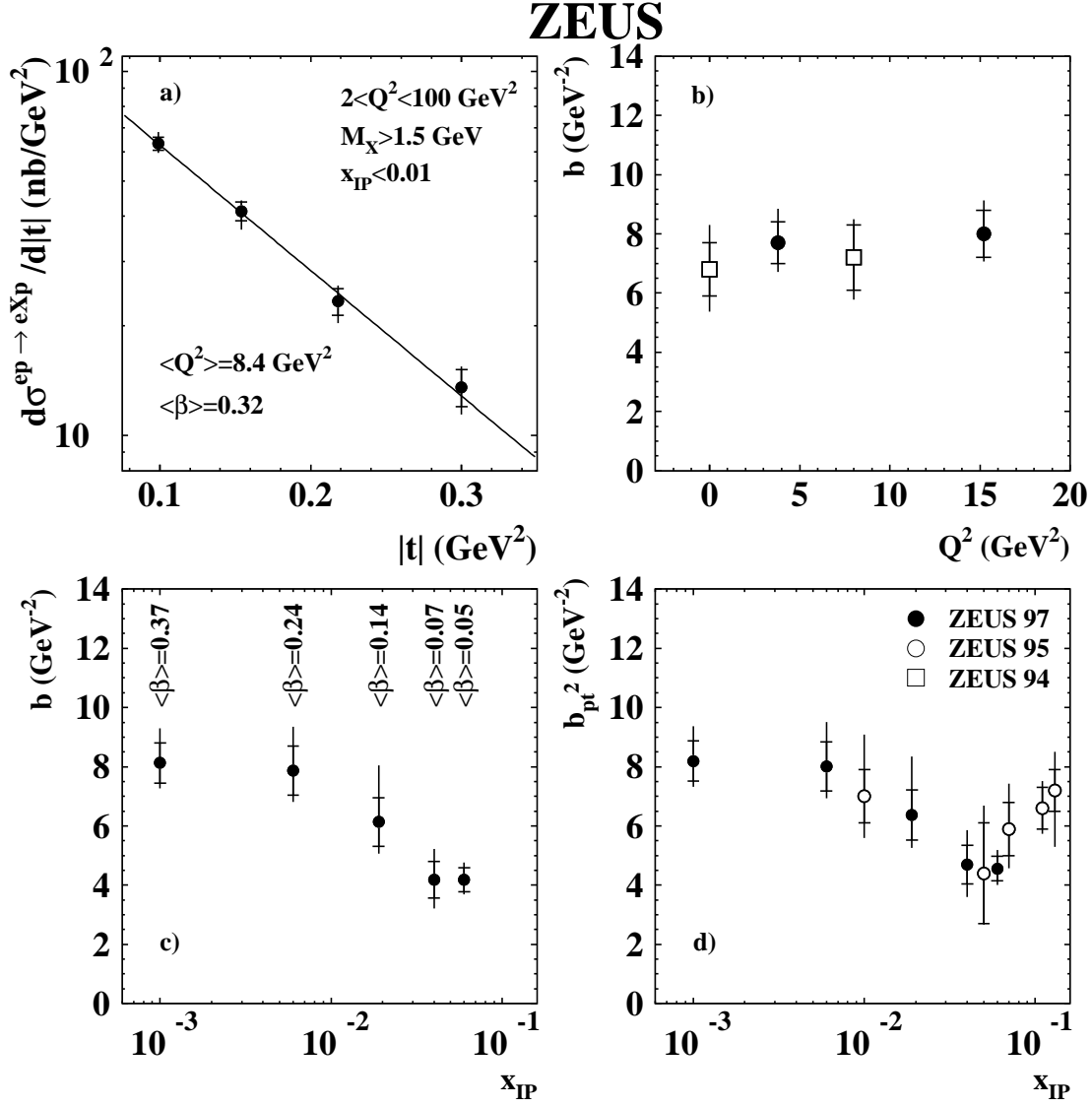}
\end{center}
\caption{(a) The differential cross-section $d\sigma^{ep\rightarrow 
eXp}/dt$ in the region $x_{\pom}<0.01$, $2<Q^2<100$ {\rm GeV}$^2$ and 
$M_X>1.5$ {\rm GeV}. The inner error bars show the statistical
uncertainties and the full bars indicate the statistical and the
systematic uncertainties added in quadrature. 
The overall normalisation uncertainty of $\pm 10$\% is not shown.
The line shows the result of the fit described in the text.
(b) The value of the slope parameter $b$ of the 
differential cross-section $d\sigma^{ep \rightarrow eXp}/dt$
as a function of $Q^2$. 
(c) The value of the slope parameter $b$ of the 
differential cross-section $d\sigma^{ep \rightarrow eXp}/dt$
as a function of $\xpom$. The mean value of $\beta$ in each bin is also 
given. (d) The value of the slope parameter $b_{p_T^2}$ of the
differential cross-section $d\sigma^{ep\rightarrow eXp}/dp_T^2$
as a function of $x_{\pom}$. 
The symbols labelled ZEUS 97 indicate the present results. 
Earlier ZEUS results are also shown: 
ZEUS 95~\protect\cite{low-xl}, 
ZEUS 94~\protect\cite{lps94b} ($Q^2=0$) and  
ZEUS 94~\protect\cite{lps94a} ($Q^2>0$).  
}
\label{fig-t}
\vfill
\end{figure}

Figure~\ref{fig-t}a presents the differential cross-section 
$d\sigma^{ep\rightarrow eXp}/dt$ in the kinematic range $2<Q^2<100$ 
GeV$^2$, $M_X>1.5$ GeV, $x_{\pom}<0.01$. For this sample,
$\langle Q^2 \rangle =8.4$ GeV$^2$ and $\langle \beta \rangle =0.32$.
The value of the slope parameter, $b$, obtained from the fit
with the function $d\sigma^{ep\rightarrow eXp}/dt \propto 
e^{-b|t|}$ in the range $0.075<|t|<0.35$~GeV$^2$ is
\begin{equation*}
b=7.9\pm0.5 (\rm stat.)^{+0.9}_{-0.5} (\rm syst.)~ \rm{GeV^{-2}}.
\end{equation*}
This agrees with and improves on the previous measurement of the
diffractive slope of
$b=7.2\pm1.1 (\rm{stat.})^{+0.7}_{-0.9} (\rm{syst.})$ 
GeV$^{-2}$~\cite{lps94a}.

The $t$ distribution was studied in two $Q^2$ bins,
$2<Q^2<7$ GeV$^2$ and $7<Q^2<100$ GeV$^2$. 
The fitted values of $b$ are
$7.7\pm0.7 (\rm{stat.})^{+0.9}_{-0.7} (\rm{syst.})~ \rm{GeV^{-2}}$ and
$8.0\pm0.8 (\rm{stat.})^{+0.9}_{-0.5} (\rm{syst.})~ \rm{GeV^{-2}}$,
respectively. These results, presented in Fig.~\ref{fig-t}b 
together with the previous ZEUS
measurements in photoproduction~\cite{lps94b} and in DIS~\cite{lps94a},
show that $b$ is independent of $Q^2$ within the errors. This behaviour is 
expected in a QCD-based model~\cite{nikolaev98}.

However, the value of $b$ decreases with $x_{\pom}$, as shown 
in Fig.~\ref{fig-t}c (see also Table~\ref{tablet}). 
Figure~\ref{fig-t}d 
shows the values of 
$b_{p_T^2}$ in bins of $x_{\pom}$, calculated from the fit to 
$d\sigma^{ep\rightarrow eXp}/dp_T^2$, so that they can be compared with 
those of a previous ZEUS publication~\cite{low-xl}. 
The relationship between $t$ and $p_T^2$ is given in Eq.~(\ref{tpt2}). 
In the region of the present data, the values of $b$ and $b_{p_T^2}$ 
differ by less than their uncertainties.
The present results are consistent with those of the previous
ZEUS publication~\cite{low-xl}. 
The $p_T^2$ slope reaches a minimum value for 
$\xpom \approx 0.05$ and then rises to $b_{p_T^2}\approx 7$~GeV$^{-2}$ for 
higher $\xpom$ values. 

In the dipole model, in which the virtual photon fluctuates into $q\bar{q}$ 
or $q\bar{q}g$ systems, the $q\bar{q}g$ contribution 
dominates for $\beta <0.2$-0.3~\cite{rev}. Different $t$ dependences for 
the $q\bar{q}$ and $q\bar{q}g$ regions are expected~\cite{nikolaev98}, 
with slopes higher by up to 3 GeV$^{-2}$ at high $\beta$ in the 
diffractive peak. In addition, in the $q\bar{q}g$ region, the slope $b$ is 
expected to decrease logarithmically with $\xpom$: $b=b_0^{\pom}- 
2\alpha^{\prime}_{\pom} \ln{\xpom}$,
the so-called shrinkage of the diffractive cone~\cite{rev}. 
A decrease is observed in the data, and so is a significant $\beta$ 
dependence. However, these dependences are visible only over  an $\xpom$
range, $\xpom~\lsim~0.07$, that goes beyond the diffractive-peak 
region (Fig.~\ref{fig-t}c), making a comparison with the predictions 
inconclusive. At 
higher $\xpom$ values, a rise of $b$ with $\xpom$ is predicted by a 
Regge-based model~\cite{SNS}, though at a rate smaller than that observed 
in the data.

\subsection{Azimuthal asymmetry}
\label{sec-phi}
The azimuthal angle $\Phi$ between the positron and proton scattering 
planes is sensitive to the helicity structure of the 
reaction $ep \to eXp$, as shown
explicitly in Eq.~(\ref{fullsigma}). The analysis of the azimuthal 
distribution is limited to the high-$Q^2$ data, since the statistics of 
the low-$Q^2$ sample is too small. For this part of the analysis, a radial 
cut of 18~cm was imposed on the impact point of the scattered electron at 
the RCAL surface, along with the restriction $Q^2>4$~GeV$^2$. This reduces 
the $\Phi$ dependence of the acceptance.

\begin{figure}[p]
\vfill
\begin{center}
\includegraphics[width=16cm,height=16cm]{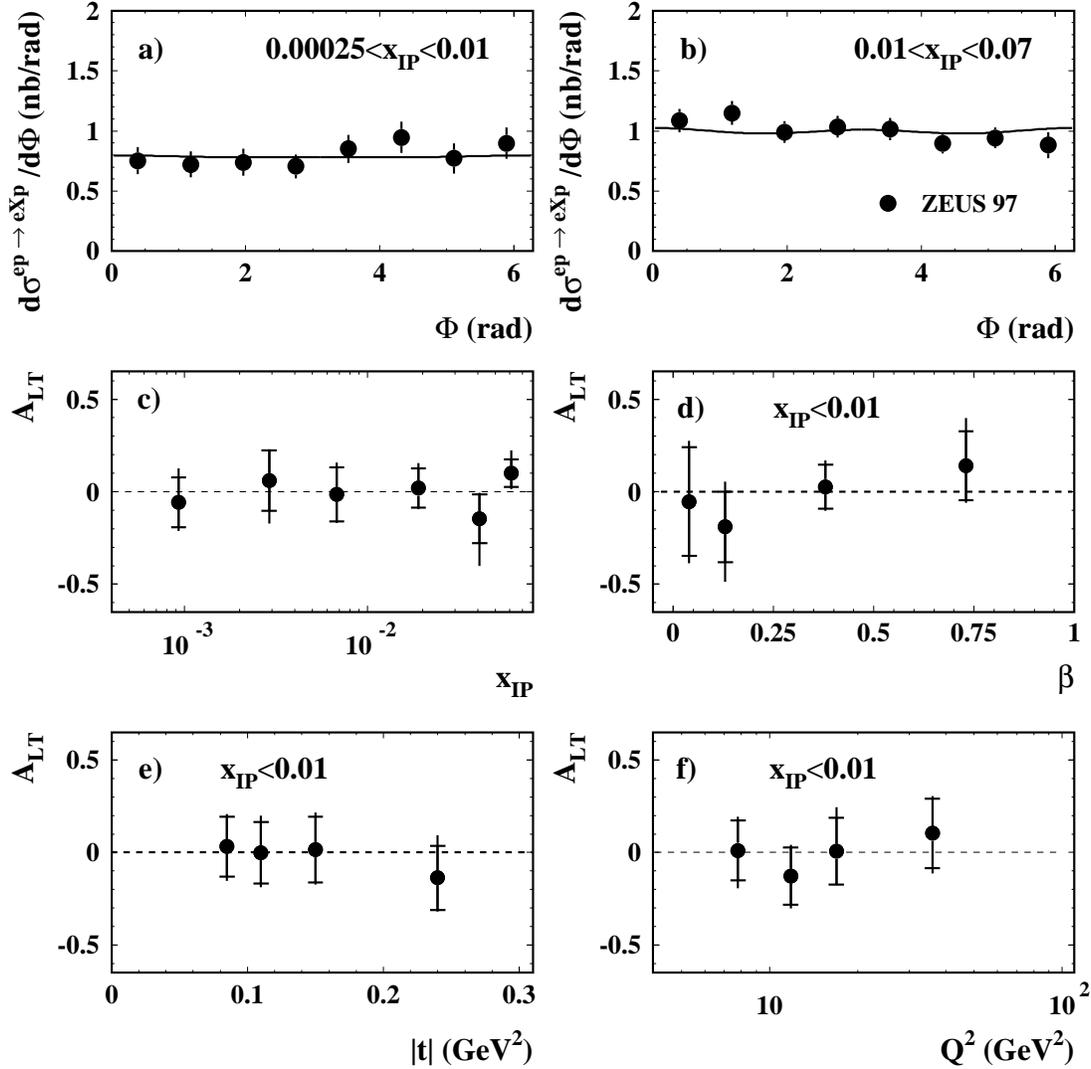}
\end{center}
\vspace{-0.1cm}
\caption{
The differential cross-section $d\sigma^{ep \rightarrow eXp}/d\Phi$
in the kinematic ranges (a) $0.00025<x_{\pom}<0.01$   and (b) 
$0.01<x_{\pom}<0.07$.
The error bars show the statistical uncertainty.
The line shows the result of the fit described in the text.
The azimuthal asymmetry $A_{LT}$ as a function of (c) $x_{\pom}$,
(d) $\beta$, (e) $|t|$ and (f) $Q^2$ for $\xpom<0.01$. The inner error 
bars show the statistical uncertainties and the full bars indicate the
statistical and the systematic uncertainties added in
quadrature.
}
\label{fig-phi}
\vfill
\end{figure}

Figures~\ref{fig-phi}a and~\ref{fig-phi}b show the 
$\Phi$ distribution for the two ranges
$0.00025<x_{\pom}<0.01$ and $0.01<x_{\pom}<0.07$. 
Here again, 
$\xpom=0.01$ was chosen as the transition between the high- and 
low-$\xpom$ bins 
such that the low-$\xpom$ bin is dominated by diffractive-peak events. 
The distributions were fitted to the form
\begin{equation*}
\frac{d\sigma^{ep\rightarrow eXp}}{d\Phi} 
\propto 1+A_{LT}\cos{\Phi}+A_{TT}\cos{2\Phi},
\end{equation*}
where $A_{LT}$ and $A_{TT}$ are proportional to 
$\sigma^{\gamma^{\star}p \rightarrow Xp}_{LT}$ and 
$\sigma^{\gamma^{\star}p \rightarrow Xp}_{TT}$, respectively.
The values of the azimuthal asymmetries obtained in the fit are
\begin{eqnarray}
A_{LT} &=& 0.009 \pm 0.073 (\rm stat.) ^{+0.076}_{-0.039} (\rm 
syst.),\nonumber\\
A_{TT}& =& 0.005 \pm 0.074 (\rm stat.) ^{+0.043}_{-0.074} (\rm 
syst.)\nonumber
\end{eqnarray}
and
\begin{eqnarray}
A_{LT} &=& 0.007 \pm 0.048 (\rm stat.)^{+0.043}_{-0.071} (\rm 
syst.),\nonumber\\
A_{TT} &=& 0.019 \pm 0.046 (\rm stat.)^{+0.026}_{-0.053} (\rm 
syst.)\nonumber
\end{eqnarray}
for  the ranges $0.00025<x_{\pom}<0.01$ and $0.01<x_{\pom}<0.07$,
respectively.
The interference terms between the longitudinal and transverse amplitudes
and between the two transverse amplitudes thus appear to be small in the 
measured kinematic range, both in the diffractive-peak region and 
at higher $\xpom$ values, suggesting that the helicity structure of the 
reaction $ep \rightarrow eXp$ is similar for Pomeron and Reggeon 
exchanges.

Figure~\ref{fig-phi}c presents $A_{LT}$ as a 
function of $x_{\pom}$ and Figs.~\ref{fig-phi}d-f present $A_{LT}$ 
as a function of $\beta$, $t$ and $Q^2$ for $\xpom <0.01$.
The asymmetry is consistent with zero in all measured bins. The  
results are summarised in Table~\ref{tablephi}.

The measured value of 
$A_{LT}$ can be compared with the results obtained in exclusive 
electroproduction of $\rho^0$ mesons, $ep \rightarrow e \rho^0 p$, in 
which the hadronic final state, $X$, consists of a $\rho^0$ meson only. In 
this case, $A_{LT}=-\sqrt{2 \epsilon 
(1+\epsilon)} \cdot (r^5_{00}+2r^5_{11})= -0.262\pm 0.038 (\rm{stat.}) \pm 
0.068 (\rm{syst.})$, where $r^5_{00}$ and $r^5_{11}$ are two 
of the $\rho^0$ spin-density matrix elements~\cite{rho}. The present data 
show that the asymmetry is smaller for inclusive scattering than for 
exclusive 
$\rho^0$ electroproduction.

There are numerous pQCD-based predictions for the behaviour of 
$A_{LT}$~\cite{gehrmann, arens, diehl, nikolaev} in the diffractive peak, 
mostly concerning the 
high-$\beta$ region ($\beta > 0.6$-0.9), where the asymmetry is expected to 
be largest, reflecting the large expected value of 
$\sigma_L^{\gamma^{\star}p\rightarrow Xp}$;
this region was not accessible due to limited 
statistics. In all calculations, back-to-back configurations, i.e. 
$A_{LT}<0$, are favoured; the asymmetry is expected to be close to zero at 
low $\beta$, in agreement with the present data. 

The measurement of the $\Phi$ dependence can, in principle, be used to 
constrain the cross section of longitudinally polarised 
photons~\cite{nikolaev, arens,diehl}, a quantity notoriously difficult to 
extract unless data at different centre-of-mass energies are available. No 
experimental results on $\sigma^{\gamma^{\star}p \rightarrow Xp}_{L}$ 
exist so far. The asymmetry $A_{LT}$ can be related to 
$R^D$~\cite{nikolaev}; however, only for $\beta>0.8$-0.9, beyond the 
region covered by the present data,  is the  determination of 
$R^D$ model-independent. More general limits 
can be obtained for $\sigma^{\gamma^{\star}p \rightarrow Xp}_L$;
using hermiticity and parity conservation, the following relations
are found~\cite{arens,diehl}:
\begin{equation}
\frac{2(\sigma^{\gamma^*p \rightarrow Xp}_{LT})^2}{\sigma_0-\sigma^{\gamma^*p \rightarrow Xp}_{TT}} \le 
\sigma^{\gamma^*p \rightarrow Xp}_L \le 
\frac{\sigma_0-\sigma^{\gamma^*p \rightarrow Xp}_{TT}}{\epsilon},
\label{inequalities}
\end{equation}
where $\sigma_0 = \sigma^{\gamma^*p \rightarrow Xp}_T 
+ \epsilon \sigma^{\gamma^*p \rightarrow Xp}_L$. The larger the values 
of the interference terms $\sigma^{\gamma^*p \rightarrow Xp}_{LT}$ and $\sigma^{\gamma^*p 
\rightarrow Xp}_{TT}$, the stronger the constraint on 
$\sigma_L^{\gamma^{\star}p \rightarrow Xp}$. The fact that in the 
present data the measured asymmetries $A_{LT}$ and $A_{TT}$ are consistent 
with zero implies that the interference terms 
$\sigma^{\gamma^*p \rightarrow Xp}_{LT}$ and $\sigma^{\gamma^*p \rightarrow Xp}_{TT}$ 
are also consistent with 
zero. In this case, the inequalities~(\ref{inequalities}) are 
trivially satisfied and give no information on 
$\sigma_L^{\gamma^{\star}p \rightarrow Xp}$.

\subsection{\boldmath{$Q^2$} dependence of 
\boldmath{$d\sigma^{\gamma^{\star}p \rightarrow Xp}/dM_X$}}
\label{sec-q2}

\begin{figure}[p] \vfill \begin{center}
\includegraphics[width=15cm,height=15cm]{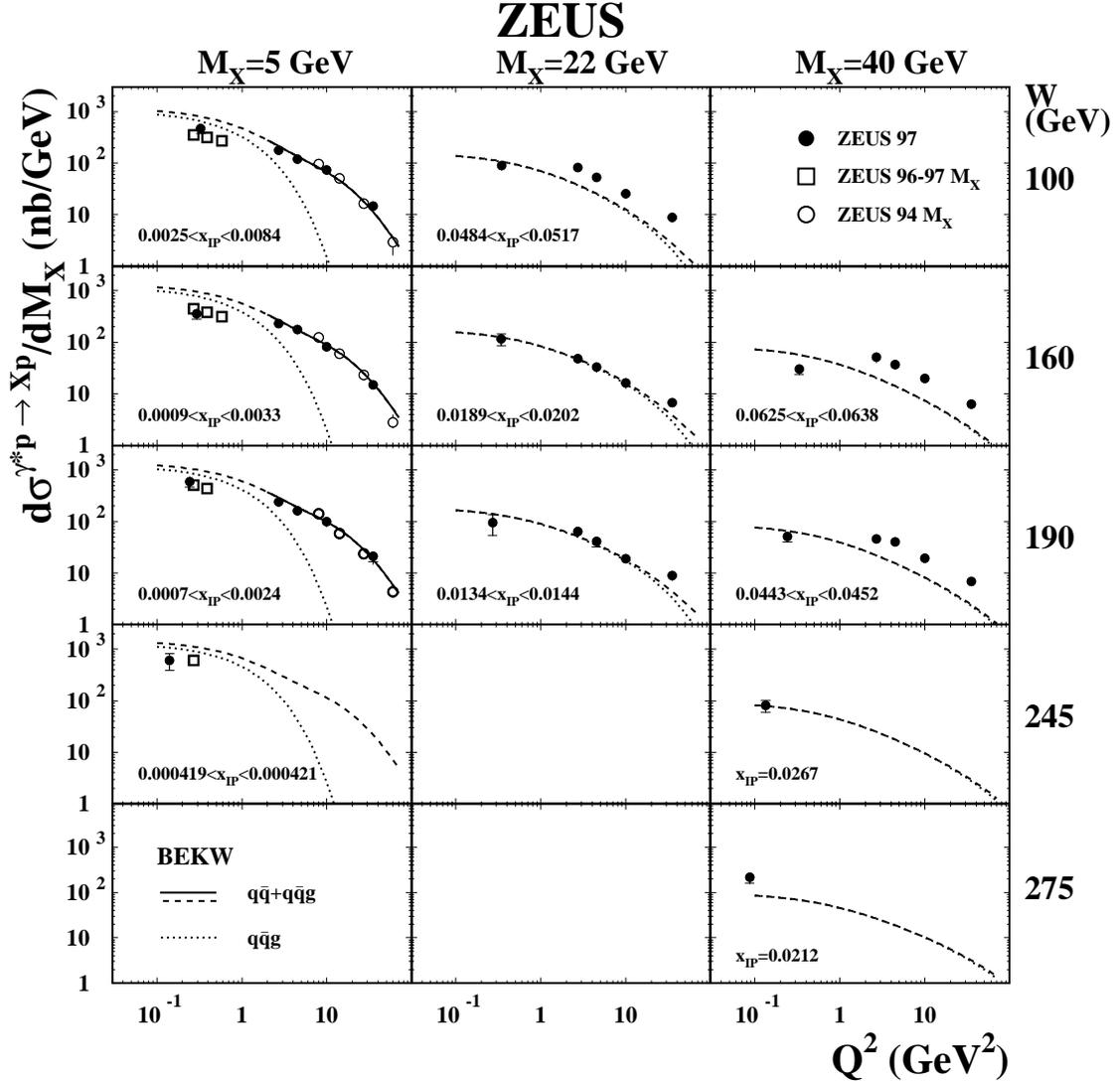} \end{center}
\vspace{2.0cm} 
\caption{ The cross-section $d\sigma^{\gamma^{\star}p
\rightarrow Xp}/dM_X$ as a function of $Q^2$ at different $M_X$ and $W$
values. The inner error bars show the statistical uncertainties and the
full bars indicate the statistical and the systematic uncertainties added
in quadrature; in several bins the size of the bars is smaller than that 
of the symbol used for the points.
The overall normalisation uncertainty of~$^{+12}_{-10}\%$
is not shown. 
The symbols labelled ZEUS 97 indicate the present results. 
Earlier ZEUS results are also shown: 
ZEUS 96-97 $M_X$~\protect\cite{lps95}, 
ZEUS 94 $M_X$~\protect\cite{zeusdiff}.
The solid lines are
the result of the BEKW fit to the present high-$Q^2$ data, described in
Section~\protect\ref{BEKWfit}; the dashed lines indicate the extrapolation
outside the fit region. The dotted lines indicate the $q{\bar q}g$
contribution. The $\xpom$ ranges given refer to the coverage of the data.
} 
\label{fig-sigma_vs_q2} \vfill
\end{figure}

Figure~\ref{fig-sigma_vs_q2} shows the cross-section 
$d\sigma^{\gamma^{\star}p \rightarrow Xp}/dM_X$ as 
a function of $Q^2$ for different $M_X$ and $W$ values. 
The data are also presented in 
Tables~\ref{tab-fig6lowq2} and~\ref{tablecrosshi}. 
The present measurements are
shown together with the previous ZEUS results at low~\cite{lps95} and 
high $Q^2$~\cite{zeusdiff}; the latter have been corrected for the 
residual double-dissociative background, taken to be $31\%$, as 
determined~\cite{zeusdiff} by comparing 
those data with the LPS results~\cite{lps94a}~\footnote{As discussed in 
Section~\ref{sec-comparison}, a higher fraction was measured at lower 
$Q^2$~\cite{lps95}.}. The present 
results are consistent with the earlier 
ZEUS measurements~\cite{zeusdiff,lps95} and cover a wider kinematic 
region; notably, 
they reach higher values
of $M_X$, lower values of $Q^2$, as well as values of $W$ close to the
kinematic limit. The points at $M_X=5$~GeV are all in
the diffractive-peak region, since $\xpom <0.01$. The other bins have 
contributions from 
$\xpom >0.01$.  In all regions of $\xpom$, the data exhibit a 
behaviour qualitatively similar to that of the total $\gamma^*p$ cross 
section, $\sigma_{\tot}^{\gamma^{\star}p}$:  $d\sigma^{\gamma^{\star}p 
\rightarrow Xp}/dM_X$ falls rapidly with $Q^2$ at high $Q^2$; 
as $Q^2 \rightarrow 0$, the cross-section dependence on $Q^2$ becomes 
weak, with 
$d\sigma^{\gamma^{\star}p \rightarrow Xp}/dM_X$ approaching a constant, 
as expected from the conservation of the electromagnetic current. 
The behaviour of $d\sigma^{\gamma^{\star}p \rightarrow Xp}/dM_X$ cannot be 
fitted with a simple, form-factor-like function of the type 
$1/(Q^2+M_X^2)^n$, but is described by a pQCD-based 
model~\cite{bekw} at $\xpom~\lsim~0.01$ and large $Q^2$ (see the 
continuous curves in Fig.~\ref{fig-sigma_vs_q2}). 
This model and the comparison between its predictions and the data, as 
well 
as the curves on Fig.~\ref{fig-sigma_vs_q2}, are discussed in 
Section~\ref{comp-models}.

\begin{figure}[p]
\vfill
\begin{center}
\includegraphics[width=15cm,height=15cm]{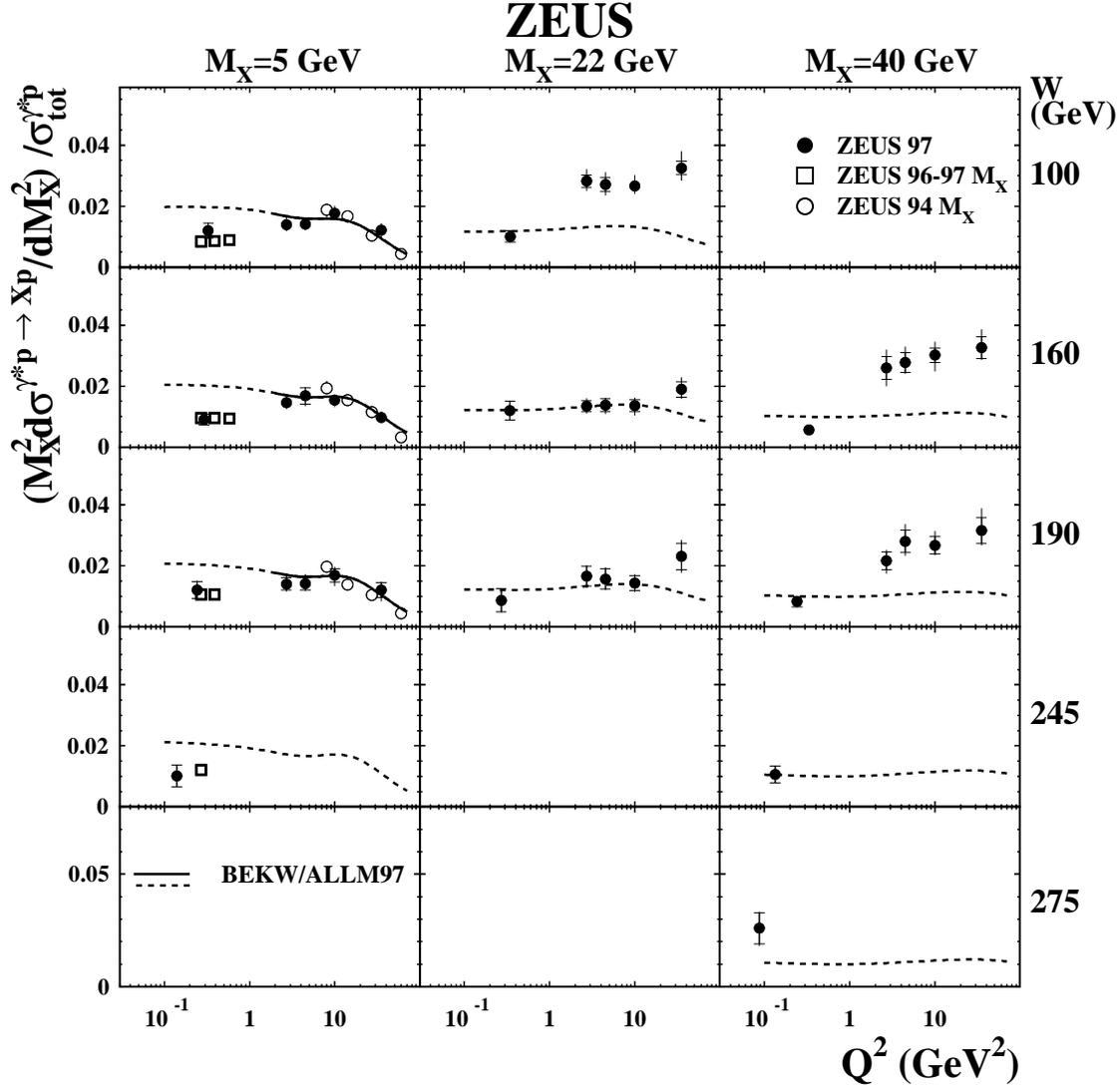}
\end{center}
\vspace{3.0cm}
\caption{
The ratio of $M_X^2 d\sigma^{\gamma^*p \rightarrow Xp}/dM_X^2$ to the 
total virtual photon proton cross section as a
function of $Q^2$ at different $M_X$ and $W$ values.
The inner error bars show the statistical
uncertainties and the full bars indicate the statistical and the
systematic uncertainties added in quadrature.
The overall normalisation uncertainty of~$^{+12}_{-10}\%$ is not shown.
Earlier ZEUS results are also shown: 
ZEUS 96-97 $M_X$~\protect\cite{lps95}, 
ZEUS 94 $M_X$~\protect\cite{zeusdiff}.
The solid lines are the result of the BEKW fit divided by the ALLM97
parameterisation, as described in Section~\protect\ref{BEKWfit}.
The dashed lines indicate the extrapolation outside the fit region.
} 
\label{fig-ratio_vs_q2}
\vfill
\end{figure}

A direct comparison between $d\sigma^{\gamma^{\star} p\rightarrow Xp}/dM_X$ and 
the total photon-proton cross section is shown in Fig.~\ref{fig-ratio_vs_q2}, 
where the ratio $(M^2_X d\sigma^{\gamma^{\star}p \rightarrow Xp}/dM^2_X)/
\sigma_{\tot}^{\gamma^{\star}p}$ is presented as a function of $Q^2$ at 
different $M_X$ and $W$ values. The values of the $\gamma^{\star}p$
total cross section were obtained from the ALLM97 
parameterisation~\cite{allm97}, which is consistent with the latest H1 
and ZEUS $F_2$ data~\cite{levy}. The plot shows that, in spite of their 
qualitative similarity, 
$d\sigma^{\gamma^{\star} p\rightarrow Xp}/dM_X$ and the total cross 
section exhibit some differences in their $Q^2$ dependences. 
In the bin at $M_X=5$~GeV, which has data from the diffractive peak 
region only, the ratio grows slowly with $Q^2$ for $Q^2 < M_X^2$ and then 
falls in the region dominated by $q\bar{q}$ fluctuations of the photon. 
At higher $M_X$ values, which correspond to $\xpom$ larger than 
0.01, $Q^2$ is always smaller than $M_X^2$ and the ratio grows, indicating 
a softer $Q^2$ dependence of  $d\sigma^{\gamma^{\star}p \rightarrow Xp}/dM^2_X$
than of  $\sigma^{\gamma^{\star}p}_{\tot}$. Here again, the low-$\xpom$ data 
in the high-$Q^2$ region can be described by pQCD-based models 
of diffraction, as argued in Section~\ref{comp-models}, where the curves 
on Fig.~\ref{fig-ratio_vs_q2} are discussed.

%
%
\subsection{The structure function \boldmath{$F_2^{D(3)}$} }
\label{sec-f2}

\begin{figure}[p]
\vfill
\begin{center}
\includegraphics[width=16cm,height=16cm]{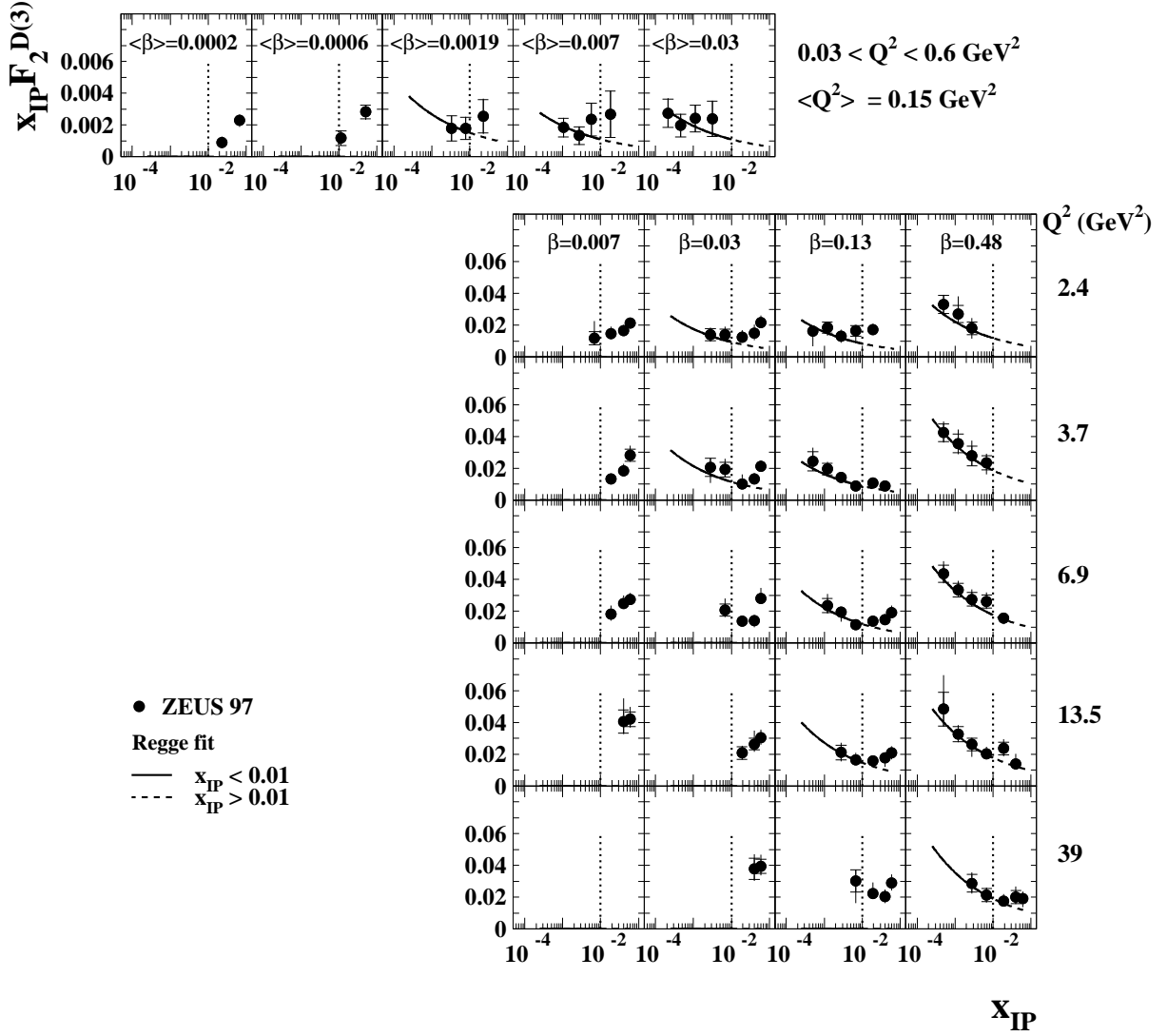}
\end{center}
\caption{
The diffractive structure function multiplied by
$\xpom$, $\xpom F_2^{D(3)}(\beta,Q^2,\xpom)$, as
a function of $\xpom$, for different values of $\beta$
and $Q^2$. The inner error bars show 
the statistical
uncertainties and the full bars indicate the statistical and the
systematic uncertainties added in quadrature. In some bins the size of 
the bars is smaller than that of the symbol used for the points.
The overall normalisation 
uncertainty of $^{+12}_{-10}\%$ is not shown. 
The vertical dotted lines indicate $\xpom=0.01$. 
For the low-$Q^2$ points, the average value of $Q^2$ in each 
$\beta$-$\xpom$ bin varies between $0.10$ and $0.22$~{\rm GeV}$^2$
(see Table~5). The solid lines show the result of the 
Regge fit described in Section~\protect\ref{intercept}. 
The dashed curves are the extension of the fit for $\xpom >0.01$.
} 
\label{fig-f2d_vs_xpom}
\vfill
\end{figure}

The data of Fig.~\ref{fig-sigma_vs_q2} are presented in 
Fig.~\ref{fig-f2d_vs_xpom} (see also Tables~\ref{tab-fig9}-\ref{tabf2dis3}) 
in terms of the structure function $F_2^{D(3)}$, evaluated under the 
assumption that $R^{D}=0$. As discussed in Section~\ref{sec-kin}, 
$F_2^{D(3)}$ is defined in this paper as the integral of $F_2^{D(4)}$ over 
the range $0<|t|<1$~GeV$^2$. The figure shows 
$x_{\pom}F_2^{D(3)}(\beta,Q^2,x_{\pom})$ as a function 
of $x_{\pom}$ for different values of $\beta$ and $Q^2$. 
In order to maximise the kinematic overlap between the low-$Q^2$ and the 
high-$Q^2$ samples, only a subset of the low-$Q^2$ sample is presented 
in Fig.~\ref{fig-f2d_vs_xpom} and Table~\ref{tab-fig9}. 
The values of $\xpom F_2^{D(3)}$ decrease with $\xpom$ at small 
$\xpom$, indicating that $F_2^{D(3)}$ falls with $\xpom$ faster than 
$1/\xpom$. At 
larger $\xpom$, $\xpom F_2^{D(3)}$ flattens and, in some bins, 
increases with $\xpom$, indicating a softer dependence of $F_2^{D(3)}$ on 
$\xpom$. 
The $\xpom$ dependence can be parameterised in terms of the 
Pomeron intercept $\alpha_{\pom}(0)$. The extraction of 
$\alpha_{\pom}(0)$
is presented in Section~\ref{intercept}, where the curves in 
Fig.~\ref{fig-f2d_vs_xpom} are discussed.

\begin{figure}[p]
\vfill
\begin{center}
\includegraphics[width=16cm,height=16cm]{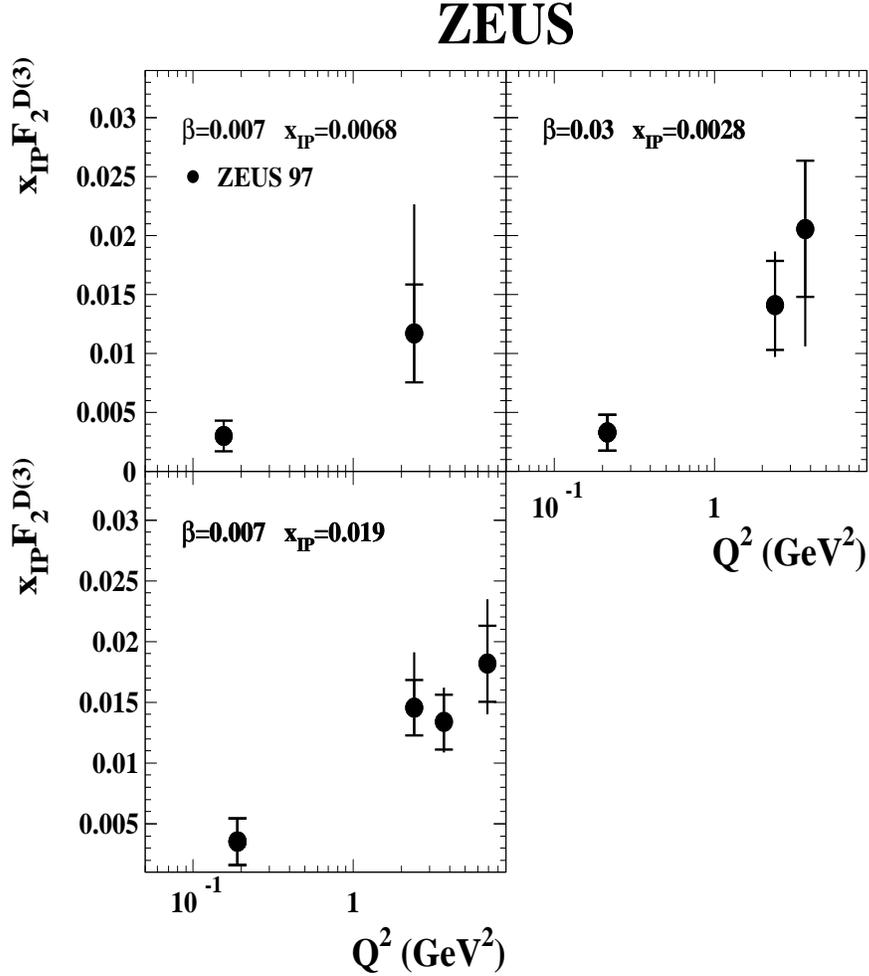}
\end{center}
\caption{The diffractive structure function multiplied by
$\xpom$, $\xpom F_2^{D(3)}(\beta,Q^2,\xpom)$, as
a function of $Q^2$, for the low- and high-$Q^2$ data, 
in the $\beta$-$\xpom$ region in which the two data-sets overlap. 
The inner error bars show the statistical uncertainties and the full bars 
indicate the statistical and the systematic uncertainties added in 
quadrature. The overall normalisation uncertainty of~$^{+12}_{-10}\%$ is 
not shown. The data shown here are those presented in Table 5 rebinned in 
$x_L$ slightly for plotting purposes.
}%
\label{f2d3bpcdisvsq2}
\vfill
\end{figure}

\begin{figure}[p]
\vfill
\begin{center}
\includegraphics[width=16cm,height=16cm]{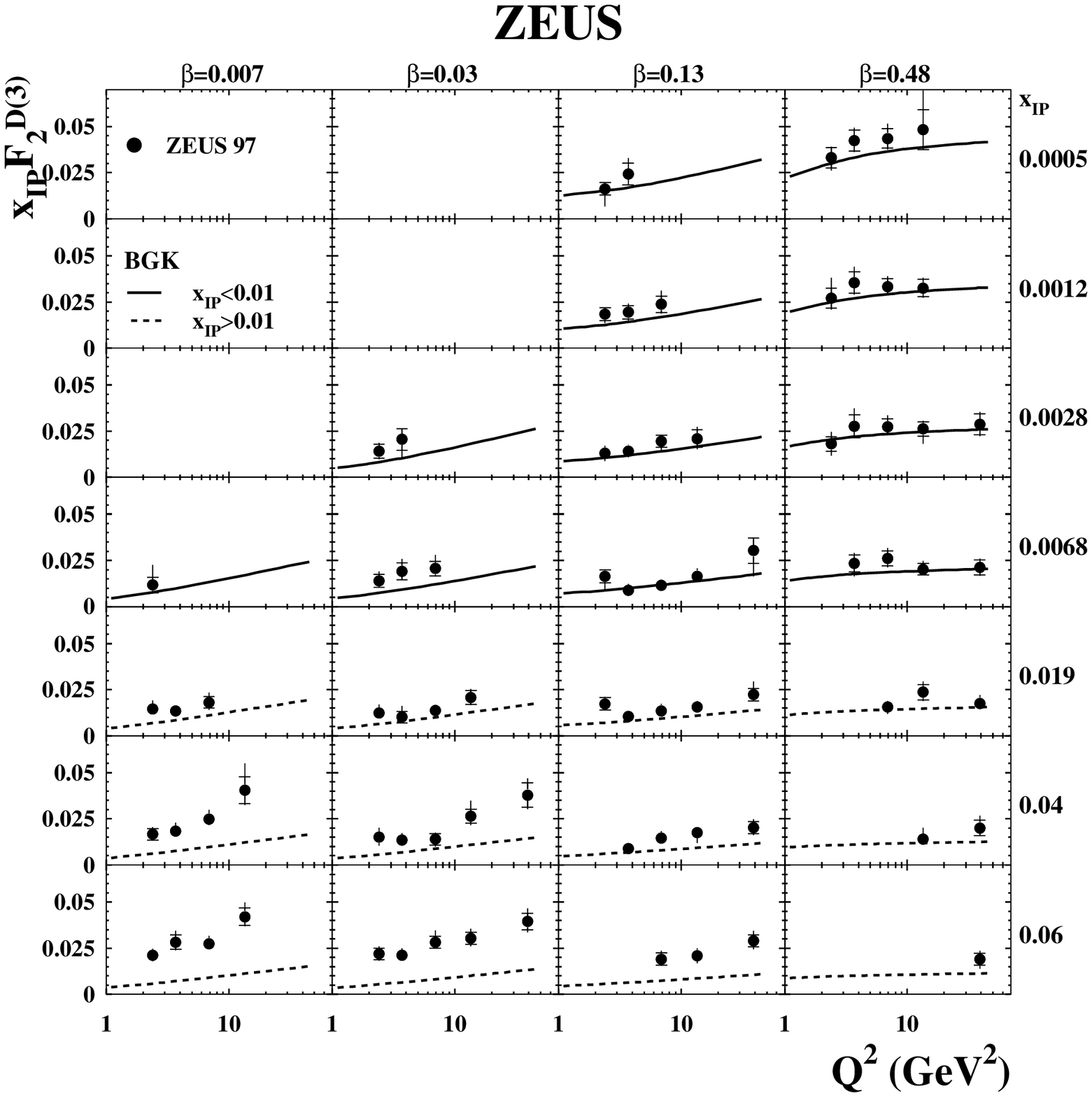}
\end{center}
\caption{The diffractive structure function multiplied by
$\xpom$, $\xpom F_2^{D(3)}(\beta,Q^2,\xpom)$, as
a function of $Q^2$, for different values of $\xpom$
and $\beta$.
The inner error bars show the statistical
uncertainties and the full bars indicate the statistical and the
systematic uncertainties added in quadrature. The overall
normalisation uncertainty of~$^{+12}_{-10}\%$ is not
shown. 
The solid lines are the prediction of the saturation
model of Bartels et al.~\protect\cite{satrap2} (BGK) discussed in 
Section~\protect\ref{saturation}; 
the dashed lines 
indicate the extrapolation of the model beyond $\xpom=0.01$.
} 
\label{fig-f2d_vs_q2}
\vfill
\end{figure}

\begin{figure}[p]
\vfill
\begin{center}
\includegraphics[width=16cm,height=16cm]{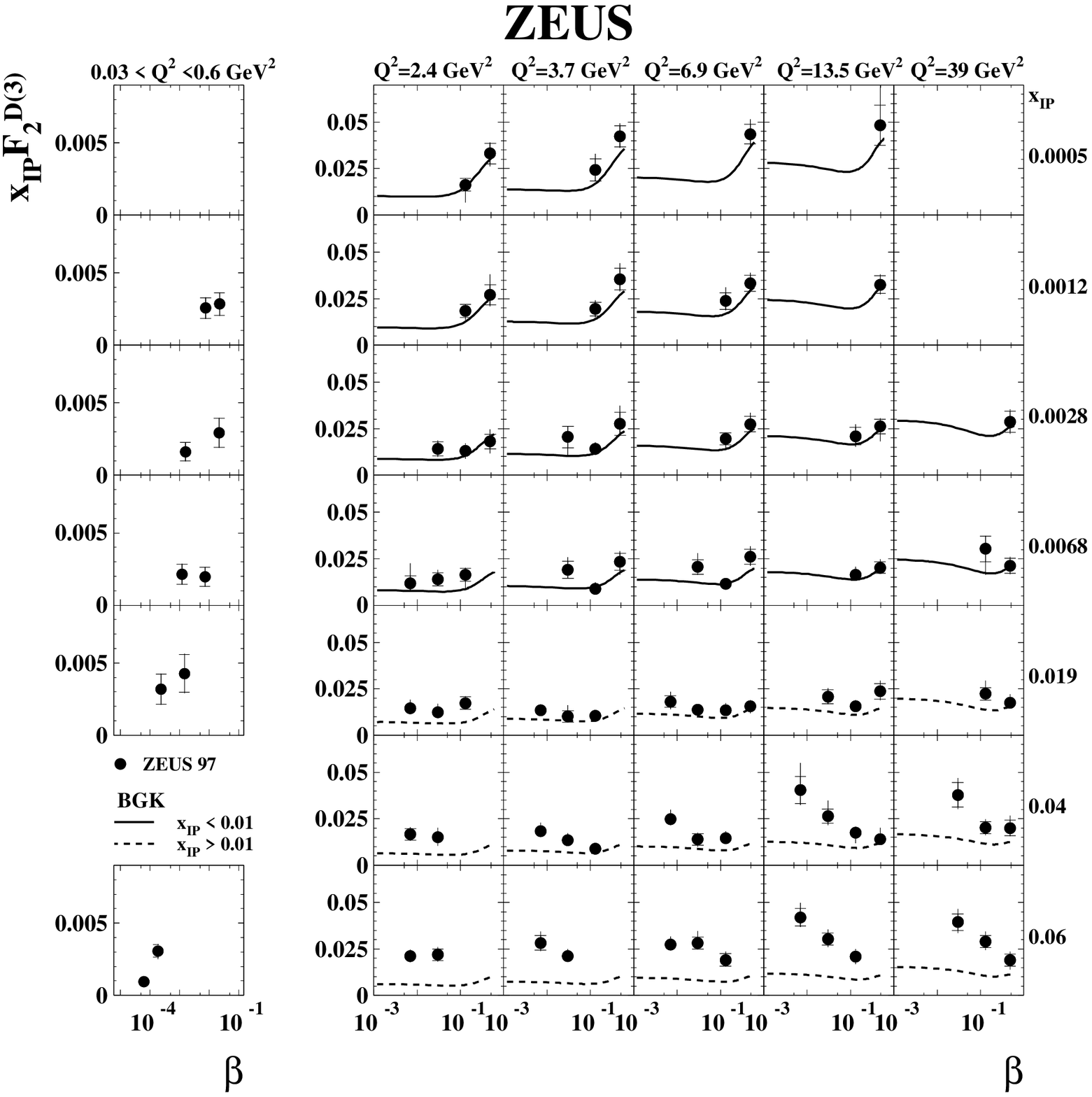}
\end{center}
\caption{The diffractive structure function multiplied by
$\xpom$, $\xpom F_2^{D(3)}(\beta,Q^2,\xpom)$, as
a function of $\beta$, for different values of $\xpom$
and $Q^2$.
The inner error bars show the statistical
uncertainties and the full bars indicate the statistical and the
systematic uncertainties added in quadrature. The overall
normalisation uncertainty of~$^{+12}_{-10}\%$ is not
shown. The low-$Q^2$ data in the bin labelled $\xpom=0.06$ have 
$\xpom=0.067$ and $\xpom=0.066$, respectively (see Table 6).
The solid lines are the prediction of the saturation
model of Bartels et al.~\protect\cite{satrap2} (BGK) discussed in 
Section~\protect\ref{saturation}; 
the dashed lines 
indicate the extrapolation of the model beyond $\xpom=0.01$.
}
\label{fig-f2d_vs_beta}
\vfill
\end{figure}

The dependences of the structure function on $Q^2$ and $\beta$
are presented in Figs.~\ref{f2d3bpcdisvsq2}-\ref{fig-f2d_vs_beta} 
for different values of $x_{\pom}$.
The structure function rises with $Q^2$ in all 
of the explored kinematic region. 
Figure~\ref{f2d3bpcdisvsq2}
compares the low-$Q^2$ and high-$Q^2$ results 
in the $\xpom$-$\beta$ region where they overlap.
Between the low- and high-$Q^2$ data, i.e. between $\langle Q^2 \rangle 
\simeq 0.15$ GeV$^2$ and $\langle Q^2\rangle \simeq 10.5$ GeV$^2$, the 
increase is  
about a factor five; this steep rise, $F_2^D \propto Q^2$, 
reflects the flattening of $d\sigma^{\gamma^{\star} p\rightarrow Xp}/dM_X$ 
for $Q^2 \rightarrow 0$ (see Eqs.~(\ref{sigma-1}),~(\ref{sigma-2})), and 
is a consequence of the conservation of the electromagnetic current. 
The $Q^2$ dependence 
becomes slower in the high-$Q^2$ data (Fig.~\ref{fig-f2d_vs_q2}). In this
region, the rise can be interpreted as a manifestation of QCD evolution; 
these positive scaling violations are due to the large gluon contribution 
to $F_2^{D}$ (see Section~\ref{qcdfits}). The behaviour is similar in all 
$\xpom$ bins, both in the 
diffractive-peak region and at larger $\xpom$ values, suggesting that the 
QCD evolution of the diffractive PDFs is largely independent of $\xpom$. 
The $\beta$ dependence of $F_2^{D(3)}$ (Fig.~\ref{fig-f2d_vs_beta}) has 
instead a different behaviour 
at different values of $\xpom$: at small $\xpom$, $F_2^{D(3)}$ has a weak 
$\beta$ dependence with a tendency to rise at large values of $\beta$; for 
$\xpom~\gsim~0.02$, $F_2^{D(3)}$ decreases as $\beta$ approaches unity, as 
expected in a hadron (bearing in mind that $\beta$ is the 
equivalent of Bjorken-$x$ for the exchange). The solid curves shown on 
Figs.~\ref{fig-f2d_vs_q2} and~\ref{fig-f2d_vs_beta} are the predictions 
from a pQCD-based model~\cite{satrap2} valid in the diffractive region  
and discussed in Section~\ref{saturation}.
 
\begin{figure}[p]
\vfill
\begin{center}
\includegraphics[width=16cm,height=16cm]{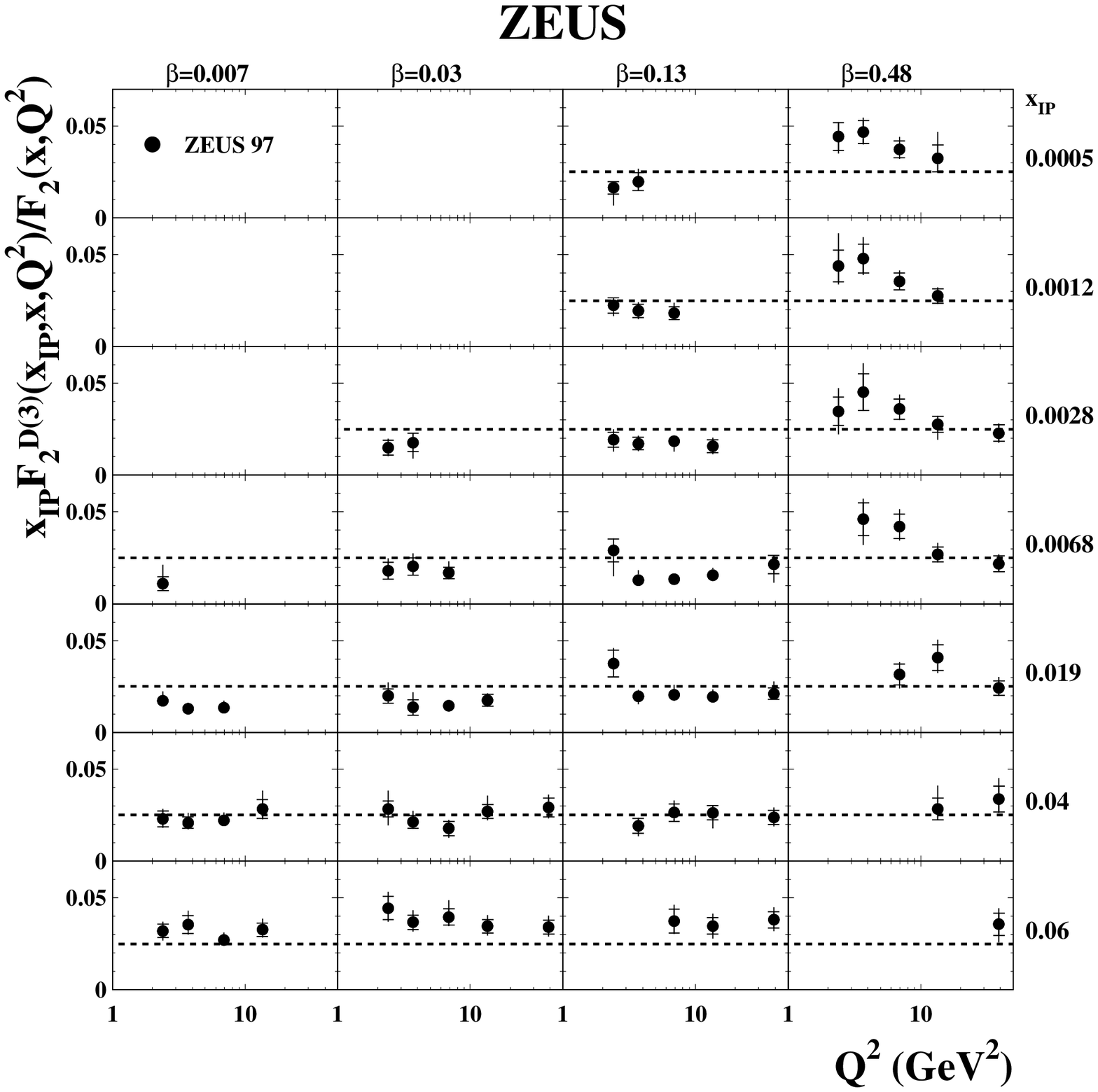}
\end{center}
\caption{
The ratio of the diffractive to the inclusive structure functions,
$\xpom F_2^{D(3)}(\xpom,x,Q^2)/F_2(x,Q^2)$, as a function of $Q^2$
at different values of $\xpom$ and $x$.
The values of $F_2(x,Q^2)$ were 
obtained from the ALLM97 parameterisation.
The inner error bars show the statistical
uncertainties and the full bars indicate the statistical and the
systematic uncertainties added in quadrature. The overall
normalisation uncertainty of~$^{+12}_{-10}\%$ is not
shown. 
The horizontal lines indicate the average value of the ratio and are 
only meant to guide the eye.
}
\label{fig-ratiof2d_vs_q2}
\vfill
\end{figure}

\begin{figure}[p]
\vfill
\begin{center}
\includegraphics[width=16cm,height=16cm]{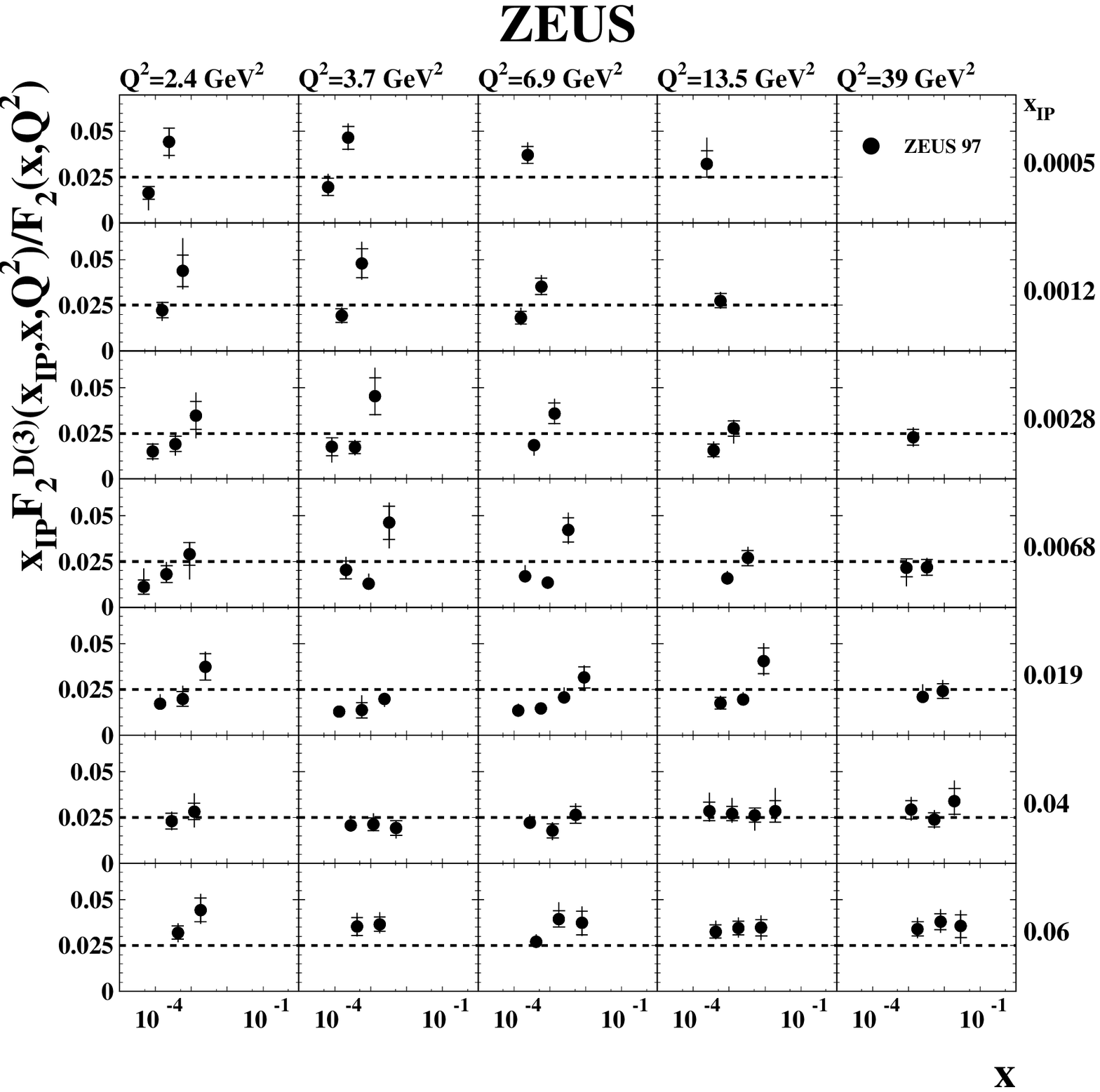}
\end{center}
\caption{
The ratio of the diffractive to the inclusive structure functions,
$\xpom F_2^{D(3)}(\xpom,x,Q^2)/F_2(x,Q^2)$, as a function of $x$
at different values of $\xpom$ and $Q^2$.
The values of $F_2(x,Q^2)$ were 
obtained from the ALLM97 parameterisation.
The inner error bars show the statistical
uncertainties and the full bars indicate the statistical and the
systematic uncertainties added in quadrature. The overall
normalisation uncertainty of~$^{+12}_{-10}\%$ is not
shown.  
The horizontal lines indicate the average value of the ratio and are 
only meant to guide the eye.
}
\label{fig-ratiof2d_vs_x}
\vfill
\end{figure}

The structure function $F_2^{D(3)}$ and the inclusive proton structure 
function $F_2$ are compared in terms of the ratio 
$x_{\pom}F_2^{D(3)}(x_{\pom},x,Q^2)/F_2(x,Q^2)$
calculated at fixed values of $x_{\pom}$. The values of $F_2(x,Q^2)$ were 
obtained from the ALLM97 parameterisation. 
The ratio is presented in Figs.~\ref{fig-ratiof2d_vs_q2} 
and~\ref{fig-ratiof2d_vs_x} as a function of $Q^2$ and $x$,  respectively. 
The $x$ range covered by the data is  $5 \times 10^{-5}~\lsim~x~\lsim~5 
\times 10^{-2}$.
The ratio is largely $Q^2$-independent, with possibly some structure at 
high $\beta$ and low $\xpom$. A $Q^2$-independent ratio would indicate 
equal scaling violations in the proton for the reaction 
$\gamma^{\star}p \rightarrow Xp$ and for inclusive DIS. 
The ratio grows with $x$ in the diffractive-peak 
region, suggesting that the $x$ dependence of the proton PDFs is different 
when the proton is probed 
in $\gamma^{\star}p \rightarrow Xp$ and in inclusive DIS. 
At higher $\xpom$, $\xpom~\gsim~0.02$, the ratio 
becomes flatter. 

\subsection{Comparison with the results of the \boldmath{$M_X$} method and  
of H1}
\label{sec-comparison}

In this section, the present results are compared with those obtained with 
the $M_X$ method~\cite{zeusdiff} and with the H1 results obtained with the  
rapidity-gap technique~\cite{h1diff}. 

The $M_X$-method analysis~\cite{zeusdiff} includes events in which the 
proton diffractively 
dissociates into a system $N$ of mass $M_N~\lsim~5.5$~GeV. In order to 
facilitate the comparison, the present data were replotted using the 
binning of the $M_X$-method publication~\cite{zeusdiff}.

\begin{figure}[p]
\vfill
\begin{center}
\includegraphics[width=16cm,height=16cm]{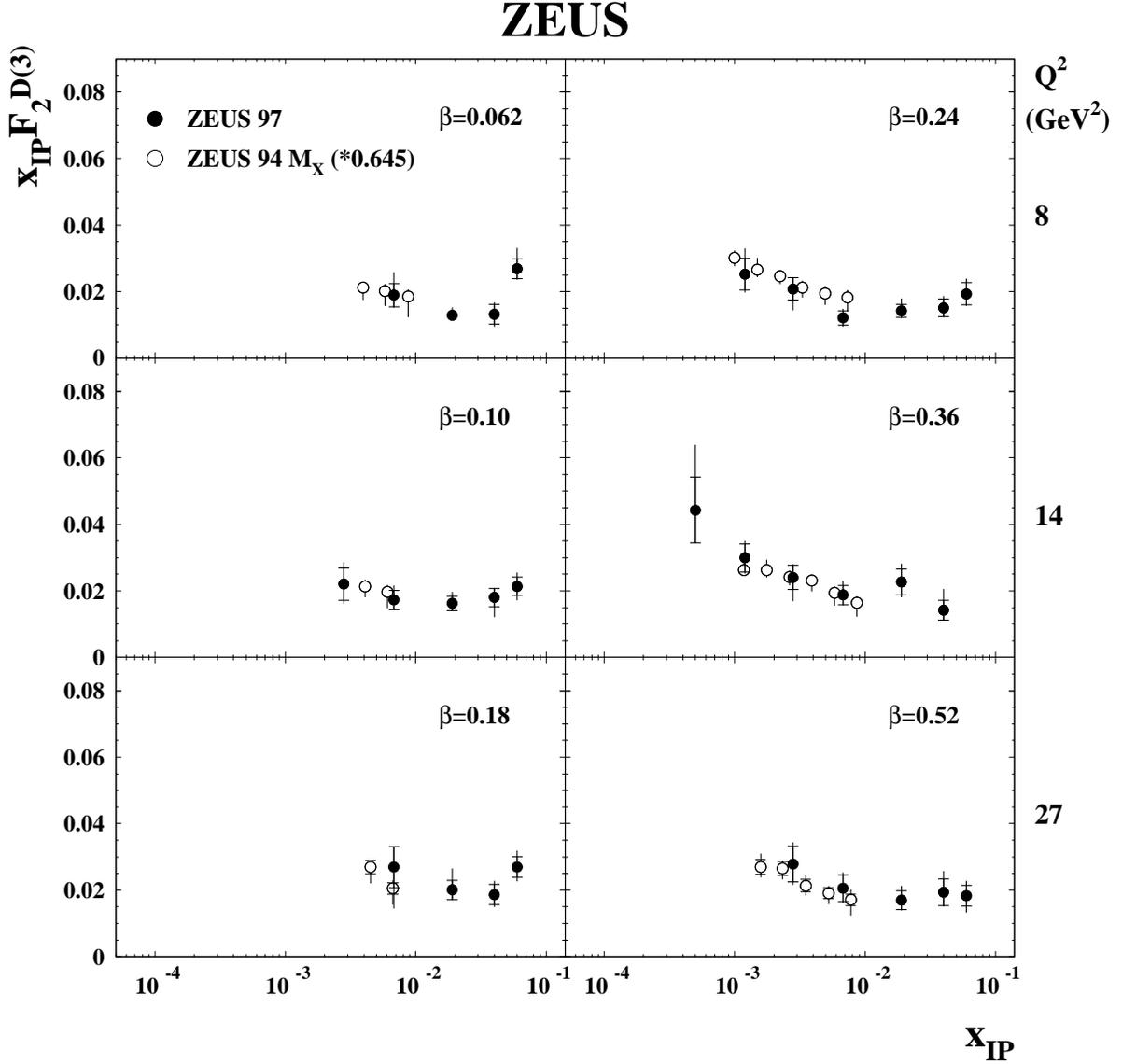}
\end{center}
\vspace{1.0cm}
\caption{
The diffractive structure function multiplied by
$\xpom$, $\xpom F_2^{D(3)}(\beta,Q^2,\xpom)$ as
a function of $\xpom$, for different values of $\beta$
and $Q^2$ for the LPS and $M_X$-method~\protect\cite{zeusdiff} 
analyses; the latter points 
are rescaled by $1/R_{M_X}=0.645$, as discussed in the text.
The inner error bars show the statistical
uncertainties and the full bars indicate the statistical and the
systematic uncertainties added in quadrature. 
The overall normalisation uncertainties of~ $^{+12}_{-10}\%$ (LPS data) 
and  $2\%$ ($M_X$-method data) are not shown.
}
\label{comparison3}
\vfill
\end{figure}

The $M_X$-method points are higher than those obtained with the 
LPS method. The difference was quantified by means of BEKW-type fits to 
the two data sets (see Section~\ref{BEKWfit}), 
which give a ratio of the $M_X$-method to LPS points of 
$R_{M_X}=1.55 \pm 0.08^{+0.15}_{-0.17}$, where the first error 
includes the statistical and systematic uncertainties and the second is 
due to the normalisation uncertainty.   
Figure~\ref{comparison3} shows $\xpom F_2^{D(3)}$ for the two data 
sets after scaling down the $M_X$-method points by $R_{M_X}$. The agreement 
between the LPS points and the renormalised $M_X$-method results is good, 
indicating that the difference is mainly in the normalisation. 
The normalisation difference can be attributed to 
the residual $p$-dissociative background in the $M_X$ method.
This background was estimated in previous studies~\cite{zeusdiff} 
by comparing the $M_X$ results~\cite{zeusdiff} and the earlier LPS 
data~\cite{lps94a}; there the ratio $R_{M_X}$ was found to be
$R_{M_X}=1.45 ^{+0.34}_{-0.23}$, consistent with the present result.
A similar study was performed~\cite{lps95} for the BPC
region, resulting in $R_{M_X}=1.85 \pm 0.38~(\rm 
stat.)$, which is also consistent with the present measurement.
The measured value of $R_{M_X}$ 
corresponds to a percentage  of $p$-dissociative events in the sample 
of $R_{\rm diss}=(1-1/R_{M_X})= [35.5 \pm 3.3^{+6.2}_{-7.1}]\%$; the first 
error corresponds  to the statistical and systematic uncertainties in 
quadrature, the second is due to the normalisation uncertainty. 

The agreement between the LPS and the 
$M_X$-method results, after taking the proton-dissociative background into 
account, lends support to the assumptions on which the $M_X$ method is 
based.

\begin{figure}[p]
\vfill
\begin{center}
\includegraphics[width=16cm,height=16cm]{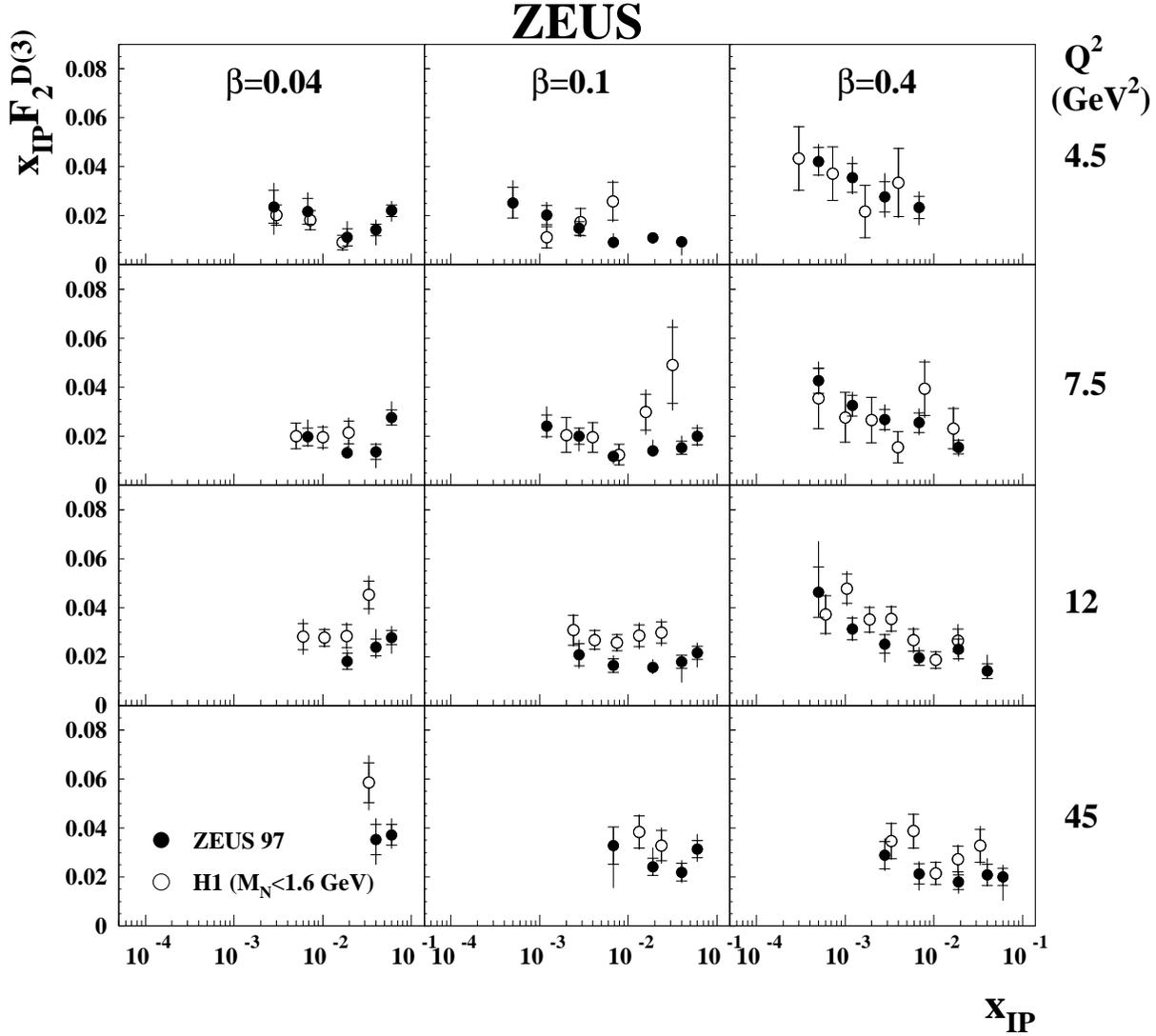}
\end{center}
\vspace{1.0cm}
\caption{
The diffractive structure function multiplied by
$\xpom$, $\xpom F_2^{D(3)}(\beta,Q^2,\xpom)$ as
a function of $\xpom$, for different values of $\beta$
and $Q^2$ for the LPS and the H1 data~\protect\cite{h1diff}.
The inner error bars show the statistical
uncertainties and the full bars are the statistical and the
systematic uncertainties added in quadrature. 
The overall normalisation uncertainties of~$^{+12}_{-10}\%$ (LPS data),
$\pm 6\%$ (H1 data, $Q^2<9$~{\rm GeV}$^2$) and $\pm 4.8\%$ (H1 data, 
$Q^2>9$~{\rm GeV}$^2$) are not shown.
}
\label{comparisonh1}
\vfill
\end{figure}

Figure~\ref{comparisonh1} shows a comparison of the present 
$F_2^{D(3)}$ results with those of the H1 collaboration obtained with the 
rapidity-gap selection~\cite{h1diff}. These data include a 
$p$-dissociative contribution with $M_N<1.6$~GeV. The data are plotted in 
terms of $\xpom F_2^{D(3)}$ as a function of $\xpom$ in different $\beta$ 
and $Q^2$ bins. The ZEUS points were extrapolated to the H1 bin centres 
using the measured dependences. At small $\xpom$, $\xpom<0.01$, the agreement 
is good, although with a tendency for the H1 points to be higher than the 
present 
results at high $Q^2$. While a normalisation difference is consistent with 
the presence of a $p$-dissociative contribution in the H1 data, a $Q^2$ 
dependence of this difference is not expected. The comparison 
indicates that the H1 data have a stronger $Q^2$ dependence than the 
present data.
For $\xpom>0.01$, the H1 data are also higher, but 
the shape is somewhat different, with a larger Reggeon-like contribution in 
the H1 data.

\subsection{Extraction of the Pomeron intercept}
\label{intercept}

In the framework of Regge phenomenology, the $\xpom$ dependence of 
$F_2^{D(3)}$ is related to the intercept of the Pomeron trajectory, 
the parameter that drives the energy dependence of the total 
hadron-hadron cross section at high energies~\cite{rev}. The Pomeron intercept 
has been determined to be $1.0964^{+0.0115}_{-0.0091}$~\cite{cudell} in soft 
hadronic interactions. However, the same parameter is significantly larger 
in the diffractive production of heavy vector mesons, notably in $J/\psi$ 
photoproduction (see e.g.~\cite{recent_review}), reflecting the rapid rise 
of the cross section with $W$. This is a consequence of the increase of 
the parton densities in the proton at low $x$, which drives the rise of the 
cross section with decreasing $x$, and hence with decreasing $\xpom$ (since 
$\xpom \propto 1/W^2 \propto x$). It is interesting to determine if such 
a deviation from the behaviour of the hadron-hadron data is also 
apparent in the inclusive diffractive dissociation of virtual photons.

The high-$Q^2$ data of Fig.~\ref{fig-f2d_vs_xpom} with $x_{\pom}<0.01$ 
were fitted to the form:
\begin{equation*}
F_2^{D(3)}(\beta,Q^2,x_{\pom})=f_{\pom}(x_{\pom})\cdot F_2^{\pom}(\beta,Q^2),
\end{equation*}
i.e. assuming ``Regge factorisation'' (see Section~\ref{sec-mc}). 
The Pomeron flux was  parameterised 
as~\cite{regge}
\begin{equation*}
f_{\pom}(x_{\pom})=\int{ 
\frac{e^{b_0^{\pom}t}}{x_{\pom}^{2\alpha_{\pom}(t)-1}}dt}.
\end{equation*}
The parameter  $\alpha_{\pom}'$  was set to $0.25$ GeV$^{-2}$, consistent
with the hadron-hadron data~\cite{dl}. The parameter $b_0^{\pom}$
was taken to be 4.67~GeV$^{-2}$, such that the relation 
$b=b_0^{\pom}-2\alpha_{\pom}'\ln{\xpom}$~\cite{rev} reproduces the
results of Section~\ref{sec-t}. The values of $F_2^{\pom}(\beta,Q^2)$
in each $\beta$ and $Q^2$ bin and the Pomeron intercept (assumed to be 
$\beta$ and $Q^2$ independent) were treated as free parameters.
The resulting Pomeron intercept is
\begin{equation*}
\alpha_{\pom}(0)=1.16\pm0.02~(\rm stat.)\pm0.02 (\rm syst.).
\end{equation*}
Varying $\alpha_{\pom}'$ in the range $0<\alpha_{\pom}'<0.4$ GeV$^{-2}$ 
causes $\alpha_{\pom}(0)$ to change by  
$^{+0.011}_{-0.030}$; the chosen range 
includes the small values of $\alpha_{\pom}'$ measured in the  diffractive 
photoproduction of $J/\psi$ mesons~\cite{jpsi}.
A variation of $R^{D}$ between 0 
and 1 produces a change of $\alpha_{\pom}(0)$ of ${+0.017}$.
The quality of the fit is good, with $\chi^2/ndf=14.2/22$ (considering 
statistical uncertainties only). 
The fact that the same value of $\alpha_{\pom}(0)$ fits the whole $Q^2$, 
$\beta$ region covered by the data indicates that, within the present 
accuracy, the hypothesis of Regge factorisation is a good 
approximation. The result does not change if the fit is extended to the 
low-$Q^2$ region; in this case $\chi^2/ndf=14.6/28$ (statistical 
uncertainties only). 
The result of this latter fit is shown in Fig.~\ref{fig-f2d_vs_xpom}.
The extrapolation of the fit for $\xpom >0.01$, where the contribution from 
the exchange of the Reggeon trajectory becomes important, is also shown; 
indeed the fit does not describe this region satisfactorily.
The present value of $\alpha_{\pom}(0)$ 
is consistent with that from H1, $\alpha_{\pom}(0)=1.203 \pm 0.020 
{\rm(stat.)} \pm  0.013 
{\rm(syst.)}^{+0.030}_{-0.035}{\rm(model)}$ (measured in the region 
$4.5<Q^2<75$~GeV$^2$)~\cite{h1diff}, 
and with the earlier ZEUS result, 
$\alpha_{\pom}(0)=1.157 \pm 0.009 {\rm(stat.)} ^{+0.039}_{-0.012}{\rm(syst.)}$ 
(measured in the region $7<Q^2<140$~GeV$^2$)~\cite{zeusdiff}; it is higher 
than that of the soft Pomeron, 
suggesting that the parton densities probed in inclusive 
diffractive $ep$ interactions also increase rapidly at small $x$ and 
that a single Pomeron trajectory cannot simultaneously describe the 
high-$Q^2$ diffractive data and the soft hadron-hadron data.

\subsection{Comparison with models}
\label{comp-models}

As discussed earlier, the diffractive dissociation of virtual photons can be 
described in pQCD since the virtuality of the photon provides a 
hard scale. In the proton rest frame, the reaction can be viewed as the 
sequence of the photon fluctuating into a $q\bar{q}$ (or  $q\bar{q}g$) 
colour dipole, the dipole scattering off the proton and producing the final 
state $X$. At high centre-of-mass energies, these processes are widely 
separated in time. The $q\bar{q}$, $q\bar{q}g$ fluctuations are described 
in terms of the 
photon wavefunction derived from QCD. The interaction of the
dipole with the proton is mediated, in the lowest order, by the exchange of 
two gluons in a colour-singlet state.

Several models of inclusive diffraction are available, which are discussed 
in review articles~\cite{recent_review, rev}. The discussion of 
this section is restricted to two approaches based on the framework 
just outlined. The data were fitted (Section~\ref{BEKWfit}) with a 
parameterisation based on the model of Bartels et al. (BEKW)~\cite{bekw}, 
which gives a satisfactory description of the earlier ZEUS 
results~\cite{zeusdiff,lps95}. It is interesting to see if the same 
parameterisation is able to describe the present data which cover a wider 
kinematic region; the fit is also a useful tool to compare the present data 
and those based on the ZEUS analysis using the $M_X$ method~\cite{zeusdiff}.

In  Section~\ref{saturation}, the results of this paper are also compared  
with the Golec-Biernat and W\"usthoff model based on the idea of the 
saturation of the dipole-proton cross section~\cite{gbw1,*gbw2,*gbw3}, which 
successfully describes both the inclusive $ep$ scattering data and earlier 
diffractive data. 

Finally, Section~\ref{qcdfits} describes the results of an 
NLO QCD fit to the present high-$Q^2$ data. In 
this approach, the $Q^2$ dependence of the data is interpreted as due to 
the QCD evolution of the diffractive PDFs. A parameterisation of the 
diffractive PDFs at a starting scale is evolved according to the QCD 
evolution equations and fitted to the data. 

\subsubsection{BEKW fit}
\label{BEKWfit}

In the BEKW model~\cite{bekw}, the dominant (leading-twist) contributions 
to the diffractive cross section in the kinematic domain of the present 
measurement come from fluctuations of transversely polarised virtual photons  
into either $q\bar{q}$ or $q\bar{q}g$ states. The $\beta$ (and hence 
$M_X$) spectra of these two components are determined by general 
properties of the photon wave-function, with the $q\bar{q}$ contribution 
to the cross section proportional to $\beta(1-\beta)$ and the $q\bar{q}g$ 
contribution proportional to $(1-\beta)^{\gamma}$, where $\gamma$ is a 
free parameter. For small values of 
$M_X$, the $q\bar{q}$ states dominate, while at large masses the 
$q\bar{q}g$ contribution becomes dominant. The model does not fix the $\xpom$ 
dependence of the $q\bar{q}$ and $q\bar{q}g$ contributions, but assumes 
for both a power-like behaviour, $\xpom^{-n(Q^2)}$, where the exponent $n$ 
is determined from fits to the data. More explicitly, in the BEKW approach, 
the diffractive structure function can be parameterised as
\begin{equation}
\xpom F_2^{D(3)}(\beta, \xpom,Q^2)= 
c_T F^T_{q\bar{q}}+c_L F^L_{q\bar{q}}+c_g F^T_{q\bar{q}g}, \nonumber
\end{equation}
where
\begin{equation}
F^T_{q\bar{q}}=(x_0/\xpom)^{n_T(Q^2)} \cdot \beta(1-\beta), \nonumber
\end{equation}
\begin{equation}
F^T_{q\bar{q}g}=(x_0/\xpom)^{n_g(Q^2)} \cdot \ln{(1+Q^2/Q_0^2)} \cdot
(1-\beta)^{\gamma}. \nonumber
\end{equation}
The contribution of longitudinal photons, $F^L_{q\bar{q}}$, which is 
relevant only at high $\beta$, was 
neglected in this analysis. A higher-twist term for $q\bar{q}$ states 
produced by transverse photons was also neglected.
In the original BEKW model~\cite{bekw}, the exponents $n_{T,g}(Q^2)$ are 
parameterised as $n_{T,g}(Q^2)=n_0^{T,g}+n_1^{T,g} \ln{(1+\ln{Q^2/Q_0^2})}$; here, 
this dependence was modified to 
$n_{T,g}(Q^2)=n_0^{T,g}+n_1^{T,g} \ln{(1+Q^2/Q_0^2)}$, which is well defined also  
when $Q^2 \rightarrow 0$. 

A fit was performed to the present high-$Q^2$ data using the 
parameterisation described above. The fit was limited 
to the region $\xpom<0.01$, well within the diffractive peak.
The parameters $Q^2_0$, $x_0$ and $n_0^{T,g}$ were taken to be
0.4~GeV$^2$, 0.01, and 0.13, respectively~\footnote{Since
$n_{T,g}(Q^2)=n_0^{T,g}+n_1^{T,g} \ln{(1+Q^2/Q_0^2)}$, $n_0$ gives the 
$\xpom$ 
dependence of $F_2^{D(3)}$ for $Q^2 \rightarrow 0$. If 
$F_2^{D(3)} \propto 1/\xpom^{2\bar{\alpha}_{\pom}-1}$ is assumed and 
$\alpha_{\pom}(t)$ is taken to be the soft Pomeron trajectory 
($\alpha_{\pom}(0)=1.096$, $\alpha_{\pom}'=0.25$~GeV$^{-2}$), then $n_0 =0.13$.}.  
The coefficients $c_T$, $c_g$, $n_1^{T,g}$ and $\gamma$ 
were determined in the fit, and have the following values:
$c_T=0.072\pm0.006 {\rm(stat.)}$, $c_g=0.008\pm 0.001 {\rm(stat.)}$,
$n_1^{T,g}=0.053\pm0.014 {\rm(stat.)}$, $\gamma=12.78\pm2.08 {\rm 
(stat.)}$.
The main features of the data are broadly reproduced by the fit, 
as shown in Fig.~\ref{fig-sigma_vs_q2}; the description 
of the $Q^2$ 
dependence of the diffractive to the inclusive cross-section ratio is also 
reasonable, as seen in Fig.~\ref{fig-ratio_vs_q2}. 
This indicates that the framework
in which the incoming virtual photon fluctuates into a colour dipole is, 
in general, adequate to describe diffractive processes in $ep$
collisions. At the same time, the data suggest the increasing importance 
of the contribution from $q\bar{q}g$ states at low $Q^2$, as indicated in 
Fig.~\ref{fig-sigma_vs_q2}. 
The fit gives only a qualitative description of the low-$Q^2$ sample, 
which is outside the region of applicability of pQCD; these points were 
not included in the fit. 
The fit is also lower than the high-$Q^2$ data in the 
high-$M_X$ bins that have
$\xpom>0.01$-0.02, suggesting that different mechanisms, such as Reggeon 
exchange, are at work in the diffractive-peak region and at high $\xpom$. 
In this region the discrepancy between the data and the fit can be taken 
as an estimate of the contribution to the cross section due to exchanges 
other than the Pomeron. 

\subsubsection{Saturation model}
\label{saturation}
In the saturation model by Golec-Biernat and 
W\"usthoff~\cite{gbw1,*gbw2,*gbw3}, diffractive DIS is also 
described in terms of the interaction of the $q\bar{q}$ ($q\bar{q}g$) 
fluctuation of the virtual photon with the proton. At high $Q^2$, the 
dipole-proton cross section is obtained from pQCD and is proportional to the 
square of the transverse size of the dipole, which is in turn proportional 
to $1/Q^2$. As $Q^2$ decreases, the rise of the cross section with the dipole 
size would violate unitarity and is tamed by requiring that it saturates at 
a typical value of the hadron-hadron cross section. The value of 
$Q^2$ at which saturation occurs is $x$-dependent.
The parameters of the model were
obtained from a fit to $F_2$ data. The latest modification of the 
model~\cite{satrap2}, denoted by BGK in the following, includes the QCD 
evolution of the gluon distribution.

\begin{figure}[p]
\vfill
\begin{center}
\includegraphics[width=16cm,height=16cm]{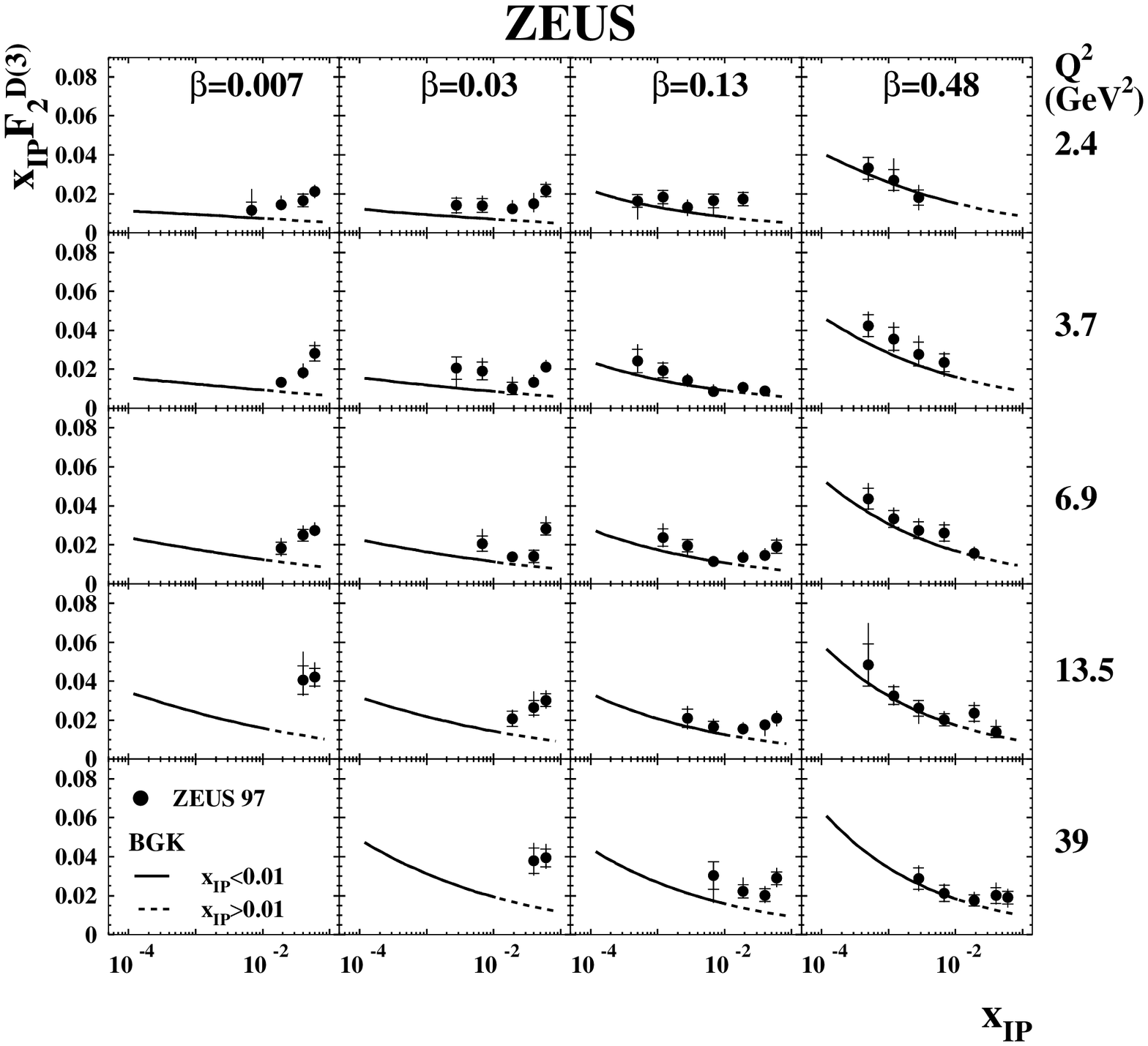}
\end{center}
\caption{The diffractive structure function multiplied by
$\xpom$, $\xpom F_2^{D(3)}(\beta,Q^2,\xpom)$, as
a function of $\xpom$, for different values of $\beta$
and $Q^2$. The inner error bars show the statistical
uncertainties and the full bars indicate the statistical and the
systematic uncertainties added in quadrature. The overall
normalisation uncertainty of~$^{+12}_{-10}\%$ is not
shown. The solid lines are the prediction of the saturation
model of Bartels et al.~\protect\cite{satrap2} (BGK) and are discussed in 
Section~\protect\ref{saturation}.
The dashed curves are the extrapolation of the model prediction
for $x_{\pom}>0.01$. 
}
\label{fig-f2d_vs_xpom_saturation}
\vfill
\end{figure}

Figures~\ref{fig-f2d_vs_q2},~\ref{fig-f2d_vs_beta} 
and \ref{fig-f2d_vs_xpom_saturation} show the comparison of the 
measured structure function $F_2^{D(3)}$ with the 
BGK prediction~\cite{satrap2}. In the region of applicability of the 
model, $x_{\pom}~\lsim~0.01$ and $Q^2$ larger than a few 
GeV$^2$, the $x_{\pom}$, 
$Q^2$ and $\beta$ dependences of $F_2^{D(3)}$ are adequately described, 
although the data are slightly higher than the model prediction.
The extrapolation of the model to large $\xpom$ 
values, beyond the Pomeron-dominated region, is significantly
lower than the data, and the discrepancy increases with $\xpom$; in this 
region, the $\beta$ dependences of the data and the model are 
also markedly different.

Both the BEKW and the saturation model imply Regge factorisation 
breaking. The fact that these models describe the data is not in 
contradiction with the possibility to fit the same data assuming Regge 
factorisation, as was done in Section~\ref{intercept},
since the magnitude of the predicted violation is smaller than
the precision of the present data.

\subsubsection{QCD fit}
\label{qcdfits}

An NLO QCD fit was performed to the present high-$Q^2$ data  together 
with the recent ZEUS results on diffractive charm production in 
DIS~\cite{charm}. The 
latter are important to constrain the gluon contribution to the 
diffractive PDFs. The fit was 
limited to the data in the region $Q^2>2$~GeV$^2$ and $\xpom<0.01$.
Regge factorisation was assumed, 
$F_2^{D(3)}(\beta,Q^2,x_{\pom})=f_{\pom}(x_{\pom})\cdot F_2^{\pom}(\beta,Q^2)$
(see Section~\ref{sec-mc}),
and the Pomeron flux was taken to be of the
Donnachie-Landshoff form~\cite{dl}
\begin{equation}
f_{\pom}(x_{\pom},t)=\frac{9 \beta^2_0}{4 \pi^2 \xpom^{2\alpha_{\pom}(t)-1}}
[F_1(t)]^2, \nonumber
\label{flux_DL}
\end{equation}
where $\beta_0=1.8$~GeV$^{-1}$, $F_1(t)$ is the elastic form factor of the 
proton,  $\alpha_{\pom}(0)$ was
fixed to the result given in Section~\ref{intercept} and $\alpha_{\pom}'$ 
was set to 0.25~GeV$^{-2}$. The results do not 
change if Eq.~(\ref{flux}) is used.

The diffractive parton distributions (quark flavour singlet and gluon)
were parameterised at the starting scale, $Q_0^2=2$ GeV$^2$, using the 
general polynomial form $zf(z)=(a_1+a_2z+a_3z^2)\cdot (1-z)^{a_4}$,
where $z$ is the parton fractional momentum. 
For the light quark distribution, 
it was assumed that $u=d=s=\bar{u}=\bar{d}=\bar{s}$; it was verified that 
setting the strange quark density to zero at the starting scale 
produces no appreciable change in the results.
Charm quarks were treated in the
Thorne-Roberts variable flavour number 
(TRVFN) scheme~\cite{thorne, *tr1, *tr2, *tr3}, with the charm-quark mass, 
$m_c$, set to 1.45 GeV. The NLO evolution package 
QCDNUM~\cite{qcdnum} was used to evolve the PDFs from the 
starting scale to the $Q^2$ values of each data point. The evolved 
PDFs were then fitted to the data.

\begin{figure}[p]
\vfill
\begin{center}
\includegraphics[width=16cm,height=16cm]{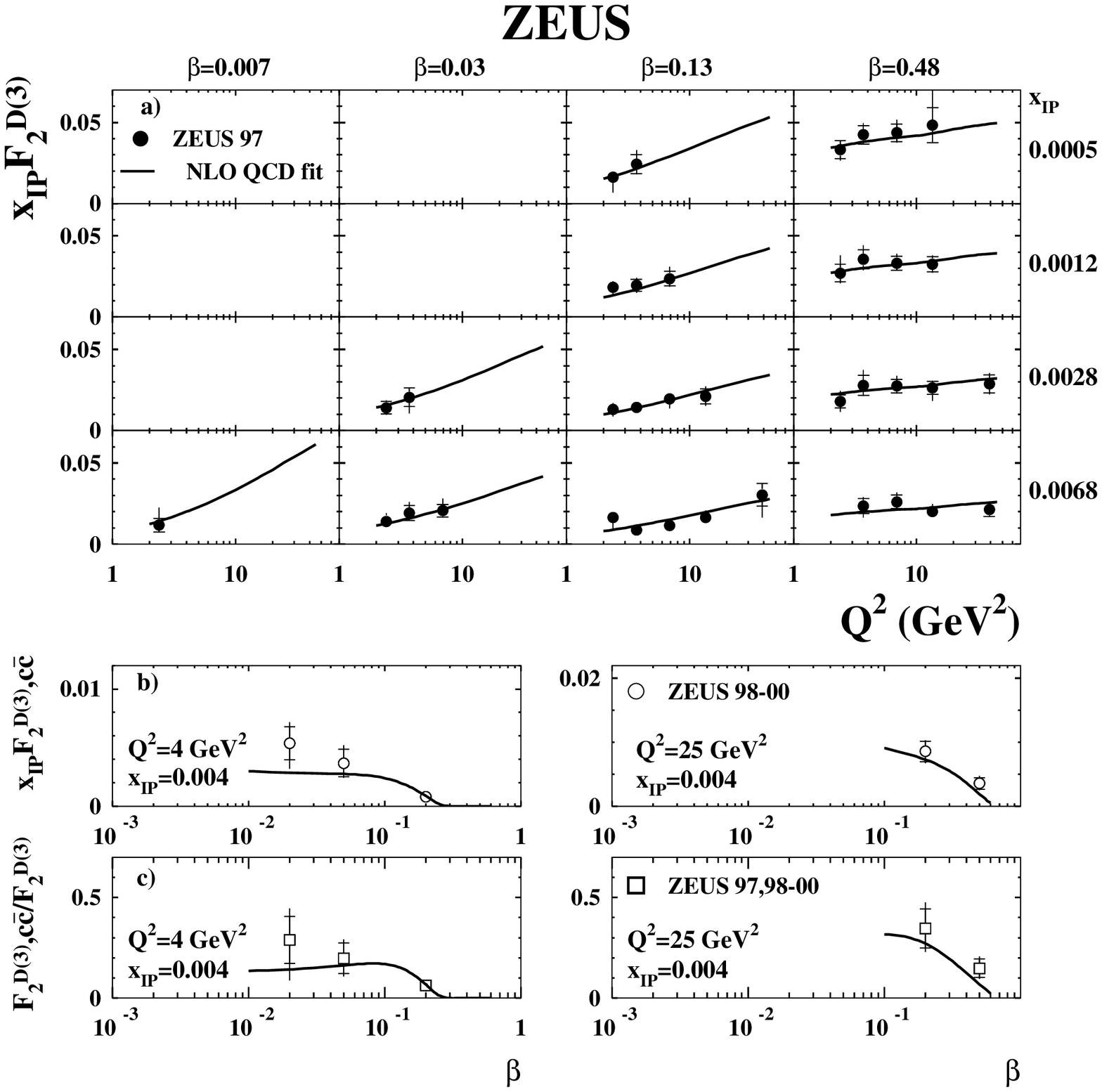}
\end{center}
\caption{
(a) The diffractive structure function multiplied by
$x_{\pom}$, $x_{\pom}F_2^{D(3)}(\beta,Q^2,x_{\pom})$, as
a function of $Q^2$, for different values of $x_{\pom}$
and $\beta$.
The inner error bars show the statistical
uncertainties and the full bars are the statistical and the
systematic uncertainties added in quadrature. The overall
normalisation uncertainty of~ $^{+12}_{-10}\%$ is not
shown.
(b) The measured charm contribution to the diffractive structure
function  multiplied by $x_{\pom}$, 
$x_{\pom}F_2^{D(3),c\bar{c}}(\beta,Q^2,x_{\pom})$, 
as
a function of $\beta$, for different values of  $Q^2$ and  
$x_{\pom}=0.004$~\protect\cite{charm}.
(c) The ratio of $F_2^{D(3),c\bar{c}}$~\protect\cite{charm}
and the present $F_2^{D(3)}$ measurement as a function of $\beta$.
The solid lines are the result of QCD NLO fit described in the
text. 
}
\label{qcdfit}
\vfill
\end{figure}

The result of the fit is shown by the lines in Fig.~\ref{qcdfit}. They 
satisfactorily reproduce the measurements, with $\chi^2/ndf=37.8/36$ 
(statistical errors only). The resulting fraction of the $t$-channel momentum
carried by gluons is $(82 \pm 8 {\rm (stat.)} ^{+5}_{-16} {\rm (syst.)})\%$ at 
$Q^2=2$~GeV$^2$, consistent with earlier ZEUS~\cite{diffr_php} and 
H1~\cite{h1diff} results, but higher than that found in a recent QCD 
analysis of the same data by Martin, Ryskin and Watt~\cite{mrw}. The 
systematic uncertainty includes the 
contributions listed in Section~\ref{sec-sys}; in addition, the charm 
quark mass was varied between 1.3 and 1.6 GeV, and the relative normalisation 
between the charm and the $F_2^{D(3)}$ data was changed by 
$^{+11}_{-13}\%$, reflecting the uncertainty on the proton-dissociative 
background and on the luminosity in the charm data, as well as the 
normalisation uncertainty of the LPS data. The fixed-flavour-number scheme
(FFNS) was used instead of TRVFN, without any significant change of
the results.
Various PDF parameterisations at the starting scale were tried, including 
the function used by H1~\cite{h1diff}. 
The shape of the fitted PDFs changes significantly depending on the 
functional form of the initial parameterisation, a consequence of
the relatively large statistical uncertainties of the 
present sample. Therefore, these data cannot 
constrain the shapes of the PDFs.
However, the integrals over $z$ of the fitted PDFs and notably 
the fraction of the $t$-channel momentum carried by gluons
are robust and change only slightly with the parameterisation chosen. 
This contribution was included in the systematic uncertainties quoted. 

Also shown in Fig.~\ref{qcdfit} is the ratio of the charm contribution to 
the diffractive structure function, $F_2^{D(3),c\bar{c}}$~\cite{charm}, 
and the present $F_2^{D(3)}$ results as a function of $\beta$ (see 
Table~\ref{charm}). The ratio increases with increasing $Q^2$ and 
decreasing $\beta$, up to values of $30\%$.
The ratio is well described by the fit.

\section{Summary}
\label{sec-con}

New measurements have been presented of the reaction $e^+p
\rightarrow e^+Xp$ in the regions $0.03 < Q^2 <0.60$~GeV$^2$ 
and $2<Q^2 <100$ GeV$^2$. The scattered proton was measured 
in the ZEUS leading proton spectrometer, and was required to carry
a fraction $x_L$  of the incoming proton momentum of at least 
$90\%$. The data cover the region $0.075<|t|<0.35$~GeV$^2$.

The results can be summarised as follows:

\begin{itemize}

\item the $t$ dependence of the cross section is exponential, with a 
$t$-slope $b=7.9 \pm0.5 {\rm (stat.)} ^{+0.8}_{-0.5} {\rm 
(syst.)}$ GeV$^{-2}$ for $\xpom<0.01$. The  slope is independent of $Q^2$ 
but decreases with $\xpom$;

\item  there is no observed $\Phi$ dependence of the cross section, 
indicating that the interference terms between the longitudinal and 
transverse amplitudes and between the two transverse amplitudes are 
consistent with zero in the measured kinematic region;

\item the cross-section $d\sigma^{\gamma^*p \rightarrow Xp}/dM_X$ falls 
rapidly with $Q^2$ at high $Q^2$ but approaches a constant as 
$Q^2 \rightarrow 0$. This behaviour is similar to that of the total 
photon-proton cross section, and is a consequence of the conservation of the 
electromagnetic current. In detail, it was found that the cross-section 
$d\sigma^{\gamma^*p \rightarrow Xp}/dM_X$ falls 
with $Q^2$ more slowly than $\sigma_{\tot}^{\gamma^{\star}p}$ 
when $Q^2<M_X^2$ and faster than $\sigma_{\tot}^{\gamma^{\star}p}$ when 
$Q^2>M_X^2$;

\item the data were also analysed in terms of the structure function 
$F_2^{D(3)}$. The $\xpom$, $Q^2$ and $\beta$ dependences of $F_2^{D(3)}$ 
were studied;

\begin{itemize}

\item $F_2^{D(3)}$ falls with $\xpom$ faster than $1/\xpom$ for 
$\xpom~\lsim~0.01$ and more slowly at larger values of $\xpom$. From the 
$\xpom$ dependence of $F_2^{D(3)}$ at low $\xpom$, the Pomeron intercept 
$\alpha_{\pom}(0)$ was measured to be $\alpha_{\pom}(0)=1.16\pm0.02~(\rm 
stat.)\pm0.02 (\rm syst.)$, higher than that of the soft Pomeron, and 
similar to that measured in the photoproduction of heavy vector mesons. 
This suggests that, also in the present reaction, the virtual photon 
probes the proton in a region where the parton density increases quickly 
with decreasing $x$.

\item $F_2^{D(3)}$ rises with $Q^2$ over the whole measured region. The 
increase is very significant, about a factor five, between the low-$Q^2$ 
and the high-$Q^2$ region. In the high-$Q^2$ region, the rise becomes 
softer, and is reminiscent of the logarithmic scaling violations of the 
proton structure function. Positive scaling violations reflect a large 
gluon density. This is confirmed by an NLO QCD analysis of the present 
data 
for $\xpom<0.01$ in conjunction with the earlier ZEUS results on 
diffractive charm production~\cite{charm}. The analysis indicates that 
the fraction of the  $t$-channel 
momentum carried by gluons is $(82 \pm 8 {\rm (stat.)} ^{+5}_{-16} {\rm 
(syst.)})\%$ at
$Q^2=2$ GeV$^2$. However, the present data are not precise enough to 
constrain the shapes of the PDFs.

\item The $\beta$ dependence of $F_2^{D(3)}$ changes with $\xpom$. For
$\xpom~\lsim~0.01$, $F_2^{D(3)}$ grows with $\beta$. 
For values of $\xpom~\gsim~0.01$-0.02, $F_2^{D(3)}$ decreases with 
$\beta$. The 
latter behaviour is similar to that of the structure functions of hadrons 
as a function of $x$, and is consistent with the hypothesis that, at large 
$\xpom$, the $t$-channel exchange mediating the photon-proton interaction 
is a meson-like object;

\end{itemize}

\item the results presented are consistent, in the small $\xpom$ region, 
with the predictions of pQCD-based models of diffraction. In particular, 
the data were compared with models in which the virtual photon fluctuates 
into $q\bar{q}$ or $q\bar{q}g$ colour dipoles which then interact with the 
proton via the exchange of a gluon pair.

\end{itemize}

\section*{Acknowledgements}
\label{sec-ack}
We thank the DESY Directorate for their support and encouragement. We are
grateful for the support of the DESY computing and network services.
We are specially grateful to the HERA machine group: collaboration with 
them was crucial to the successful installation and operation of the 
leading proton spectrometer. 
The design, construction and installation of the ZEUS detector have
been made possible by the ingenuity and effort of many people 
who are not listed as authors. It is also
a pleasure to thank M.~Diehl, K.~Golec-Biernat, O.~Nachtmann, 
N.N.~Nikolaev, M.G. Ryskin and G. Watt for many useful discussions. We are 
grateful to K.~Golec-Biernat for providing the prediction of his model for the 
kinematic range covered by our data.

\vfill\eject

\include{DESY-04-131-ref}
\providecommand{\etal}{et al.\xspace}
\providecommand{\coll}{Coll.\xspace}
\catcode`\@=11
\def\@bibitem#1{%
\ifmc@bstsupport
  \mc@iftail{#1}%
    {;\newline\ignorespaces}%
    {\ifmc@first\else.\fi\orig@bibitem{#1}}
  \mc@firstfalse
\else
  \mc@iftail{#1}%
    {\ignorespaces}%
    {\orig@bibitem{#1}}%
\fi}%
\catcode`\@=12


\clearpage
\begin{table}
\begin{center}
\begin{tabular}{|c|c|c|c|}
\hline
$\langle Q^2 \rangle$ (GeV$^2$) & $\langle \beta \rangle$ & $\langle \xpom 
\rangle$ & $b$ (GeV$^{-2}$) \\
\hline
$ 7.1$&$ 0.37$&$ 0.001 $&$ 8.13\pm 0.68^{+ 0.95}_{- 0.53}$ \\
$ 10.2$&$ 0.24$&$ 0.006$&$ 7.87\pm 0.83^{+ 1.21}_{- 0.66}$ \\
$ 11.5$&$ 0.14$&$ 0.019$&$ 6.14\pm 0.82^{+ 1.73}_{- 0.68}$ \\
$ 11.7$&$ 0.07$&$ 0.04 $&$ 4.19\pm 0.62^{+ 0.83}_{- 0.75}$ \\
$ 12.2$&$ 0.05$&$ 0.06 $&$ 4.19\pm 0.40^{+ 0.42}_{- 0.30}$ \\
\hline
\end{tabular}
\caption{
Fitted values of the $t$-slopes. 
The first uncertainty is statistical, the 
second systematic.
}
\label{tablet}
\end{center}
\end{table}

\clearpage
\begin{table}
\begin{center}
\begin{tabular}{|c|c|c|c|c|}
\hline
$\langle Q^2 \rangle$ (GeV$^2$) &
$\langle \beta \rangle$ &
$\langle \xpom \rangle$ &
$\langle |t| \rangle $ (GeV$^2$) &
$A_{LT}$ \\
\hline
$13.0$&$ 0.48$& $0.0009$&$0.145  $&$            -0.06  \pm 0.13^{+0.12}_{-0.08}$ \\
$17.7$&$ 0.34$& $0.0029$&$0.145  $&$ \phantom{+} 0.06  \pm 0.16^{+0.04}_{-0.17}$ \\
$20.3$&$ 0.27$& $0.0068$&$0.145  $&$            -0.02  \pm 0.15^{+0.09}_{-0.04}$ \\
$21.1$&$ 0.18$& $0.0190$&$0.147  $&$ \phantom{+} 0.02  \pm 0.11^{+0.09}_{-0.04}$ \\
$22.5$&$ 0.11$& $0.041 $&$0.151  $&$            -0.15  \pm 0.13^{+0.05}_{-0.22}$ \\
$23.0$&$ 0.08$& $0.061 $&$0.161  $&$ \phantom{+}0.10  \pm 0.08^{+0.10}_{-0.04}$ \\
\hline
$9.6 $&$0.04  $&$0.0064$&$0.145  $&$          -0.05\pm 0.29^{+0.15}_{-0.16}$ \\
$14.4 $&$0.13 $&$0.0042$&$0.145  $&$          -0.19\pm 0.19^{+0.15}_{-0.23}$ \\
$17.7 $&$0.38 $&$0.0030$&$0.145  $&$ \phantom{+}0.03\pm 0.12^{+0.08}_{-0.05}$ \\
$20.1 $&$0.73 $&$0.0024$&$0.145  $&$ \phantom{+}0.14\pm 0.19^{+0.18}_{-0.08}$ \\
\hline
$17.0  $&$0.37 $&$0.0035$&$0.085  $&$  \phantom{+}0.03\pm  0.16^{+0.07}_{-0.09}$ \\
$17.0  $&$0.37 $&$0.0035$&$0.11   $&$  \phantom{+}0.00\pm  0.17^{+0.11}_{-0.09}$ \\
$17.0  $&$0.37 $&$0.0035$&$0.15   $&$  \phantom{+}0.02\pm  0.18^{+0.09}_{-0.07}$ \\
$17.0  $&$0.37 $&$0.0035$&$0.24   $&$           -0.14\pm  0.17^{+0.15}_{-0.05}$ \\
\hline
$7.8  $&$0.28 $&$0.0025$&$0.145  $&$ \phantom{+}0.01 \pm 0.16^{+0.08}_{-0.13}$ \\
$11.8 $&$0.37 $&$0.0033$&$0.145  $&$          -0.13\pm 0.15^{+0.07}_{-0.08}$ \\
$16.9 $&$0.42 $&$0.0039$&$0.145  $&$ \phantom{+}0.01 \pm 0.18^{+0.16}_{-0.02}$ \\
$36.0 $&$0.44 $&$0.0046$&$0.145  $&$ \phantom{+}0.10 \pm 0.19^{+0.07}_{-0.11}$ \\
\hline
\end{tabular}
\caption{Fitted values of the asymmetry parameter $A_{LT}$. 
The first uncertainty is statistical, the second systematic. 
}
\label{tablephi}
\end{center}
\end{table}

\clearpage
\begin{table}
\renewcommand{\arraystretch}{1.3}
\def\largestrut{\vrule width 0pt height 30pt depth 20pt\relax}
\begin{center}
\begin{tabular}{|c|c|c|r@{$\pm$}l|}
\hline
$\langle Q^2 \rangle$ (GeV$^2$) &
$\langle M_X \rangle$ (GeV) &
$\langle W \rangle$ (GeV) &
\multicolumn{2}{c|}{$d\sigma^{\gamma^*p\rightarrow Xp}/dM_X$ ($\mu$b/GeV) } \\
\hline
$ 0.33 $&$ 5  $&$ 100 $&$ ~~~~~~~0.469  $&$ 0.095 $$^{+ 0.037}_{- 0.031}$ \\
$ 0.29 $&$ 5  $&$ 160 $&$ 0.355   $&$ 0.075 $$^{+ 0.035}_{- 0.021}$ \\
$ 0.24 $&$ 5  $&$ 190 $&$ 0.60\phantom{0}$&$ 0.14\phantom{0}$$^{+ 0.07\phantom{0}} _{- 0.04\phantom{0}}  $ \\
$ 0.14 $&$ 5  $&$ 245 $&$ 0.61\phantom{0}$&$ 0.22\phantom{0}$$^{+ 0.05\phantom{0}} _{- 0.04\phantom{0}} $ \\
\hline
$ 0.34 $&$ 22 $&$ 100 $&$ 0.113   $&$ 0.020 $$^{+ 0.009}_{- 0.010}$ \\
$ 0.34 $&$ 22 $&$ 160 $&$ 0.115   $&$ 0.029 $$^{+ 0.011}_{- 0.007}$ \\
$ 0.27 $&$ 22 $&$ 190 $&$ 0.095   $&$ 0.041 $$^{+ 0.009}_{- 0.004}$ \\
\hline
$ 0.33 $&$ 40 $&$ 160 $&$ 0.046   $&$ 0.010 $$^{+ 0.003}_{- 0.004}$ \\
$ 0.24 $&$ 40 $&$ 190 $&$ 0.055   $&$ 0.012 $$^{+ 0.004}_{- 0.005}$ \\
$ 0.13 $&$ 40 $&$ 245 $&$ 0.081   $&$ 0.021 $$^{+ 0.006}_{- 0.005}$ \\
$ 0.09 $&$ 40 $&$ 275 $&$ 0.218   $&$ 0.059 $$^{+ 0.025}_{- 0.030}$ \\
\hline
\end{tabular}
\caption{
The diffractive cross section, $d\sigma^{\gamma^*p\rightarrow Xp}/dM_X$, 
for the low-$Q^2$ sample. The first uncertainty given is statistical, 
the second systematic. 
}
\label{tab-fig6lowq2}
\end{center}
\end{table}

\clearpage
\begin{table}
\begin{center}
\begin{tabular}{|c|c|c|c|c|}
\hline
$Q^2$ (GeV$^2$) &
$M_X$ (GeV)  &
$W$ (GeV) &
$d\sigma^{\gamma^{\star}p \to Xp}/dM_X$ ($\mu$b/GeV) \\
\hline
$     2.7$&$     5$&$   100$&$  0.179\pm  0.022^{+  0.030}_{-  0.006}$\\
$     4.5$&$     5$&$   100$&$  0.119\pm  0.013^{+  0.013}_{-  0.008}$\\
$    10.0$&$     5$&$   100$&$  0.074\pm  0.006^{+  0.009}_{-  0.005}$\\
$    35.0$&$     5$&$   100$&$  0.015\pm  0.002^{+  0.002}_{-  0.001}$\\
\hline
$     2.7$&$    22$&$   100$&$  0.082\pm  0.006^{+  0.010}_{-  0.004}$\\
$     4.5$&$    22$&$   100$&$  0.052\pm  0.004^{+  0.007}_{-  0.004}$\\
$    10.0$&$    22$&$   100$&$  0.025\pm  0.001^{+  0.003}_{-  0.001}$\\
$    35.0$&$    22$&$   100$&$  0.009\pm  0.001^{+  0.001}_{-  0.001}$\\
\hline
$     2.7$&$     5$&$   160$&$  0.230\pm  0.027^{+  0.021}_{-  0.019}$\\
$     4.5$&$     5$&$   160$&$  0.178\pm  0.028^{+  0.011}_{-  0.011}$\\
$    10.0$&$     5$&$   160$&$  0.082\pm  0.008^{+  0.007}_{-  0.005}$\\
$    35.0$&$     5$&$   160$&$  0.015\pm  0.002^{+  0.002}_{-  0.001}$\\
\hline
$     2.7$&$    22$&$   160$&$  0.048\pm  0.006^{+  0.007}_{-  0.002}$\\
$     4.5$&$    22$&$   160$&$  0.033\pm  0.005^{+  0.003}_{-  0.004}$\\
$    10.0$&$    22$&$   160$&$  0.016\pm  0.002^{+  0.002}_{-  0.001}$\\
$    35.0$&$    22$&$   160$&$  0.007\pm  0.001^{+  0.001}_{-  0.001}$\\
\hline
$     2.7$&$    40$&$   160$&$  0.051\pm  0.007^{+  0.010}_{-  0.007}$\\
$     4.5$&$    40$&$   160$&$  0.037\pm  0.004^{+  0.005}_{-  0.002}$\\
$    10.0$&$    40$&$   160$&$  0.020\pm  0.002^{+  0.003}_{-  0.001}$\\
$    35.0$&$    40$&$   160$&$  0.006\pm  0.001^{+  0.001}_{-  0.000}$\\
\hline
$     2.7$&$     5$&$   190$&$  0.240\pm  0.035^{+  0.021}_{-  0.026}$\\
$     4.5$&$     5$&$   190$&$  0.163\pm  0.024^{+  0.016}_{-  0.012}$\\
$    10.0$&$     5$&$   190$&$  0.098\pm  0.012^{+  0.009}_{-  0.016}$\\
$    35.0$&$     5$&$   190$&$  0.021\pm  0.004^{+  0.002}_{-  0.003}$\\
\hline
$     2.7$&$    22$&$   190$&$  0.064\pm  0.012^{+  0.007}_{-  0.008}$\\
$     4.5$&$    22$&$   190$&$  0.041\pm  0.009^{+  0.005}_{-  0.004}$\\
$    10.0$&$    22$&$   190$&$  0.019\pm  0.003^{+  0.002}_{-  0.002}$\\
$    35.0$&$    22$&$   190$&$  0.009\pm  0.002^{+  0.001}_{-  0.001}$\\
\hline
$     2.7$&$    40$&$   190$&$  0.046\pm  0.006^{+  0.005}_{-  0.004}$\\
$     4.5$&$    40$&$   190$&$  0.040\pm  0.005^{+  0.004}_{-  0.004}$\\
$    10.0$&$    40$&$   190$&$  0.019\pm  0.002^{+  0.003}_{-  0.001}$\\
$    35.0$&$    40$&$   190$&$  0.007\pm  0.001^{+  0.001}_{-  0.001}$\\
\hline
\end{tabular}
\caption{
The diffractive cross section, $d\sigma^{\gamma^*p\rightarrow Xp}/dM_X$, 
for the high-$Q^2$ sample. The first uncertainty given is statistical, 
the second systematic. 
}
\label{tablecrosshi}
\end{center}
\end{table}

\clearpage
\begin{table}
\renewcommand{\arraystretch}{1.3}
\def\largestrut{\vrule width 0pt height 30pt depth 20pt\relax}
\begin{center}
\begin{tabular}{|c|c|c|r@{$\pm$}l|}
\hline
$\langle Q^2 \rangle$ (GeV$^2$) &
$\langle \beta \rangle$&
$\langle \xpom \rangle$&
\multicolumn{2}{c|}{$\xpom F_2^{D(3)}$ } \\
\hline
$ 0.10 $&$ 0.0002  $&$ 0.023   $&$ 0.00122 $&$ 0.00045 $$^{+ 0.00015}_{- 0.00007}$ \\
$ 0.14 $&$ 0.0002  $&$ 0.066   $&$ 0.00307 $&$ 0.00044 $$^{+ 0.00022}_{- 0.00037}$ \\
\hline
$ 0.12 $&$ 0.0006  $&$ 0.011   $&$ 0.00156 $&$ 0.00062 $$^{+ 0.00016}_{- 0.00011}$ \\
$ 0.15 $&$ 0.0006  $&$ 0.050   $&$ 0.00376 $&$ 0.00055 $$^{+ 0.00028}_{- 0.00036}$ \\
\hline
$ 0.12 $&$ 0.0019  $&$ 0.0033  $&$ 0.0024\phantom{0}   $&$ 0.0010\phantom{0}  $$^{+ 0.0003\phantom{0}} _{-0.0001\phantom{0}}$ \\
$ 0.15 $&$ 0.0019  $&$ 0.0077  $&$ 0.00239  $&$ 0.00092 $$^{+ 0.00027}_{- 0.00011}$ \\
$ 0.16 $&$ 0.0019  $&$ 0.022   $&$ 0.0034\phantom{0}   $&$ 0.0014\phantom{0}  $$^{+ 0.0004\phantom{0}}_{-0.0005\phantom{0}}$ \\
\hline
$ 0.12 $&$ 0.008   $&$ 0.0010  $&$ 0.00246  $&$ 0.00079 $$^{+ 0.00026}_{- 0.00034}$ \\
$ 0.15 $&$ 0.007   $&$ 0.0027  $&$ 0.00177  $&$ 0.00075 $$^{+ 0.00017}_{- 0.00020}$ \\
$ 0.16 $&$ 0.007   $&$ 0.0057  $&$ 0.0031\phantom{0}   $&$ 0.0014\phantom{0}  $$^{+ 0.0003\phantom{0}} _{-0.0002\phantom{0}}$ \\
$ 0.19 $&$ 0.006   $&$ 0.018   $&$ 0.0036\phantom{0}   $&$ 0.0019\phantom{0}  $$^{+ 0.0004\phantom{0}} _{-0.0003\phantom{0}}$ \\
\hline
$ 0.17 $&$ 0.03    $&$ 0.00021 $&$ 0.0036\phantom{0}   $&$ 0.0012\phantom{0}  $$^{+ 0.0004\phantom{0}} _{-0.0002\phantom{0}}$ \\
$ 0.20 $&$ 0.03    $&$ 0.00046 $&$ 0.00262  $&$ 0.00094 $$^{+ 0.00025}_{- 0.00012}$ \\
$ 0.21 $&$ 0.03    $&$ 0.0011  $&$ 0.0032\phantom{0}   $&$ 0.0011\phantom{0}  $$^{+ 0.0002\phantom{0}}_{-0.0003\phantom{0}}$ \\
$ 0.22 $&$ 0.03    $&$ 0.0031  $&$ 0.0032\phantom{0}   $&$ 0.0015\phantom{0}  $$^{+ 0.0003\phantom{0}}_{-0.0002\phantom{0}}$ \\
\hline
\end{tabular}
\caption{
The diffractive structure function multiplied by $\xpom$, $\xpom 
F_2^{D(3)}(\beta,Q^2,\xpom)$, for the low-$Q^2$ sample.
The first uncertainty given is statistical, the second systematic. 
}
\label{tab-fig9}
\end{center}
\end{table}

\clearpage
\begin{table}
\renewcommand{\arraystretch}{1.3}
\def\largestrut{\vrule width 0pt height 30pt depth 20pt\relax}
\begin{center}
\begin{tabular}{|c|c|c|r@{$\pm$}l|}
\hline
$\langle Q^2   \rangle$ (GeV$^2$) &
$\langle \beta \rangle$ &
$\langle \xpom \rangle$&
\multicolumn{2}{c|}{$\xpom F_2^{D(3)}$ } \\
\hline
$0.13$&$ 0.0077  $&$ 0.0012  $&$ 0.00257 $&$ 0.00071$$^{+ 0.00029}_{- 0.00028}$ \\
$0.16$&$ 0.023   $&$ 0.0012  $&$ 0.00284 $&$ 0.00077$$^{+ 0.00025}_{- 0.00017}$ \\
\hline
$0.11$&$ 0.0016  $&$ 0.0028  $&$ 0.00162 $&$ 0.00065$$^{+ 0.00011}_{- 0.00018}$ \\
$0.19$&$ 0.022   $&$ 0.0028  $&$ 0.0029\phantom{0}  $&$ 0.0010\phantom{0}$$^{+ 0.0002\phantom{0}}  _{- 0.0002\phantom{0}} $ \\
\hline
$0.13$&$ 0.0012  $&$ 0.0068  $&$ 0.00214 $&$ 0.00069$$^{+ 0.00031}_{- 0.00010}$ \\
$0.16$&$ 0.0073  $&$ 0.0068  $&$ 0.00197 $&$ 0.00067$$^{+ 0.00019}_{- 0.00013}$ \\
\hline
$0.10$&$ 0.00024  $&$ 0.019   $&$ 0.0032\phantom{0} $&$ 0.0010\phantom{0}$$^{+ 0.0005\phantom{0}}  _{- 0.0002\phantom{0}} $ \\
$0.15$&$ 0.0015   $&$ 0.019   $&$ 0.0043\phantom{0} $&$ 0.0013\phantom{0}$$^{+ 0.0003\phantom{0}}  _{- 0.0005\phantom{0}} $ \\
\hline
$0.10$&$ 0.000062 $&$ 0.067   $&$ 0.00094 $&$ 0.00018$$^{+0.00007}_{- 0.00005}$ \\
$0.14$&$ 0.00019  $&$ 0.066   $&$ 0.00307 $&$ 0.00044$$^{+ 0.00022}_{- 0.00037}$ \\
\hline
\end{tabular}
\caption{
The diffractive structure function multiplied by $\xpom$, $\xpom 
F_2^{D(3)}(\beta,Q^2,\xpom)$, for the low-$Q^2$ sample.
The first uncertainty given is statistical, the second systematic.
These are the points plotted in Fig. 12. The data in this table are not 
independent of those in Table~5.
}
\label{tab-fig10-12}
\end{center}
\end{table}

\footnotesize
\clearpage
\begin{table}
\renewcommand{\arraystretch}{1.3}
\def\largestrut{\vrule width 0pt height 30pt depth 20pt\relax}
\begin{center}
\begin{tabular}{|c|c|c|r@{$\pm$}l|}
\hline
$Q^2$ (GeV$^2$) &$\beta$ &$\xpom$ &
\multicolumn{2}{c|}{$\xpom F_2^{D(3)}$ }\\
\hline
$ 2.4$&$ 0.0070$&$ 0.0068$&$ 0.0117$&$ 0.0041$$^{+ 0.0101}_{- 0.0006}$ \\
$ 2.4$&$ 0.0070$&$ 0.0190$&$ 0.0146$&$ 0.0023$$^{+ 0.0039}_{- 0.0000}$ \\
$ 2.4$&$ 0.0070$&$ 0.0400$&$ 0.0166$&$ 0.0032$$^{+ 0.0021}_{- 0.0014}$ \\
$ 2.4$&$ 0.0070$&$ 0.0600$&$ 0.0212$&$ 0.0023$$^{+ 0.0023}_{- 0.0025}$ \\
\hline
$ 2.4$&$ 0.0300$&$ 0.0028$&$ 0.0141$&$ 0.0038$$^{+ 0.0025}_{- 0.0022}$ \\
$ 2.4$&$ 0.0300$&$ 0.0068$&$ 0.0139$&$ 0.0035$$^{+ 0.0038}_{- 0.0003}$ \\
$ 2.4$&$ 0.0300$&$ 0.0190$&$ 0.0123$&$ 0.0025$$^{+ 0.0038}_{- 0.0002}$ \\
$ 2.4$&$ 0.0300$&$ 0.0400$&$ 0.0150$&$ 0.0023$$^{+ 0.0047}_{- 0.0040}$ \\
$ 2.4$&$ 0.0300$&$ 0.0600$&$ 0.0218$&$ 0.0031$$^{+ 0.0030}_{- 0.0019}$ \\
\hline
$ 2.4$&$ 0.1300$&$ 0.0005$&$ 0.0163$&$ 0.0033$$^{+ 0.0000}_{- 0.0089}$\\
$ 2.4$&$ 0.1300$&$ 0.0012$&$ 0.0184$&$ 0.0035$$^{+ 0.0011}_{- 0.0034}$\\
$ 2.4$&$ 0.1300$&$ 0.0028$&$ 0.0131$&$ 0.0028$$^{+ 0.0029}_{- 0.0033}$\\
$ 2.4$&$ 0.1300$&$ 0.0068$&$ 0.0164$&$ 0.0035$$^{+ 0.0013}_{- 0.0071}$\\
$ 2.4$&$ 0.1300$&$ 0.0190$&$ 0.0173$&$ 0.0033$$^{+ 0.0019}_{- 0.0007}$\\
\hline
$ 2.4$&$ 0.4800$&$ 0.0005$&$ 0.0331$&$ 0.0056$$^{+ 0.0019}_{- 0.0039}$\\
$ 2.4$&$ 0.4800$&$ 0.0012$&$ 0.0271$&$ 0.0054$$^{+ 0.0097}_{- 0.0030}$\\
$ 2.4$&$ 0.4800$&$ 0.0028$&$ 0.0181$&$ 0.0040$$^{+ 0.0052}_{- 0.0051}$\\
\hline
$ 3.7$&$ 0.0070$&$ 0.0190$&$ 0.0134$&$ 0.0022$$^{+ 0.0017}_{- 0.0011}$\\
$ 3.7$&$ 0.0070$&$ 0.0400$&$ 0.0183$&$ 0.0027$$^{+ 0.0037}_{- 0.0014}$\\
$ 3.7$&$ 0.0070$&$ 0.0600$&$ 0.0282$&$ 0.0039$$^{+ 0.0046}_{- 0.0017}$\\
\hline
$ 3.7$&$ 0.0300$&$ 0.0028$&$ 0.0206$&$ 0.0058$$^{+ 0.0008}_{- 0.0081}$\\
$ 3.7$&$ 0.0300$&$ 0.0068$&$ 0.0191$&$ 0.0046$$^{+ 0.0050}_{- 0.0007}$\\
$ 3.7$&$ 0.0300$&$ 0.0190$&$ 0.0102$&$ 0.0032$$^{+ 0.0051}_{- 0.0001}$\\
$ 3.7$&$ 0.0300$&$ 0.0400$&$ 0.0134$&$ 0.0022$$^{+ 0.0031}_{- 0.0005}$\\
$ 3.7$&$ 0.0300$&$ 0.0600$&$ 0.0211$&$ 0.0022$$^{+ 0.0031}_{- 0.0018}$\\
\hline
\end{tabular}
\caption{
The diffractive structure function multiplied by $\xpom$, $\xpom 
F_2^{D(3)}(\beta,Q^2,\xpom)$, for the high-$Q^2$ sample, part I.
The first uncertainty given is statistical, the second systematic.
} \label{tabf2dis1}
\end{center}
\end{table}

\clearpage
\begin{table}
\renewcommand{\arraystretch}{1.3}
\def\largestrut{\vrule width 0pt height 30pt depth 20pt\relax}
\begin{center}
\begin{tabular}{|c|c|c|r@{$\pm$}l|}
\hline
$Q^2$ (GeV$^2$) &
$\beta$ &
$\xpom$ &
\multicolumn{2}{c|}{$\xpom F_2^{D(3)}$ } \\
\hline
$ 3.7$&$ 0.1300$&$ 0.0005$&$ 0.0243$&$ 0.0060$$^{+ 0.0064}_{- 0.0012}$\\
$ 3.7$&$ 0.1300$&$ 0.0012$&$ 0.0194$&$ 0.0037$$^{+ 0.0029}_{- 0.0024}$\\
$ 3.7$&$ 0.1300$&$ 0.0028$&$ 0.0142$&$ 0.0027$$^{+ 0.0022}_{- 0.0019}$\\
$ 3.7$&$ 0.1300$&$ 0.0068$&$ 0.0087$&$ 0.0019$$^{+ 0.0030}_{- 0.0000}$\\
$ 3.7$&$ 0.1300$&$ 0.0190$&$ 0.0106$&$ 0.0016$$^{+ 0.0009}_{- 0.0016}$\\
$ 3.7$&$ 0.1300$&$ 0.0400$&$ 0.0089$&$ 0.0019$$^{+ 0.0007}_{- 0.0018}$\\
\hline
$ 3.7$&$ 0.4800$&$ 0.0005$&$ 0.0423$&$ 0.0056$$^{+ 0.0045}_{- 0.0024}$\\
$ 3.7$&$ 0.4800$&$ 0.0012$&$ 0.0355$&$ 0.0058$$^{+ 0.0064}_{- 0.0033}$\\
$ 3.7$&$ 0.4800$&$ 0.0028$&$ 0.0277$&$ 0.0062$$^{+ 0.0074}_{- 0.0008}$\\
$ 3.7$&$ 0.4800$&$ 0.0068$&$ 0.0234$&$ 0.0046$$^{+ 0.0034}_{- 0.0055}$\\
\hline
$ 6.9$&$ 0.0070$&$ 0.0190$&$ 0.0182$&$ 0.0031$$^{+ 0.0043}_{- 0.0027}$\\
$ 6.9$&$ 0.0070$&$ 0.0400$&$ 0.0250$&$ 0.0030$$^{+ 0.0040}_{- 0.0013}$\\
$ 6.9$&$ 0.0070$&$ 0.0600$&$ 0.0274$&$ 0.0024$$^{+ 0.0033}_{- 0.0019}$\\
\hline
$ 6.9$&$ 0.0300$&$ 0.0068$&$ 0.0206$&$ 0.0038$$^{+ 0.0065}_{- 0.0017}$\\
$ 6.9$&$ 0.0300$&$ 0.0190$&$ 0.0138$&$ 0.0017$$^{+ 0.0020}_{- 0.0003}$\\
$ 6.9$&$ 0.0300$&$ 0.0400$&$ 0.0139$&$ 0.0031$$^{+ 0.0020}_{- 0.0025}$\\
$ 6.9$&$ 0.0300$&$ 0.0600$&$ 0.0281$&$ 0.0032$$^{+ 0.0057}_{- 0.0013}$\\
\hline
$ 6.9$&$ 0.1300$&$ 0.0012$&$ 0.0237$&$ 0.0044$$^{+ 0.0058}_{- 0.0025}$\\
$ 6.9$&$ 0.1300$&$ 0.0028$&$ 0.0196$&$ 0.0032$$^{+ 0.0010}_{- 0.0051}$\\
$ 6.9$&$ 0.1300$&$ 0.0068$&$ 0.0114$&$ 0.0020$$^{+ 0.0016}_{- 0.0014}$\\
$ 6.9$&$ 0.1300$&$ 0.0190$&$ 0.0136$&$ 0.0018$$^{+ 0.0031}_{- 0.0011}$\\
$ 6.9$&$ 0.1300$&$ 0.0400$&$ 0.0147$&$ 0.0026$$^{+ 0.0025}_{- 0.0006}$\\
$ 6.9$&$ 0.1300$&$ 0.0600$&$ 0.0190$&$ 0.0033$$^{+ 0.0031}_{- 0.0012}$\\
\hline
$ 6.9$&$ 0.4800$&$ 0.0005$&$ 0.0436$&$ 0.0053$$^{+ 0.0060}_{- 0.0026}$\\
$ 6.9$&$ 0.4800$&$ 0.0012$&$ 0.0333$&$ 0.0043$$^{+ 0.0037}_{- 0.0013}$\\
$ 6.9$&$ 0.4800$&$ 0.0028$&$ 0.0275$&$ 0.0042$$^{+ 0.0044}_{- 0.0014}$\\
$ 6.9$&$ 0.4800$&$ 0.0068$&$ 0.0260$&$ 0.0041$$^{+ 0.0042}_{- 0.0024}$\\
$ 6.9$&$ 0.4800$&$ 0.0190$&$ 0.0156$&$ 0.0028$$^{+ 0.0017}_{- 0.0024}$\\
\hline
\end{tabular}
\caption{
The diffractive structure function multiplied by $\xpom$, $\xpom 
F_2^{D(3)}(\beta,Q^2,\xpom)$, for the high-$Q^2$ sample, part II. 
The first uncertainty given is statistical, the second systematic. 
}
\label{tabf2dis2}
\end{center}
\end{table}

\clearpage
\begin{table}
\renewcommand{\arraystretch}{1.3}
\def\largestrut{\vrule width 0pt height 30pt depth 20pt\relax}
\begin{center}
\begin{tabular}{|c|c|c|r@{$\pm$}l|c|}
\hline
$Q^2$ (GeV$^2$) &
$\beta$ &
$\xpom$ &
\multicolumn{2}{c|}{$\xpom F_2^{D(3)}$ } \\
\hline
$13.5$&$ 0.0070$&$ 0.0400$&$ 0.0406$&$ 0.0073$$^{+ 0.0126}_{- 0.0036}$\\
$13.5$&$ 0.0070$&$ 0.0600$&$ 0.0420$&$ 0.0046$$^{+ 0.0061}_{- 0.0028}$\\
\hline
$13.5$&$ 0.0300$&$ 0.0190$&$ 0.0207$&$ 0.0039$$^{+ 0.0025}_{- 0.0014}$\\
$13.5$&$ 0.0300$&$ 0.0400$&$ 0.0264$&$ 0.0037$$^{+ 0.0074}_{- 0.0032}$\\
$13.5$&$ 0.0300$&$ 0.0600$&$ 0.0303$&$ 0.0032$$^{+ 0.0041}_{- 0.0030}$\\
\hline
$13.5$&$ 0.1300$&$ 0.0028$&$ 0.0210$&$ 0.0046$$^{+ 0.0043}_{- 0.0033}$\\
$13.5$&$ 0.1300$&$ 0.0068$&$ 0.0165$&$ 0.0028$$^{+ 0.0032}_{- 0.0010}$\\
$13.5$&$ 0.1300$&$ 0.0190$&$ 0.0157$&$ 0.0021$$^{+ 0.0026}_{- 0.0008}$\\
$13.5$&$ 0.1300$&$ 0.0400$&$ 0.0175$&$ 0.0027$$^{+ 0.0010}_{- 0.0050}$\\
$13.5$&$ 0.1300$&$ 0.0600$&$ 0.0210$&$ 0.0027$$^{+ 0.0030}_{- 0.0032}$\\
\hline
$13.5$&$ 0.4800$&$ 0.0005$&$ 0.0483$&$ 0.0108$$^{+ 0.0185}_{- 0.0011}$\\
$13.5$&$ 0.4800$&$ 0.0012$&$ 0.0326$&$ 0.0046$$^{+ 0.0032}_{- 0.0023}$\\
$13.5$&$ 0.4800$&$ 0.0028$&$ 0.0262$&$ 0.0040$$^{+ 0.0017}_{- 0.0068}$\\
$13.5$&$ 0.4800$&$ 0.0068$&$ 0.0202$&$ 0.0031$$^{+ 0.0034}_{- 0.0009}$\\
$13.5$&$ 0.4800$&$ 0.0190$&$ 0.0236$&$ 0.0041$$^{+ 0.0040}_{- 0.0026}$\\
$13.5$&$ 0.4800$&$ 0.0400$&$ 0.0140$&$ 0.0030$$^{+ 0.0056}_{- 0.0002}$\\
\hline
$39.0$&$ 0.0300$&$ 0.0400$&$ 0.0379$&$ 0.0066$$^{+ 0.0063}_{- 0.0038}$\\
$39.0$&$ 0.0300$&$ 0.0600$&$ 0.0394$&$ 0.0045$$^{+ 0.0057}_{- 0.0037}$\\
\hline
$39.0$&$ 0.1300$&$ 0.0068$&$ 0.0303$&$ 0.0070$$^{+ 0.0007}_{- 0.0122}$\\
$39.0$&$ 0.1300$&$ 0.0190$&$ 0.0223$&$ 0.0033$$^{+ 0.0063}_{- 0.0005}$\\
$39.0$&$ 0.1300$&$ 0.0400$&$ 0.0203$&$ 0.0033$$^{+ 0.0031}_{- 0.0023}$\\
$39.0$&$ 0.1300$&$ 0.0600$&$ 0.0289$&$ 0.0033$$^{+ 0.0041}_{- 0.0033}$\\
\hline
$39.0$&$ 0.4800$&$ 0.0028$&$ 0.0287$&$ 0.0056$$^{+ 0.0040}_{- 0.0032}$\\
$39.0$&$ 0.4800$&$ 0.0068$&$ 0.0212$&$ 0.0041$$^{+ 0.0028}_{- 0.0019}$\\
$39.0$&$ 0.4800$&$ 0.0190$&$ 0.0176$&$ 0.0029$$^{+ 0.0033}_{- 0.0012}$\\
$39.0$&$ 0.4800$&$ 0.0400$&$ 0.0200$&$ 0.0042$$^{+ 0.0052}_{- 0.0017}$\\
$39.0$&$ 0.4800$&$ 0.0600$&$ 0.0190$&$ 0.0033$$^{+ 0.0033}_{- 0.0041}$\\
\hline
\end{tabular}
\caption{
The diffractive structure function multiplied by $\xpom$, $\xpom 
F_2^{D(3)}(\beta,Q^2,\xpom)$, for the high-$Q^2$ sample, part III.
The first uncertainty given is statistical, the second systematic. 
}
\label{tabf2dis3}
\end{center}
\end{table}

\clearpage
\begin{table}
\renewcommand{\arraystretch}{1.3}
\def\largestrut{\vrule width 0pt height 30pt depth 20pt\relax}
\begin{center}
\begin{tabular}{|c|c|c|c|}
\hline
$Q^2$ (GeV$^2$) &
$\beta$ &
$\xpom$ &
$F_2^{D(3),c\bar{c}}/F_2^{D(3)}$ \\
\hline
$ 4.0 $&$ 0.02 $&$ 0.004 $&$ 0.29 \pm 0.12 ^{+ 0.10}_{- 0.16}$ \\
$ 4.0 $&$ 0.05 $&$ 0.004 $&$ 0.20 \pm 0.08 ^{+ 0.06}_{- 0.05}$ \\
$ 4.0 $&$ 0.2  $&$ 0.004 $&$ 0.06 \pm 0.02 ^{+ 0.02}_{- 0.02}$ \\
$ 25.0$&$ 0.2  $&$ 0.004 $&$ 0.35 \pm 0.10 ^{+ 0.09}_{- 0.06}$ \\
$ 25.0$&$ 0.5  $&$ 0.004 $&$ 0.15 \pm 0.05 ^{+ 0.04}_{- 0.03}$ \\
\hline
\end{tabular}
\caption{The ratio of $F_2^{D(3),c\bar{c}}$~\protect\cite{charm}
and the present $F_2^{D(3)}$ measurement. The first uncertainty given is 
statistical, the second systematic.
}
\label{charm}
\end{center}
\end{table}

%


\clearpage

\clearpage

%
%
\end{document}